\documentclass[11pt]{article}

\pdfoutput=1

\usepackage[dvipsnames]{xcolor}
\usepackage{amssymb,amsmath,amscd}
\usepackage{epsfig}
\usepackage{epstopdf}
\usepackage{latexsym}
\usepackage{graphicx}
\usepackage{booktabs}
\usepackage{bbm}
\usepackage[margin=15pt,small]{caption}
\usepackage{subcaption}
\usepackage{enumitem}
\usepackage{float}
\usepackage[toc]{appendix}
\usepackage[numbers,compress]{natbib}
\usepackage{tikz}
\usepackage[labelfont=bf]{caption}
\usepackage[export]{adjustbox}

\usepackage{braket,cellspace}
\usepackage{mathrsfs}
\usepackage{braket}
\usepackage{chngpage}

\allowdisplaybreaks

\newcommand{\mcite}[1]{\mbox{\cite{#1}}}

\newcommand*{\colorboxed}{}
\def\colorboxed#1#{\colorboxedAux{#1}}
\newcommand*{\colorboxedAux}[3]
{\begingroup
    \colorlet{cb@saved}{.}
    \color#1{#2}
    \boxed{\color{cb@saved}#3}
  \endgroup}

\usetikzlibrary{positioning}
\definecolor{darkspringgreen}{rgb}{0.09, 0.45, 0.27}

\usepackage{datetime}
\usepackage[
      colorlinks=true,
      linkcolor=darkspringgreen,  
      urlcolor=darkspringgreen,    
      filecolor=darkspringgreen,     
      citecolor=darkspringgreen,
      linktocpage=true,
      pdfstartview=FitV,
      bookmarksopen=true     
      ]{hyperref}

\setlist{nolistsep}
\ifpdf
\fi

\let\oldbibliography\thebibliography
\renewcommand{\thebibliography}[1]{\oldbibliography{#1}
\setlength{\itemsep}{0pt}} 

\numberwithin{equation}{section} 

\begin{document}  

\begin{titlepage}

\begin{center} 

\vspace*{20mm}

{\LARGE \bf 
Phase transitions for deformations 
\\ \vspace{3mm}
of JT supergravity and matrix models
}

\bigskip
\bigskip
\bigskip
\bigskip

{\bf Felipe Rosso${}^{1}$ and Gustavo J. Turiaci${}^{2}$}
\\ 
\vskip 2mm
${}^1$ Department of Physics and Astronomy 
\\
University of British Columbia \\
Vancouver, BC V6T 1Z1, Canada \\
\vskip 2mm
${}^2$ Institute for Advanced Study \\
Princeton, NJ 08540, USA \\
\bigskip
\tt{
feliperosso6@gmail.com,
turiaci@ias.edu}  \\

\end{center}

\bigskip

\begin{abstract}
\noindent 

\noindent We analyze deformations of $\mathcal{N}=1$ Jackiw-Teitelboim (JT) supergravity by adding a gas of defects, equivalent to changing the dilaton potential. We compute the Euclidean partition function in a topological expansion and find that it matches the perturbative expansion of a random matrix model to all orders. The matrix model implements an average over the Hamiltonian of a dual holographic description and provides a stable non-perturbative completion of these theories of $\mathcal{N}=1$ dilaton-supergravity. For some range of deformations, the supergravity spectral density becomes negative, yielding an ill-defined topological expansion. To solve this problem, we use the matrix model description and show the negative spectrum is resolved via a phase transition analogous to the Gross-Witten-Wadia transition. The matrix model contains a rich and novel phase structure that we explore in detail, using both perturbative and non-perturbative techniques.
\end{abstract}

\vfill

\end{titlepage}



\setcounter{tocdepth}{2}

\tableofcontents

\pagebreak

\section{Introduction}
\label{sec:1}

Models of two dimensional dilaton-gravity in asymptotically AdS$_2$~\mcite{Jackiw:1984je,Teitelboim:1983ux,Almheiri:2014cka,Jensen:2016pah, Maldacena:2016upp,Engelsoy:2016xyb} provide a very simple theoretical laboratory to study interesting questions about quantum gravity and black holes, with Jackiw-Teitelboim (JT) gravity as a prototypical example which is, among other things, exactly solvable~\mcite{Bagrets:2016cdf, Stanford:2017thb, Mertens:2017mtv, Lam:2018pvp, Saad:2019lba}. These theories are also relevant in very different contexts. For example, they describe a sector of strongly coupled condensed matter quantum mechanical systems such as SYK~\mcite{PhysRevLett.70.3339, Kitaev:2017awl, Maldacena:2016hyu}, and capture classical and quantum effects for near-extremal black holes in higher dimensions~\mcite{Sachdev:2015efa, Nayak:2018qej,Ghosh:2019rcj, Iliesiu:2020qvm}. They also serve as a fruitful toy model to study the role of spacetime wormholes in quantum gravity~\mcite{Saad:2019lba, Stanford:2019vob}, among other things.

In this paper we study the following problem. In two space-time dimensions, the quantum gravity partition function $Z(\beta)$ at fixed inverse temperature $\beta$ is naturally organized in terms of a topological expansion
\begin{equation}\label{eq:235}
Z(\beta)=\sum_{g=0}^{\infty}(e^{-S_0})^{2g-1}
Z_g(\beta)\ ,
\end{equation}
where~$S_0$ is a non-negative real number. The quantity~$Z_g(\beta)$ can be computed from a Euclidean path integral (with appropriate boundary conditions) that only includes contributions from surfaces of fixed genus~$g$ (thus the term `topological expansion'). For a class of dilaton theories, recent progress has enabled the exact computation of~$Z_g(\beta)$, providing unprecedented control over the partition function. From the inverse Laplace transform of~$Z(\beta)$ one obtains a density of states~$\varrho(E)$ that is expected to capture the microstates of the corresponding black hole solution of the dilaton-gravity theory.\footnote{For systems with finite entropy we would expect this density of states to be a sum over discrete delta functions with integer weights. It is now understood this expectation is too restrictive, since the gravity theory can be dual to an ensemble of theories, such that a continuous spectrum can arise after averaging. This seems to be the generic situation in two dimensions.} The issue we want to explore arises from the fact that for certain theories the spectral density $\varrho(E)$ derived from gravity becomes negative, signaling a breakdown of unitarity of the underlying black hole microstates. The purpose of this paper is to study several examples in which this puzzle arises and show that whenever a negativity of the spectrum appears, the problem is resolved by properly accounting for a phase transition. This resolution is made possible by a non-perturbative completion of the gravity theory via a random matrix model. In the rest of the introduction we describe in more detail the setup we study and summarize the main results. 

Our focus is on an $\mathcal{N}=1$ supersymmetric extension of a dilaton-gravity theory described by the following Euclidean action
\begin{equation}
I[g,\phi]=-S_0\chi(\mathcal{M})-
\frac{1}{2}\int_{\mathcal{M}}d^2x\sqrt{g}\left(\phi R+U(\phi) \right)\ + I_{\rm GHY}\ ,
\end{equation}
where $g_{\mu\nu}$ denotes the metric on the two dimensional space $\mathcal{M}$ and $\phi$ the dilaton field. To have a well defined variational principle one needs to include $I_{\rm GHY}$, the appropriate Gibbons-Hawking-York term. The Euler characteristic $\chi(\mathcal{M})$ is controlled by $S_0$ and gives rise to the topological expansion in (\ref{eq:235}). In this work we consider the following dilaton potential 
\begin{equation}
    U(\phi)=2\phi+2\xi e^{-2\pi(1-\alpha)\phi}\ ,
\end{equation}
which depends on the parameters $(\xi,\alpha)\in \mathbb{R}\times [0,1]$. The discussion can be easily generalized to a sum of multiple such exponentials. 

When $\xi=0$, both the bosonic and $\mathcal{N}=1$ cases correspond to pure Jackiw-Teitelboim (JT) gravity and supergravity respectively. The whole topological expansion of the partition function has been computed exactly \cite{Saad:2019lba,Stanford:2019vob} and the associated spectral density $\varrho(E)$ is positive definite. For the supersymmetric case one needs to distinguish between two distinct types of theories, that we denote as Type 0A and Type 0B.\footnote{These theories can be obtained as limits of the minimal superstring theory (understood as a worldsheet CFT, which is equivalent to a combination of multicritical models)  of Type 0A and 0B, hence the name. Even thought in this context they are usually defined through their GSO projection, an equivalent definition is to determine how the sum over spin structure is performed. See \cite{Johnson:2020heh,Johnson:2020lns} and footnote 79 in Appendix~F of \cite{Stanford:2019vob} for more details on this connection.} This choice appears since, when we sum over topologies, one also has to sum over spin structures. The partition function in Type 0B is defined by summing over all spin structures with equal weight, while in Type 0A one takes the difference between even and odd spin structures (see around above (\ref{eq:29}) for more details on their distinct definitions).

Deformations of JT (super)gravity are obtained by taking $\xi\neq 0$. As shown in \cite{Maxfield:2020ale,Witten:2020wvy}, the effect of  $\xi$ in the dilaton potential can be equivalently incorporated by contributions from surfaces with conical defects in the topological expansion (\ref{eq:235}) of pure JT (super)gravity. In this context, $2\pi \alpha$ is the opening angle of the defect and $\xi$ its weight in the path integral. Building on this observation, the topological expansion of the deformations of bosonic JT gravity was computed in \cite{Maxfield:2020ale,Witten:2020wvy}. The first result of this work is to show how this can be readily extended to obtain the partition function for the deformations of the Type 0A/0B JT supergravity theories.\footnote{Similarly as in \cite{Maxfield:2020ale,Witten:2020wvy}, a crucial step involves relating the volumes of certain moduli space corresponding to surfaces with and without defects via an analytic continuation. While for the bosonic case this has been established in \cite{10.4310/jdg/1143593126}, we conjecture an analogous relation for $\mathcal{N}=1$ and provide evidence by matching to results from random matrix models.} As explained in \cite{Maxfield:2020ale,Witten:2020wvy} the computation performed in this way is sensible only for $\alpha\in(0,1/2)$, corresponding to `sharp' conical defects (we leave the case of `blunt' defects with $\alpha\in[1/2,1]$ for future work, possibly extending the analysis of \cite{Turi}). After the dust settles, one obtains the spectral density from the inverse Laplace transform of (\ref{eq:235}) and finds $\varrho(E)$ is positive definite only when $\xi\ge -1$. The aim of this work is to understand the fate of the quantum theories for $\xi<-1$, which naively have negative $\varrho(E)$.\footnote{There are two types of negativities that appear in the bosonic case. One type appears from including only a single defect and is resolved by summing over the gas of defects. The negativity we are interested in this paper corresponds to one that survives even after summing over defects.} 

To solve the puzzle one needs to have better control of the theory, beyond the topological expansion (\ref{eq:235}). In other words, one has to properly understand non-perturbative effects in the parameter~${e^{-S_0}}$. To our knowledge, there is currently no method for computing such non-perturbative contributions using the gravitational description. Fortunately, a non-perturbative completion for the deformations of Type 0A/0B theories can be constructed using a random matrix model. Consider the following operators
\begin{equation}\label{eq:237}
\begin{aligned}
{\rm Type\,\,0A:}\qquad \qquad 
Z_{{\rm MM}}^{\rm -}(\beta) & \equiv  
2\,{\rm Tr}\,e^{-\beta MM^\dagger}\ ,
\qquad \qquad M\in \mathbb{C}^{N\times N} \ , \\[4pt]
 {\rm Type\,\,0B:}\qquad \qquad 
Z_{{\rm MM}}^{\rm +}(\beta) & \equiv \sqrt{2}\,{\rm Tr}\,e^{-\beta Q^2}\ ,
\qquad \qquad Q^\dagger=Q\in \mathbb{C}^{N\times N}
\ ,
\end{aligned}
\end{equation}
where the index `MM' reminds us these are matrix operators. In each case, the role of the Hamiltonian is played by~$MM^\dagger$ and~$Q^2$, with~$M$ and~$Q$ related to the supercharges. As explained in \cite{Stanford:2019vob}, the different way in which the Type 0A/0B theories sum over spin structures requires a different kind of matrix ensemble:~$M$ an arbitrary complex matrix (non-diagonalizable) and~$Q$ Hermitian (diagonalizable), see above (\ref{eq:236}) for more details. Using the loop equations and a properly defined measure for the ensemble average, we show the equivalence between the average of these operators and the topological expansion of the deformed Type~0A/0B theories to all orders in the perturbative expansion in $e^{-S_0}$. This is the natural extension of the $\xi=0$ matching previously obtained in \cite{Stanford:2019vob} for the undeformed case.\footnote{The non-perturbative completiton of pure JT gravity using matrix models is more subtle, see \cite{Johnson:2019eik,Johnson:2021zuo} and \cite{Gao:2021uro}.}

\begin{figure}
    \centering
    \includegraphics[scale=0.32]{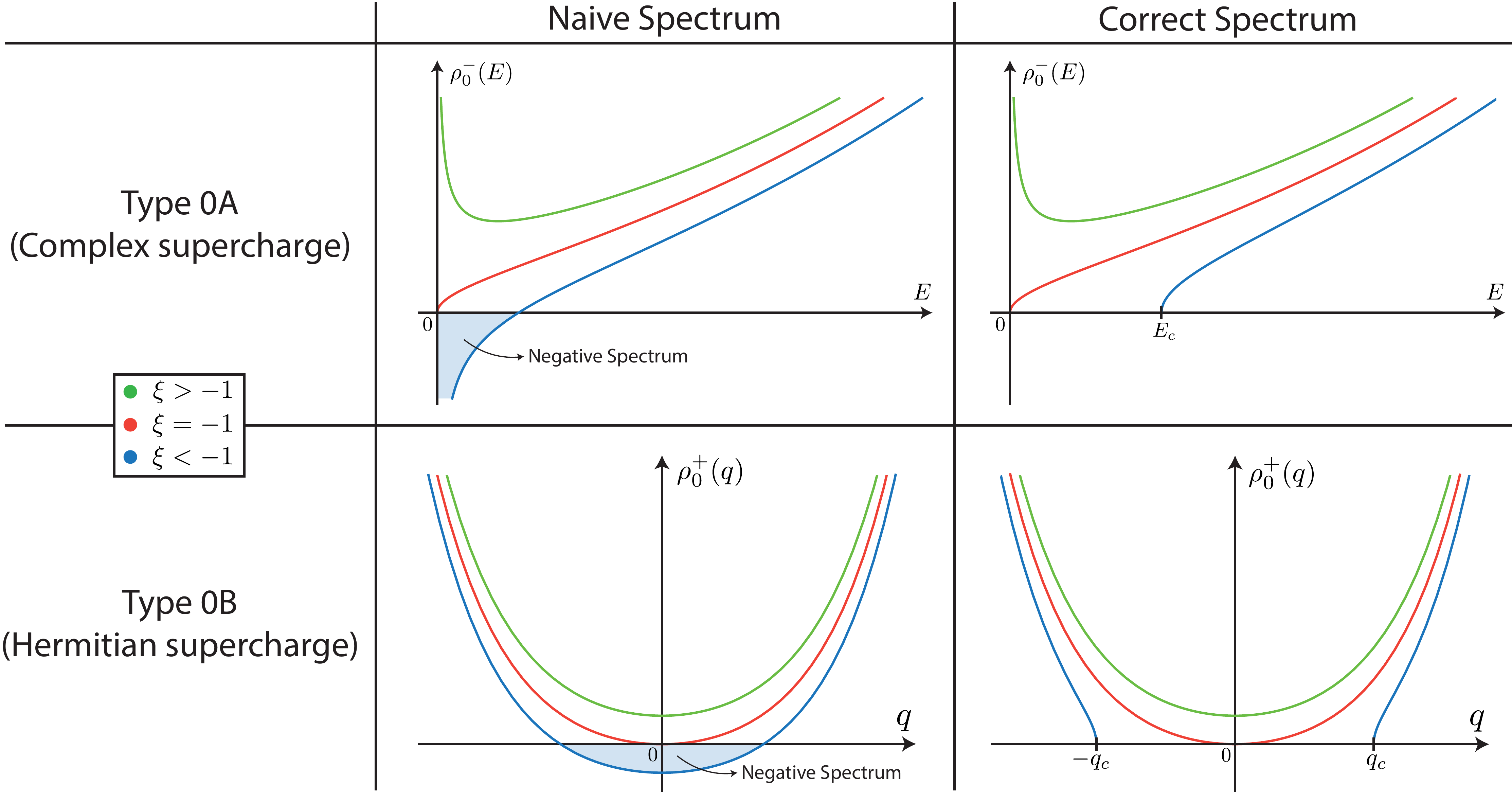}
    \caption{Leading spectral densities $\rho_0^-(E)$ and $\rho_0^+(q)$ of the complex and Hermitian random matrix models which provide a non-perturbative completion of the deformations of Type 0A/0B JT supergravity. The relevant eigenvalues of $MM^\dagger$ and $Q$ are respectively given by $E\ge 0$ and $q\in \mathbb{R}$. The negative spectrum appearing for $\xi<-1$ obtained from a naive computation is cured by a phase transition in the matrix model.}
    \label{fig:21}
\end{figure}

Equipped with the matrix model completion of the deformed Type 0A/0B theories we can properly address the negativity of the spectral density when $\xi<-1$. Non-perturbative effects can be explicitly computed and are well under control in the matrix model description, as already shown in~\mcite{Johnson:2020heh,Johnson:2021owr,Johnson:2020mwi} for the~${\xi=0}$ case. We find that whenever a negativity develops in the spectral density there is a phase transition which resolves the issue in an interesting and non-trivial way. The diagrams in figure~\ref{fig:21} summarize the mechanism for the leading densities of the matrices $MM^\dagger$ and $Q$, with eigenvalues $E\ge 0$ and $q\in \mathbb{R}$ respectively. The equivalence between the supergravity partition function $Z(\beta)$ and the matrix operators (\ref{eq:237}) implies analogous phase transition for the gravity spectral density~$\varrho(E)$ which, as characterized by the gravitational free energy $F(\beta)=-(1/\beta)\ln Z(\beta)$, is of second order.

\begin{figure}
\centering
\includegraphics[scale=0.42]{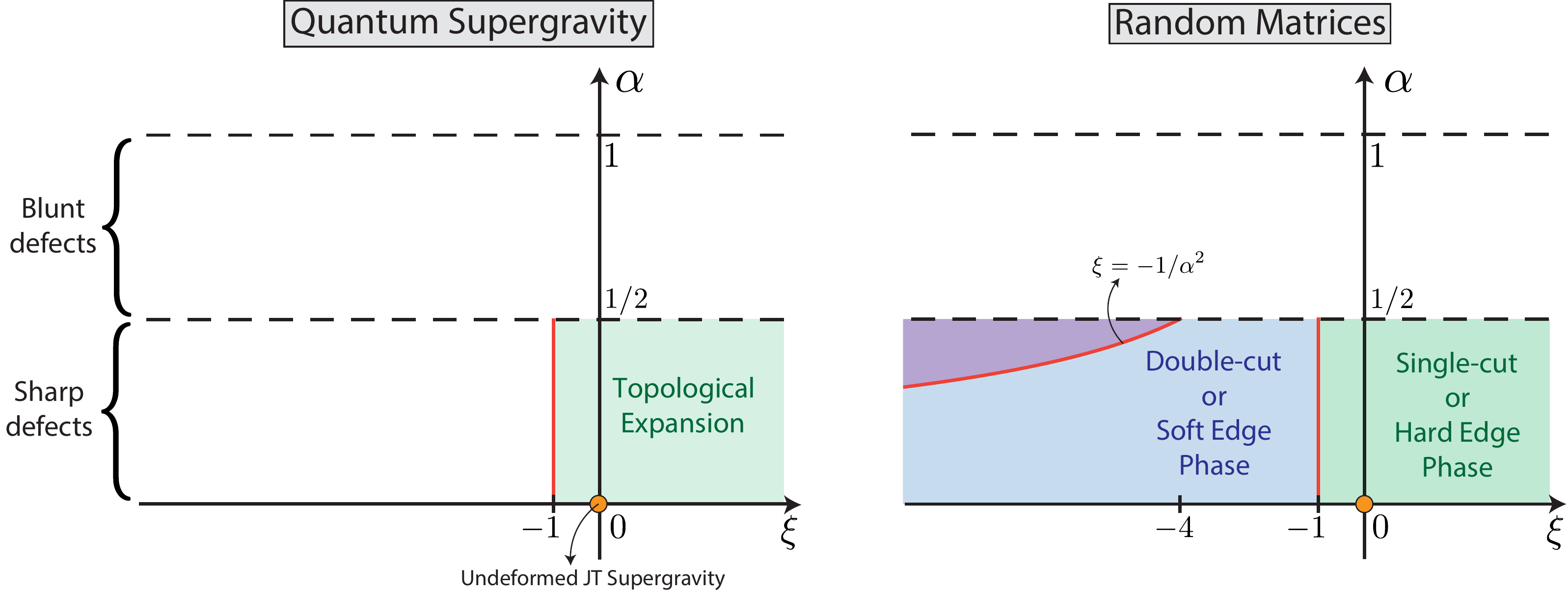}
\caption{The left diagram shows the phase diagram for the deformations of Type 0A/0B JT supergravity theories. The topological expansion is well defined for $\xi>-1$ and $\alpha\in (0,1/2)$. On the right, we show the phase diagram of the random matrix models which provide a non-perturbative completion of the supergravity theories. The matrix model is defined for $\alpha\in(0,1/2)$ and arbitrary values of $\xi \in \mathbb{R}$.}\label{fig:11}
\end{figure}

The transition in the Type 0A theory is from a hard-edge density of states behaving as $~{1/\sqrt{E}}$, to a soft-edge phase with behavior~$\sqrt{E-E_c}$. In the Type 0B case, there is a transition from a single to a double-cut phase, i.e. the supercharge density of states supported on one or two disjoint intervals in~${q\in\mathbb{R}}$. This transition is in the same universality class as the Gross-Witten-Wadia transition for unitary matrices~\cite{Gross:1980he,Wadia:2012fr}. The phase diagram is summarized in figure~\ref{fig:11}. While the supergravity computation only makes sense for~${\xi> -1}$ (left diagram), the non-perturbative completion provided by the matrix model exhibits an interesting and non-trivial phase structure (right diagram). At~${\xi=-1/\alpha^2}$ we discover a transition into a new phase (shaded in purple) which arises from a non-perturbative instability in the matrix model when~${\xi<-1/\alpha^2}$.

This paper is organized as follows. After reviewing in section \ref{sec:2.1} the pure $\mathcal{N}=1$ JT supergravity theories \cite{Stanford:2019vob}, we deform them by adding conical defects and in section \ref{sec:2.2} compute (in a topological expansion) their path integrals with an arbitrary number of boundaries. Following \cite{Stanford:2019vob}, in section \ref{sec:2.3} we appropriately construct a complex and a Hermitian matrix model and use the loop equations to show the average of (\ref{eq:237}) matches the Type~0A/0B partition functions to all orders in perturbation theory. Assuming the non-perturbative completion provided by the matrix models, in section \ref{sec:3} we present the phase transitions which cure the negativity of the spectral density. We carefully and systematically characterize the matrix models, both perturbatively and non-perturbatively in the parameter $e^{-S_0}$. In section \ref{sec:4} we study a novel phase transition at $\xi=-1/\alpha^2$ that arises from a non-perturbative instability in the matrix model when $\xi<-1/\alpha^2$. We finish in section \ref{sec:5} by discussing further questions we are interested in exploring in future work. 

Four appendices contain important technical details used in the main text. In particular, Appendix \ref{zapp:3} contains a detailed description of the double scaling of Hermitian and complex matrix models using the method of orthogonal polynomials, which is the crucial formalism that allows us to capture non-perturbative aspects of the models. Apart from putting together many results scattered in the literature, this Appendix includes some new technical results, like the precise method for computing the matrix model kernel and observables for double scaled and double-cut Hermitian matrix models.

\section{Deformations of JT supergravity and random matrices}
\label{sec:2}

We begin this section with a brief review of $\mathcal{N}=1$ JT supergravity including a sum over topologies, as described in \cite{Stanford:2019vob}. Depending on how the sum over spin structures is performed, one can define two different theories that we call Type 0A and 0B.\footnote{Using these names in the context of JT gravity is not standard practice. We hope the motivation for this convention is clear. The theory we define as Type 0A (0B) JT is a limit of Type 0A (0B) superminimal string, see \cite{Johnson:2020heh,Johnson:2020lns} or Appendix~F of \cite{Stanford:2019vob}.} We then deform these theories by adding a gas of defects, as introduced in \cite{Maxfield:2020ale,Witten:2020wvy} for the bosonic case, and study its properties. Finally, we review how Type 0A (0B) JT supergravities are dual to a complex (Hermitian) random matrix ensemble, and extend this duality to their deformations by these defects.

\subsection{Review: \texorpdfstring{$\mathcal{N}=1$}{N1} JT supergravity}
\label{sec:2.1}

The $\mathcal{N}=1$ JT supergravity theory was first defined in~\mcite{Teitelboim:1983uy,Chamseddine:1991fg} using two-dimensional superspace formalism of~\cite{Howe:1978ia}.\footnote{It can also be defined as a BF theory with supergroup ${\rm OSp}(1|2)$, see \cite{Cardenas:2018krd}.} After integrating over the fermionic superspace variables and solving the equations of motion for the auxiliary fields, one is left with the bosonic fields $(g_{\mu \nu},\phi)$, the metric and dilaton, and their supersymmetric partners $(\psi_\mu^a,\chi^a)$, the gravitino and dilatino, where $\mu,\nu=1,2$ are spacetime indices and $a=1,2$ is a spinor index. The explicit action can be found in, for example, equation~(2.3) of~\cite{Cangemi:1993mj}. After turning off the fermionic fields $\psi_\mu^a,\chi^a\to0$ one is left with the usual JT gravity action~\cite{Teitelboim:1983ux,Jackiw:1984je}
\begin{equation}\label{eq:2}
I_{\rm SJT}
\big|_{{\rm fermions}=0}
=
-\frac{1}{2}\int_{\mathcal{M}}d^2x\sqrt{g}\phi(R+2)\ ,
\end{equation}
where we are omitting (important) boundary terms that ensure a finite on-shell action and a well defined variational problem. Here $\mathcal{M}$ denotes the two dimensional space where the theory lives. The Euclidean path integral is formally defined as\footnote{We add the subscript `SJT' to denote quantities defined for the undeformed JT supergravity theory.}
\begin{equation}\label{eq:1}
Z_{{\rm SJT}}(\beta_1,\dots,\beta_n)=
\int Dg_{\mu \nu}D\phi D\Psi\,e^{-S_0\chi(\mathcal{M})-I_{\rm SJT}[g_{\mu \nu},\phi,\Psi]}\ ,
\end{equation}
where we have allowed for $n$ asymptotic boundaries and all fields apart from the metric and dilaton are indicated by $\Psi$. We make the conventional choice of boundary conditions at each of the $n$ boundaries by fixing the dilaton to $\phi|_{\partial\mathcal{M}}=\gamma/\epsilon$ and the boundary length to be $\beta/\epsilon$, with $\epsilon \to 0$. The parameter $\gamma$ is dimensionful and we can make a choice of units that sets $\gamma=1/2$. The partition function depends only on the renormalized lengths $\beta_i$, $i=1,\ldots,n$. See \cite{Forste:2017kwy,Cardenas:2018krd} for the detailed boundary conditions for the gravitino and dilatino. We have also included the Euler characteristic~$\chi(\mathcal{M})$ which does not modify the equations of motion. Its contribution to the action is controlled by the parameter $S_0$. In (\ref{eq:1}) we integrate over all manifolds $\mathcal{M}$ consistent with the boundary conditions, which in two dimensions are naturally classified in terms of their genus~$g$ and number of boundaries~$n$. Using that the Euler characteristic is $\chi(\mathcal{M})=2(g-1)+n$, one arrives at the following topological expansion
\begin{equation}\label{eq:3}
Z_{{\rm SJT}}(\beta_1,\dots,\beta_n)=
\sum_{g=0}^{\infty}
(e^{-S_0})^{2(g-1)+n}
Z_{{\rm SJT},g}(\beta_1,\dots,\beta_n)\ ,
\end{equation}
where $Z_{{\rm SJT}, g}(\beta_1,\dots,\beta_n)$ is the same path integral as in (\ref{eq:1}) but with the restriction that only manifolds of genus $g$ with $n$ boundaries are included in the integral.\footnote{While one can also include the contribution from unorientable surfaces with half-integer genus, we shall mostly restrict to the orientable case, see Appendix \ref{zapp:2}.} When computing this expansion, we will restrict to contributions which are connected spacetimes. Disconnected geometries can be easily included in the end to obtain the full answer. We will comment below on the issue of how to sum over spin structures 

The crucial feature that simplifies the calculation of the partition function is the path integral over the dilaton field $\phi(x^\mu)$, which appears linearly in the action~(\ref{eq:2}). After rotating the integration contour from the real line to the imaginary direction, one gets a Dirac delta $\delta(R+2)$ which localizes the integral over manifolds with constant negative curvature, making the evaluation of the path integral tractable. 


We begin by describing the evaluation of the path integral on the disk with a single boundary and without handles. In the notation above this evaluates  $Z_{{\rm SJT},0}(\beta)$. After integrating over the dilaton, the only metric that contributes is Euclidean AdS$_2$
\begin{equation}\label{eq:107}
ds^2=\sinh^2(r)d\tau^2+dr^2\ ,
\qquad {\rm where} \qquad
\tau\sim \tau+2\pi\ ,
\end{equation}
and $r\ge 0$. Even though the metric in the bulk localizes to the hyperbolic disk, there is still a non-trivial gravitational dynamics coming from boundary modes \cite{Almheiri:2014cka, Jensen:2016pah, Maldacena:2016upp,Engelsoy:2016xyb}. They can be thought of as associated to the freedom of picking the location of the curve $\partial \mathcal{M}$ consistent with the boundary conditions, and can be parametrized by a boundary proper time $u$ according to
\begin{equation}\label{eq:4}
x^\mu(u)=\big(
\tau(u),r(u)
\big)\ ,
\qquad \qquad
u\in [0,\beta]\ .
\end{equation}
The constraint of fixed renormalized boundary length~$\beta$ determines the function~$r(u)$ in terms of~$\tau(u)$. Inserting this in the JT gravity action with proper boundary terms gives the Schwarzian action for the boundary graviton~$\tau(u)$, as derived for the bosonic case in~\cite{Maldacena:2016upp}. Including the fermionic modes, the partition function~$Z_{{\rm SJT},0}(\beta)$ reduces to a path integral of a supersymmetric quantum mechanical system~\cite{Forste:2017kwy}
\begin{equation}\label{eq:5}
Z_{{\rm SJT},0}(\beta)=
\int 
D\tau D\eta\,
e^{-I_{\rm Sch}^{\mathcal{N}=1}[\tau(u),\eta(u)]}\ ,
\end{equation}
where $\eta(u)$ is a Grassman field, the super-partner of $\tau(u)$, arising from the gravitino. The path integral is performed over ${\rm Diff}(S^{1|1})/{\rm OSp}(1|2)$, where ${\rm Diff}(S^{1|1})$ are the diffeomorphisms of the~${\mathcal{N}=1}$ super-circle and we mod out by the isometries of the space ${\rm OSp}(1|2)$ to avoid overcounting. When computing the partition function we choose anti-periodic boundary conditions on the fermions~${\eta(u+\beta)=-\eta(u)}$.\footnote{We could also compute the partition function with periodic boundary conditions for the fermions. In this case the answer is exactly zero due to fermion zero-modes, unless we deform the theory by adding Ramond punctures. In order to get a non-zero answer for the undeformed theory it is necessary to consider extended supersymmetry.} The action in the exponent is the $\mathcal{N}=1$ super-Schwarzian \cite{Fu:2016vas}. As shown in \cite{Stanford:2017thb}, the integral is one-loop exact and can therefore be computed exactly. A one-loop computation gives the following result
\begin{equation}\label{eq:102}
Z_{\rm Disk}(\beta)= e^{S_0} Z_{{\rm SJT},0}(\beta)=e^{S_0}\sqrt{\frac{2}{\pi \beta}}e^{\pi^2/\beta}\ .
\end{equation}
The exponential term is the classical on-shell action, the same as in bosonic JT gravity. The prefactor~$\beta^{-1/2}$ comes from the one-loop determinant around the classical solution. More specifically, there is a factor of $\beta^{-3/2}$ from removing three bosonic zero-modes and a factor of $\beta^{2/2}$ from removing two fermionic zero-modes, which make up the ${\rm OSp}(1|2)$ isometries. 

From the leading contribution to the one boundary partition function we can extract the leading density of states $\varrho_{{\rm SJT}, 0}(E)$ defined through 
\begin{equation}\label{N1dos}
Z_{{\rm SJT},0}(\beta)=\int_0^{\infty}dE\,\varrho_{{\rm SJT},0}(E)e^{-\beta E} ~~~~~\Rightarrow~~~~\varrho_{{\rm SJT},0}(E)=
\frac{\sqrt{2}\cosh(2\pi\sqrt{E})}{\pi \sqrt{E}}
\ .
\end{equation}
This leading contribution to the density of states is common to both the 0A and 0B supergravity theories, since no sum over spin structures is involved so far. It presents a~$1/\sqrt{E}$ divergence near the edge of the spectrum at~$E=0$. It is instructive to rewrite the density of states as a function of~${q=\pm\sqrt{E}}$
\begin{equation}\label{chargedos}
\widehat{\varrho}_{{\rm SJT},0}(q)
\equiv |q|\,\varrho_{{\rm SJT},0}(E=q^2)
=\sqrt{2} \cosh (2\pi q)/\pi\ ,
\end{equation}
which is smooth at~${q=0}$. While~$E$ can be interpreted as the eigenvalue of the boundary Hamiltonian,~$q$ is the eigenvalue of the boundary supercharge operator \cite{Stanford:2019vob}.\footnote{This interpretation is strictly valid for the Type 0B theory, since for Type 0A the supercharge operator is not diagonalizable.} 

\subsubsection*{Sum over spin structure}
Since the supergravity theory contains fermions, each manifold we sum over in the path integral must be supplemented with a spin structure. One can choose between periodic (Ramond) or anti-periodic (Neveu-Schwarz) boundary conditions for the fermions as they are transported across closed loops in the manifold. Spin structures can be classified in terms of their parity $(-1)^\zeta$, where $\zeta$ is the number of chiral zero modes for a given spin structure, see section 3.2 in \cite{Witten:2015aba} for a precise definition. For instance, a fermion in $S^1$ has a single component and its Dirac equation is given by $\partial_{\varphi} \psi=0$, where~${\varphi \sim \varphi+2\pi}$ parametrizes the circle. While for the Neveu-Schwarz spin structure there is no zero mode $(-1)^\zeta=1$, for Ramond there is one zero mode, meaning $(-1)^\zeta=-1$. 

There are two theories one can define, depending whether or not $(-1)^\zeta$ is included in the path integral. The partition function for fixed genus in each case is schematically given by
\begin{equation}\label{eq:29}
Z_{{\rm SJT}, g}^\pm(\beta_1,\dots,\beta_n)=
({\rm even})\pm ({\rm odd})\ .
\end{equation}
For the disk result (\ref{eq:102}) there is no distinction between the two choices, given there is a single spin structure. However, as shown in \cite{Stanford:2019vob}, the sign difference has deep consequences when studying higher genus contributions. We will refer to the theory without the $(-1)^\zeta$ as Type 0B, computing $Z^+_{\rm SJT}(\beta_1,\dots,\beta_n)$, and the theory with the $(-1)^\zeta$ as Type 0A, computing $Z^-_{\rm SJT}(\beta_1,\dots,\beta_n)$.

\subsubsection*{Topological Expansion}
\begin{figure}
\centering
\includegraphics[scale=0.38]{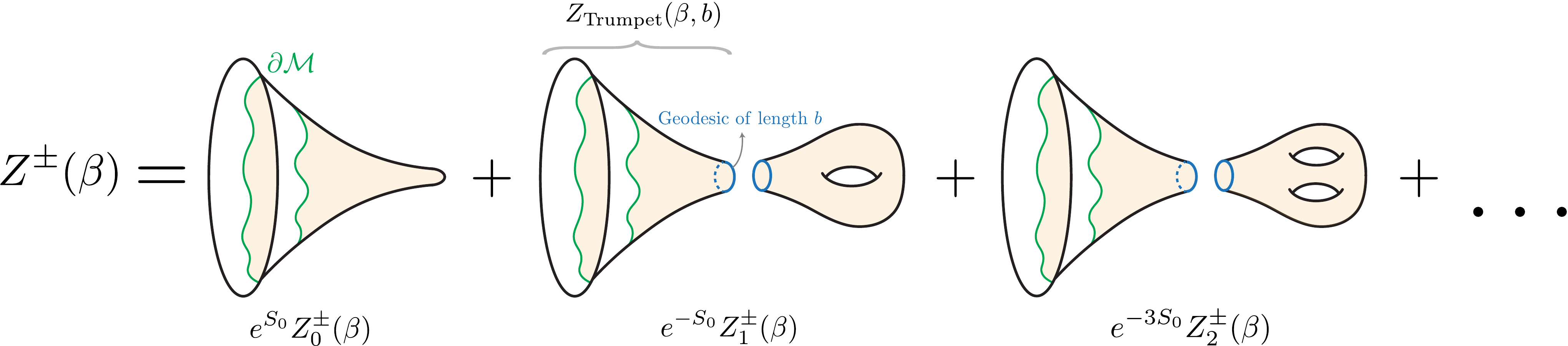}
\caption{Topological expansion developed in \cite{Saad:2019lba,Stanford:2019vob} for the Euclidean partition function of JT supergravity with a single boundary of renormalized length $\beta$, indicated in green. For each of the surfaces shown here one has to sum over the spin structures as indicated in (\ref{eq:29}).}\label{fig:6}
\end{figure}

To calculate $Z_{{\rm SJT},g}^\pm(\beta_1,\dots,\beta_n)$ with $g>0$ or $n>1$ one uses the decomposition of the surfaces developed in \cite{Saad:2019lba} for the bosonic theory and later generalized to the supersymmetric case \cite{Stanford:2019vob}. The first step involves picking a geodesic of length $b$ which separates the asymptotic boundary curve~(\ref{eq:4}) from the rest of the hyperbolic manifold that contributes to the path integral. See figure \ref{fig:6} for a sketch of the decomposition for the single boundary case. The piece that is connected to the boundary $\partial\mathcal{M}$ is called the `trumpet' and its path integral can be computed exactly, similarly as for the disk (\ref{eq:5}). The final result is given by \cite{Stanford:2019vob}
\begin{equation}\label{eq:30}
Z_{\rm Trumpet}(\beta,b)=
\frac{1}{\sqrt{2\pi\beta}}e^{-b^2/4\beta}\ .
\end{equation}
It is independent of how the sum over spin structures is defined, since there is a unique spin-structure on trumpet, that is taken to be Neveu-Schwarz (same as for the disk). 

The contribution coming from the surface to the right of the geodesic in figure \ref{fig:6} only involves an integral over the moduli of genus $g$ surfaces with constant negative curvature and a fixed geodesic boundary $b$, including the sum over spin structures as in (\ref{eq:29}). The result of this integral in each case is denoted as
\begin{equation}\label{eq:31}
V^\pm_{g,n}(b_1,\dots,b_n)\ ,
\end{equation}
and called the Weil-Petersson supervolumes. They are defined and carefully studied in \cite{Stanford:2019vob}.\footnote{The supervolumes are defined with Neveu-Schwarz spin structure at the boundaries $b_i$, see appendices A and D of~\cite{Stanford:2019vob} for further details. For the bosonic theory, the Weil-Petersson volumes are computed to arbitrary orders using a recursion relation derived by Mirzakhani \cite{Mirzakhani:2006fta}. } The complete contribution to the path integral is obtained by ``gluing" (\ref{eq:30}) and (\ref{eq:31}), and then integrating over all possible geodesic lengths $b_i$ with the appropriate measure \cite{Saad:2019lba}
\begin{equation}\label{eq:35}
Z_{{\rm SJT}, g}^\pm(\beta_1,\dots,\beta_n)=
\left[\prod_{i=1}^{n} \int_0^{\infty}db_i\,b_i\,
Z_{\rm Trumpet}(\beta_i,b_i)\right]
V_{g,n}^\pm(b_1,\dots,b_n)\ .
\end{equation}
This formula does not apply for $(g,n)=(0,1)$, where we instead have the disk result (\ref{eq:102}). Another special case is $(g,n)=(0,2)$ where the corresponding supervolume can be formally defined as $V^\pm_{0,2}(b_1,b_2)=2\delta(b_1-b_2)/b_1$ such that $Z_{0}(\beta_1,\beta_2)$ computed by \eqref{eq:35} reproduces the correct answer.\footnote{The factor of two comes from summing over the spin structures of the `double trumpet', that is the only manifold that contributes to $Z_{{\rm SJT},0}^\pm(\beta_1,\beta_2)$. Although both boundaries $\beta_1$ and $\beta_2$ have Neveu-Schwarz boundary conditions for the fermions, when gluing the two trumpets there is a freedom when identifying the fermions in each trumpet~${\psi_{\rm left}^\alpha=\pm \psi_{\rm right}^\alpha}$. Irrespectively of whether we have an insertion of $(-1)^\zeta$ or not, both cases contribute in the same way (see section 2.4.3 in \cite{Stanford:2019vob}). See footnote 56 and 72 of \cite{Stanford:2019vob} for a subtlety in the definition of the volume when~${(g,n)=(1,1)}$.}

After the dust settles, the entire topological expansion is determined by the supervolumes $V^\pm_{g,n}(b_1,\dots,b_n)$. To leading genus it can be shown all the supervolumes vanish, irrespective of how the sum over spin structures is defined
\begin{equation}\label{eq:106}
V^\pm_{g=0,n}(b_1,\dots,b_n)=0\ ,
\qquad n\ge 3\ ,
\end{equation}
see appendix A.3 of \cite{Stanford:2019vob}. Remarkably, this property generalizes to arbitrary genus for the Type 0B theory which sums even and odd spin structures
\begin{equation}\label{eq:108}
V_{g,n}^+(b_1,\dots,b_n)=0\ , \qquad g\ge 1\ .
\end{equation}
This leads to the vanishing of the topological expansion of $Z^+(\beta_1,\dots,\beta_n)$ to all orders, except for the two special cases $(g,n)=(0,1)$ and $(g,n)=(0,2)$.

The higher genus supervolumes do not vanish for the Type 0A theory which takes the difference between even and odd spin structures. As shown in appendix D of~\cite{Stanford:2019vob}, they satisfy the recursion relation~(\ref{eq:34}), that generalizes Mirzakhani's recursion relation for the corresponding bosonic case~\cite{Mirzakhani:2006fta}. The following nice analytic expressions have been worked in~\cite{norbury2020enumerative} for the first few values of $g$
\begin{align}\label{eq:10}
V_{g=1,n}^-(b_1,\dots,b_n) & =
\frac{1}{2}
\frac{(-1)^n(n-1)!}{4}\ , \nonumber\\
V_{g=2,n}^-(b_1,\dots,b_n) & =
3
\frac{(-1)^n(n+1)!}{4^{5}}
\left[
(2\pi)^2(n+2)+
\sum_{i=1}^nb_i^2
\right]\ , \\
V_{g=3,n}^-(b_1,\dots,b_n) & = 
\frac{1}{5}
\frac{(-1)^n(n+3)!}{4^{9}}
\left[
(2\pi)^4(n+4)(42n+185)
+84(2\pi)^2(n+4)\sum_{i=1}^nb_i^2+
\right. \nonumber \\
& \hspace{80	mm} \left. +25 \sum_{i=1}^nb_i^4
+84\sum_{i\neq j}^nb_i^2b_j^2
\right]. \nonumber   
\end{align}
In appendix \ref{zapp:1} we show how the $g=1$ case can be readily derived from the recursion relation satisfied by the supervolumes.

\subsection{Deformations of JT supergravity by a gas of defects}
\label{sec:2.2}

So far we have constrained ourselves to pure Type 0A and 0B $\mathcal{N}=1$ JT supergravity theories which only get contributions from smooth manifolds. We now study deformations of these theories by the inclusion of a gas of defects, as introduced in \cite{Maxfield:2020ale,Witten:2020wvy}. As in the bosonic case, these deformations are parametrized by the opening angle of defect $2\pi \alpha$ and the defect weighting parameter $\xi$. In the bosonic case these deformations can be thought of as arising from changing the dilaton potential in the action. The same interpretation can be given in the supergravity case using the techniques of~\cite{Fan:2021bwt}, although we are not going to pursue it here. 

The rules to compute the gravitational path integral for these deformed theories are clear. For any genus~$g$ in the topological expansion (\ref{eq:3}), one must include the contribution from surfaces with an arbitrary number of defects in the following way\footnote{This is the same approach that was previously developed when computing the JT gravity path integral with conical defects \cite{Maxfield:2020ale,Witten:2020wvy,Turi,Mertens:2019tcm}. The factor $1/k!$ is included to avoid overcounting, assuming the defects are indistinguishable. Compared to \cite{Maxfield:2020ale}, we find it convenient to change the notation to $\lambda\rightarrow \xi$, as $\lambda$ will be used to denote the matrix eigenvalues.}
\begin{equation}\label{eq:37}
Z^{\pm}_{g}(\beta_1,\dots,\beta_n)=\sum_{k=0}^{\infty}
\frac{\xi^k}{k!}
Z_{g,k}^{\pm}(\beta_1,\dots,\beta_n;\alpha)\ ,
\end{equation}
where $\xi \in \mathbb{R}$ is a parameter that controls the weight of the inclusion of each defect in the path integral. Each term $Z_{g,k}^\pm(\beta_1,\dots,\beta_n;\alpha)$ is the partition function including manifolds of genus $g$ and~$k$ conical defects, with the sum over spin structures as in (\ref{eq:29}). This path integral also depends on the opening angle of the defect, which we parametrize as $2\pi \alpha$. More precisely, there is a coordinate patch parametrized by $(\tau,r)$ close to the defect, located at $r=0$ in these coordinates, where the metric takes the form $ds^2 = r^2 d\tau^2 + dr^2 +\ldots$ with $\tau \sim \tau + 2\pi \alpha$. This expansion can be generalized to the case of multiple species of defects in a trivial way. 

To compute $Z_{g,k}^\pm(\beta_1,\dots,\beta_n;\alpha)$ one would like to use the decomposition for the surfaces shown in figure \ref{fig:6}. However, as explained in \cite{Maxfield:2020ale,Witten:2020wvy} when including conical defects the decomposition only works if we restrict to sharp defects, \textit{i.e.} for $\alpha\in(0,1/2)$. This means the opening angle is smaller than $\pi$. If the defects are blunt $\alpha\in[1/2,1)$, or opening angle bigger than $\pi$, there is no geodesic homologous to the asymptotic boundary that one can choose in order to construct the trumpet. For this reason, we restrict our analysis to sharp defects.\footnote{In \cite{Turi}, this obstruction was overcome for JT gravity by taking a different approach that does not require the decomposition of the surfaces according to figure \ref{fig:6}. It would be interesting to explore whether the same methods also apply to the supersymmetric case.}

With this in mind, the genus $g$ contribution (\ref{eq:37}) to the topological expansion (\ref{eq:3}) can be computed as
\begin{equation}\label{eq:39}
\begin{aligned}
Z_{g}^{\pm}(\beta_1,\dots,\beta_n)=
\sum_{k=0}^{\infty}
\frac{\xi^k}{k!}
\left[ 
\prod_{i=1}^n
\int_0^{\infty}db_i\,b_i\,
Z_{\rm Trumpet}(\beta_i,b_i)
\right] 
V_{g,n,k}^{\pm}(b_1,\dots,b_n;\underbrace{\alpha,\dots,\alpha}_{k})\ .
\end{aligned}
\end{equation}
where $V_{g,n,k}^{\pm}(b_1,\dots,b_n;\alpha_1,\dots,\alpha_k)$ is the supervolume as defined for each of Type 0A/0B JT supergravity theories but including $k$ defects with deficit angle $2\pi(1-\alpha)$. Same as before, for the term in~(\ref{eq:39}) with $(g,n,k)=(0,1,0)$ one should instead use the disk partition function (\ref{eq:102}). 

This discussion can be generalized in a straightforward way to the case where we deform JT supergravity by a gas of defects coming in a number $N_F$ of different flavors with their $\{ \alpha_i,\xi_i\}$, with~$i=1,\ldots,N_F$. 

\subsubsection{The disk with defects}

To start, let us compute the leading spectral density $\varrho_0^\pm(E)$, defined from the leading genus $g=0$ and single boundary $n=1$ partition function
\begin{equation}\label{eq:44}
Z_{0}^\pm(\beta)=\int_0^{\infty}dE\,\varrho_0^\pm(E)e^{-\beta E}\ .
\end{equation}
Since we need to sum over multiple number of defects, we might need to sum over spin structures, which will distinguish between deformations of Type 0A or 0B supergravity. From equation (\ref{eq:39}) $Z_{0}^\pm(\beta)$ can be written as
\begin{equation}\label{eq:42}
\begin{aligned}
Z_{0}^{\pm}(\beta)=
Z_{{\rm SJT},0}(\beta)
+
\sum_{k=1}^{\infty}
\frac{\xi^k}{k!}
\int_0^{\infty}db\,b\,
Z_{\rm Trumpet}(\beta,b)
V_{0,1,k}^{\pm}(b ;\underbrace{\alpha,\dots,\alpha}_{k})\ .
\end{aligned}
\end{equation}
This decomposition of space into a trumpet and a hyperbolic surface with cone points, valid for~${k>1}$, does not hold in the case $k=1$. In this case the metric is Euclidean AdS$_2$ with a single conical defect in the center. The metric is 
\begin{equation}\label{eq:40}
ds^2=\sinh^2(r)d\tau^2+dr^2\ ,
\qquad \qquad
\tau\sim \tau+2\pi \alpha\ ,
\end{equation}
where $\alpha\in(0,1)$ controls the angular deficit produced by the defect. When $\alpha\rightarrow 1$ we recover the smooth Euclidean AdS$_2$ metric (\ref{eq:107}). This path integral of JT supergravity can be explicitly computed from a slight variation of the calculation in appendix C of~\cite{Stanford:2019vob}. More details on the description of these defects in JT supergravity can be found in \cite{Fan:2021wsb}. The final result can be written as
\begin{equation}
Z_{0,1}(\beta;\alpha)=
\frac{1}{\sqrt{2\pi \beta}}e^{\pi^2 \alpha^2/\beta}\ .
\end{equation}
The exponential term is the classical action evaluated on the spacetime with one defect. The prefactor is the one-loop determinant. The power of $\beta$ is the same in the case of a disk with and without a defect. In the latter case it comes from removing $3$ and $2$ bosonic and fermionic zero-modes respectively, while in the former case there are no fermionic zero-modes and only one bosonic zero-mode corresponding to rotations around the defect. Comparing with the $k=1$ expression in (\ref{eq:42}) allows us to identify the following formal expression for the $(g,n,k)=(0,1,1)$ supervolume
\begin{equation}
V^\pm_{g=0,n=1,k=1}(b;\alpha)=
\frac{1}{b}\delta(b-2\pi i\alpha)=
\frac{1}{2}V^\pm_{g=0,n=2}(b,2\pi i \alpha)\ .
\end{equation}
This gives an interesting result that relates these (formal) supervolumes with and without defects through an analytic continuation of the geodesic boundary. This trick is not entirely new, as it is familiar to the bosonic Weil-Petersson volumes. In that case, the corresponding volumes with defects have been proven to be recovered from the ordinary volumes via the analytic continuation $b\rightarrow 2\pi i\alpha$ for arbitrary genus and number of boundaries \cite{10.4310/jdg/1143593126}. The analogous relation for the supervolumes would be given by
\begin{equation}\label{eq:46}
V^\pm_{g,n,k}(b_1,\dots,b_n;\alpha_1,\dots,\alpha_k)=
\frac{1}{2^k}
V^\pm_{g,n+k}(b_1,\dots,b_n,2\pi i\alpha_1,\dots,2\pi i \alpha_k)\ .
\end{equation}
The extra factor of $2^k$ compared to the bosonic counterpart appears to avoid an overcounting of spin structures. In section 3 of \cite{Witten:2020wvy} the relation between the analytic continuation of geodesic boundaries and conical defects was already established from the point of view of the monodromy for a theory with both bosons and fermions. In this work, we conjecture (\ref{eq:46}) and provide evidence for it by comparing and matching with results from random matrix models. 

Using \eqref{eq:46}, higher $k$ contributions to the leading genus partition function in (\ref{eq:42}) are determined by the supervolumes $V^\pm_{g=0,1+k}$. However, from (\ref{eq:106}) we know they vanish whenever $k\ge 2$. As a result, the infinite series over the number of defects in (\ref{eq:39}) collapses and only gets contributions two terms, the disk with either none or one defects
\begin{equation}\label{eq:105}
Z^\pm_{0}(\beta)=Z_{{\rm SJT},0}(\beta)
+\xi Z_{0,1}(\beta,\alpha)=
\sqrt{\frac{2}{\pi \beta}}
e^{\pi^2/\beta}+\xi 
\frac{1}{\sqrt{2\pi \beta}}e^{\pi^2\alpha^2/\beta}\ .
\end{equation}
This is exact in $\xi$, in contrast with the bosonic counterpart for which generically one needs to sum over an arbitrary number of defects even to leading order in genus expansion \cite{Maxfield:2020ale,Witten:2020wvy}. To simplify expressions below from now on we will rescale the defect weight by $\xi \to 2 \xi$ (this will also remove the extra factors of $2$ in \eqref{eq:46} when considering higher genus). Taking the inverse Laplace transform of this result, the leading spectral density (\ref{eq:44}) is readily computed
\begin{equation}\label{eq:47}
\begin{aligned}
\varrho_0^\pm(E)&=
\frac{\sqrt{2}
\cosh(2\pi\sqrt{E})}{\pi \sqrt{E}}+\xi \frac{\sqrt{2}
\cosh(2\pi\alpha\sqrt{E})}{\pi \sqrt{E}}\ ,\\
&=
\sqrt{\frac{2}{\pi^2E}}
\left(
1+\xi
\right)
+\mathcal{O}(\sqrt{E})\ .
\end{aligned}
\end{equation}
At large energies, the contribution from JT supergravity always dominates since ${0<\alpha<1/2<1}$. From the low energy expansion we observe two interesting features. First, both the density of states coming from the hyperbolic disk with and without the defect have a $1/\sqrt{E}$ divergence at low energies. This is in contrast with the bosonic counterpart where the leading behavior without defects goes as $\sqrt{E}$, and therefore the contribution from one defect can become arbitrarily large at low energies regardless of how small $\xi$ is chosen to be. In the bosonic theory the resummation of the gas of defects results again in a square root edge, but with a shifted ground state energy. In the supergravity case this is not a problem since both contributions go as $1/\sqrt{E}$. The second feature, which will be important later, is that this gravitational computation breaks down when $\xi<-1$, given that $\varrho^\pm_0(E)$ stops being positive definite. Later we will explain how this issue is resolved. In order to do this we first show an equivalence between these theories of gravity and random matrix models, and then show that when $\xi<-1$ the matrix model presents a phase transition to another phase free of pathologies. 

It is easy to extend this result to multiple defect species. Since only configurations with one defect contribute at genus zero for the disk, the only modification in \eqref{eq:47} is to replace the second term by a sum over contributions from the multiple species. For a set of $\{\alpha_i,\xi_i\}$ species the final answer is 
\begin{equation}\label{N1dosdef}
\varrho_0^\pm(E)=
\frac{\sqrt{2}
\cosh(2\pi\sqrt{E})}{\pi \sqrt{E}}+\sum_i \xi_i \frac{\sqrt{2}
\cosh(2\pi\alpha_i\sqrt{E})}{\pi \sqrt{E}}.
\end{equation}
This formula is valid for any $\xi_i$ as long as $\alpha_i\in(0,1/2)$. We would like to stress that this is exact (at genus zero). This is different from the bosonic case where an analogous formula which truncates to linear order in $\xi$ is only valid for special choices of parameters, as pointed out in \cite{Witten:2020wvy} (see also section 3 of \cite{Johnson:2020lns}).

Before moving on to higher genus contributions we compute the $g=0$ partition function with a higher number of boundaries. After using (\ref{eq:39}), (\ref{eq:46}) and (\ref{eq:106}) one easily obtains
\begin{equation}\label{eq:71}
Z^\pm_{0}(\beta_1,\beta_2)=
\frac{4}{2\pi}\frac{\sqrt{\beta_1\beta_2}}{\beta_1+\beta_2}\ ,
\qquad \qquad
Z^\pm_{0}(\beta_1,\dots,\beta_n)=0\ ,
\qquad n\ge 3\ .
\end{equation}
We note these are the same results one finds for the two JT supergravity theories without any defects.

\subsubsection{Higher genus}

Higher genus contributions to the partition function depend on how the sum over spin structures~(\ref{eq:29}) is defined. If we sum over both even and odd spin structures with the same sign, meaning we focus on Type 0B supergravity, the vanishing of the supervolumes for all~$g\ge 1$~(\ref{eq:108}) together with~(\ref{eq:46}) means the infinite series~(\ref{eq:39}) over the number of defects identically vanishes. The only two cases with non-zero contributions are the cases~$(g,n)=(0,1)$ and~$(g,n)=(0,2)$. Putting everything together, we find
\begin{equation}\label{eq:55}
\begin{aligned}
Z^+(\beta)& \simeq 
e^{S_0}\sqrt{\frac{2}{\pi \beta}}
\left[
e^{\pi^2/\beta}+\xi 
e^{\pi^2\alpha^2/\beta}
\right]\ , \\ 
Z^+(\beta_1,\beta_2)& \simeq
\frac{4}{2\pi}
\frac{\sqrt{\beta_1\beta_2}}{\beta_1+\beta_2}\ , \\
Z^+(\beta_1,\dots,\beta_n)& \simeq  0\ ,
\end{aligned}
\end{equation}
where $\simeq$ means the results hold to all orders in perturbation theory with respect to $e^{-S_0}$ (we will analyze non-perturbative effects later after showing the relation with the random matrix model). For this theory, the only correction generated by the defects is to the leading genus and single boundary partition function. We will see in the next section that this simplification does not hold in other phases of the theory.

For the theory which takes the difference between even and odd spin structures (\ref{eq:29}), which we call Type 0A supergravity, the higher genus supervolumes are non-trivial. The explicit expressions given in (\ref{eq:10}) are particularly useful for computing the infinite sum over defects in (\ref{eq:39}) for the first few values of $g$. As a warm up, let us consider the $(g,n)=(1,1)$ case which can be written as~(\ref{eq:39})
\begin{equation}
\begin{aligned}
Z_{1}^-(\beta) & =
\sum_{k=0}^{\infty}
\frac{(2\xi)^k}{k!}
\int_0^{\infty}db\,b\,
Z_{\rm Trumpet}(\beta,b)
V_{g=1,n=1,k}^{\pm}(b;\underbrace{\alpha,\dots,\alpha}_{k{-}{\rm times}}) \\
& =
\sum_{k=0}^{\infty}
\frac{\xi^k}{k!}
\int_0^{\infty}db\,b\,
Z_{\rm Trumpet}(\beta,b)
V_{g=1,n=1+k}^{\pm}(b,2\pi i \alpha,\dots,2\pi i \alpha) \\
& =-
\frac{1}{8}
\sum_{k=0}^{\infty}
\left(-\xi\right)^k
\int_0^{\infty}db\,b\,
Z_{\rm Trumpet}(\beta,b)=
\frac{-1}{8(1+\xi)}
\sqrt{\frac{2\beta}{\pi}}\ ,
\end{aligned}
\end{equation}
where in the second line we used the analytic continuation (\ref{eq:46}). In the last line we used the genus one volume given in (\ref{eq:10}) and solved the infinite series over the defects. While the geometric series only converges for $|\xi|<1$, after the sum is performed it becomes well defined in an extended domain of $\xi$. Requiring the domain is connected to the origin along the real line, the $g=1$ partition function is well defined for $\xi>-1$. This precisely agrees with the region in which the leading spectral density~(\ref{eq:47}) is positive definite.

Using the same procedure one can use (\ref{eq:10}) to derive the $g=1,2,3$ results with an arbitrary number of boundaries. A careful calculation gives
\begin{align}\label{eq:48}
Z_{1}^{-}(\beta_1,\dots,\beta_n) &  = 
\frac{1}{2}
\frac{(-1)^n(n-1)!}{4(1+\xi)^n}
\prod_{j=1}^n
\sqrt{\frac{2\beta_j}{\pi}}\nonumber \\[4pt]
Z_{2}^-(\beta_1,\dots,\beta_n) & =
3
\frac{(-1)^n(n+1)!}{4^{4}(1+\xi)^{n+3}}
\left[
(1+\xi)
\sum_{i=1}^n \beta_i+
\pi^2(n+2)(1+\xi \alpha^2)
\right]\prod_{j=1}^n
\sqrt{\frac{2\beta_j}{\pi}}
 \nonumber\\[4pt]
Z_{3}^-(\beta_1,\dots,\beta_n) &  = 
\frac{1}{5}
\frac{(-1)^n(n+3)!}{4^{7}(1+\xi)^{n+6}}
\prod_{j=1}^n\sqrt{\frac{2\beta_j}{\pi}}
\bigg[
(1+\xi)^2
\bigg(
50
\sum_{i=1}^n\beta_i^2
+84
\sum_{i\neq j}\beta_i\beta_j
\bigg)
\\
&+84\pi^2(1+\xi)(1+\xi\alpha^2)(n+4)\sum_{i=1}^n\beta_i
+\pi^4(n+4)
\bigg(
42(n+5)(1+\xi \alpha^2)^2
\nonumber \\
& \qquad \qquad \qquad \qquad \qquad \qquad \qquad \qquad \qquad \qquad \qquad
-25(1+\xi)(1+\xi \alpha^4)
\bigg)
\bigg] \nonumber
\end{align}
Each of these expressions, for fixed $g$ and $n$ is obtained by solving the infinite series over the number of defects $k$ in (\ref{eq:39}), assuming the analytic continuation (\ref{eq:46}) for the supervolumes. In every case the infinite series gives a factor of $(1+\xi)^{-\#}$ which diverges as $\xi\rightarrow -1$, yielding the perturbative expansion ill defined. In the following sections we shall use the matrix model description to show the divergence is not real physics but merely a breakdown of the perturbative expansion close to $\xi=-1$ due to a phase transition. More precisely, we shall see that after including non-perturbative effects (as defined by the matrix model), the full partition function $Z^-(\beta)$ is no longer divergent at $\xi=-1$.

\subsection{Constructing the dual random matrix models}
\label{sec:2.3}

According to holography, we expect theories of gravity in asymptotically AdS$_2$ spacetimes to be related to a quantum mechanical (QM) system living in the one dimensional boundary. It was realized in \cite{Saad:2019lba} that for the case of pure bosonic JT gravity, the theory is not dual to a single QM system, but to an ensemble where one averages over Hamiltonians with a particular measure, using random matrix model techniques. The Hamiltonian acts on a Hilbert space of dimension $N$, which in the double scaling limit is related to the parameter $S_0$ in gravity. This was later extended to the case of deformations of pure JT gravity by adding a gas of sharp defects \cite{Maxfield:2020ale,Witten:2020wvy} and \cite{Turi} for general defects. A stable non-perturbative completion of pure JT gravity was constructed in \cite{Johnson:2019eik} and later extended to the case with defects in \cite{Johnson:2020lns}.

Similarly, the theories of pure $\mathcal{N}=1$ supergravity we study here are dual to an ensemble of supersymmetric QM systems. This means there exists a Hermitian operator $Q$ acting the Hilbert space, such that the Hamiltonian is $H=Q^2$. This duality was worked out in \mcite{Stanford:2019vob} for the case of pure JT supergravity. We first summarize their results. There are two theories of supergravity depending on how one sums over spin structures, and they are dual to different matrix ensembles. When we sum over spin structures (Type 0B) we expect the boundary theory not to have any $(-1)^F$ symmetry, making $Q$ a random $N\times N$ Hermitian matrix described by the Dyson ensemble, and $H=Q^2$. On the other hand, when we include the topological term $(-1)^\zeta$ (Type 0A) we expect the dual theory to have a $(-1)^F$ symmetry. The Hilbert space can be decomposed in two $N$-dimensional blocks with~$(-1)^F=1$ or $-1$. The supercharge now acts on these blocks as 
\begin{equation}\label{eq:236}
 Q=\bigg(\,
 \begin{matrix}
0 & M^\dagger \\
M & 0 
\end{matrix}\,
\bigg)\ ,
\end{equation}
where $M$ is an arbitrary $N\times N$ complex matrix described by the $(\boldsymbol \alpha,\boldsymbol \beta)=(1,2)$ Altland-Zirnbauer ensemble, and now on each block the Hamiltonian acts as $H=MM^\dagger$. A non-perturbative formulation of these matrix models in the context of JT supergravity was then given in \cite{Johnson:2020heh,Johnson:2021owr}.

The supergravity path integral depends on the renormalized length $\beta$ of each boundary. In the holographically dual description this corresponds to inserting a quantum mechanical partition function. For each of the theories above, this corresponds to
\begin{equation}\label{eq:61}
\begin{aligned}
Z_{{\rm MM}}^+(\beta) & =\sqrt{2}\,{\rm Tr}\,e^{-\beta Q^2}\ ,
\qquad \qquad Q^\dagger=Q\in \mathbb{C}^{N\times N} \ , \\
Z_{{\rm MM}}^-(\beta) & = 
2\,{\rm Tr}\,e^{-\beta MM^\dagger}\ ,
\qquad \qquad M\in \mathbb{C}^{N\times N}\ ,
\end{aligned}
\end{equation}
where the index `MM' indicates this is a matrix operator. The origin of the additional factors of~$\sqrt{2}$ and~$2$ that are required in each case are explained in~\cite{Stanford:2019vob}. To match with supergravity we need to insert one factor~$Z_{{\rm MM}}^\pm$ for each boundary, and finally average over either~$Q$ or~$M$ with a specific probability distribution. As in the bosonic case, the dimension~$N$ is related to~$S_0$ after a suitable double scaling limit. In this limit, the 't Hooft expansion of the matrix model corresponds to the topological expansion in gravity. 

In this section we use the loop equations of each matrix model to show how this result can be extended beyond zero $\xi$, i.e. to the supergravity theories with defects. The loop equations are a powerful method for computing ensemble averages of a matrix model in an asymptotic expansion in~${1/N}$. The approach, first introduced by Migdal \cite{Migdal:1984gj} and further developed in \cite{Ambjorn:1992gw,Ambjorn:1992xu}, was solved to all orders for a Hermitian random matrix by Eynard \cite{Eynard:2004mh}. Here, we follow the discussion in \cite{Stanford:2019vob}, which extended the full analysis to the other random matrix ensembles.

\subsubsection{Type 0B: Hermitian Ensemble}
In this section we consider Type 0B supergravity, which is dual to a Hermitian ensemble for the supercharge $Q$. We will begin by reviewing basic facts about the Dyson ensemble of Hermitian matrices before describing in detail the connection with gravity, since they will be useful for the discussion of phase transitions in the next section.  

For reasons explained above, we are interested in computing averages over $Q$ of $n$ products of partition functions, which we define as  
\begin{equation}\label{eq:56}
\langle Z_{{\rm MM}}^+(\beta_1)\ldots Z_{{\rm MM}}^+(\beta_n) \rangle \equiv 
\frac{2^{\frac{n}{2}}}{\mathcal{Z}}
\int dQ\,
e^{-N\,{\rm Tr}\,V(Q)}
\,{\rm Tr}\,e^{-\beta_1 Q^2}\ldots {\rm Tr}\,e^{-\beta_n Q^2} ,
\end{equation}
where $dQ$ is the $U(N)$ invariant measure and $V(Q)$ is a potential which determines a particular probability distribution for the supercharge. The matrix partition function $\mathcal{Z}$ is the normalization of this probability distribution and its given by
\begin{equation}\label{eq:115}
\mathcal{Z}=\int dQ\,e^{-N\,{\rm Tr}\,V(Q)}=
\prod_{i=1}^N
\int_{-\infty}^{+\infty}
dq_i
\Delta(q_1,\dots,q_N)^2e^{-NV(q_i)}\ ,
\end{equation}
where in the second equality we have written the integral in terms of the eigenvalues of the supercharge $q_i$.\footnote{There is a proportionality factor relating the expressions in the last equality, coming from the integral over the unitary matrix required to diagonalize $Q$. This additional constant is inconsequential when computing expectation values of observables.} The factor $\Delta(q_1,\dots,q_N)=\prod_{i<j}(q_i-q_j)$ is the Vandermonde determinant arising from the Jacobian associated to the change of variables. We restrict ourselves to ensembles with probability distributions that depend only on eigenvalues, and not on eigenvectors, since that is all we will need to describe the theories of gravity we are interested in.   

\paragraph{Topological Expansion:} To compare with gravity we will take the double scaling limit, which involves also a large $N$ limit. Two observables that will play a central role when discussing the large~$N$ limit are the eigenvalue spectral density and resolvent, respectively given by
\begin{equation}\label{eq:52}
\rho^+(q)=\frac{1}{N}{\rm Tr}\,\delta(q-Q)\ ,
\qquad \qquad
W^+(z)={\rm Tr}\,\frac{1}{z-Q}\ .
\end{equation}
This is the spectral density of the supercharge eigenvalues, not the Hamiltonian. To simplify the expressions below define the product of observables, for example resolvants, by the following notation~${W^+(I)=\prod_{i=1}^nW^+(z_i)}$ where $I=\lbrace z_1,\dots,z_n \rbrace$. At large $N$, the expectation value of $W^+(I)$, defined by an expression analogous to \eqref{eq:56}, can be written in a perturbative expansion in $1/N$
\begin{equation}\label{eq:51}
\langle W^+(I) \rangle_c=\sum_{g=0}^{\infty}
\frac{W^+_{g}(I)}{N^{2(g-1)+|I|}}\ ,
\end{equation}
where $|I|=n$ is the number of elements in $I$ and the subscript $c$ indicates the connected expectation value. The parameter $g$ can be interpreted as the genus of a surface triangulated by the matrix Feynman graphs. The loop equations provide a recursion relation that allows us to compute higher genus contributions in terms of lower ones. They are derived from the following trivial identity
\begin{equation}\label{eq:53}
\int_{-\infty}^{+\infty}
dq_1 \dots dq_N
\frac{\partial }{\partial q_a}
\left[
\frac{W^+(I)}{z-q_a}
\Delta(q_1,\dots,q_N)^2e^{-N\sum_{i=1}^NV(q_i)}
\right]=0\ ,
\end{equation}
which give a set of closed recursion relations that determine all the coefficients $W_g(I)$ in (\ref{eq:51}). In the following we describe the recursion relations obtained from (\ref{eq:53}), see section 4.1 of \cite{Stanford:2019vob} for details.

\paragraph{Spectral curve:} For the case with $g=0$ and $I=\lbrace z \rbrace$ the solution to the loop equations gives
\begin{equation}\label{eq:119}
W_0^+(z)=\frac{1}{2}\left(
V'(z)-h(z)\sqrt{\sigma(z)}
\right)\  ,
\end{equation}
where $h(z)$ and $\sigma(z)$ are analytic functions. The function $\sigma(z)$ is an even polynomial with simple roots. Given the potential $V(z)$, the functions $h(z)$ and $\sigma(z)$ can be computed from the knowledge of the analytic structure of $W^+_0(z)$. In the large $N$ limit the singularities of the resolvent $W^+(z)$ at the spectrum of the matrix $Q$ condense into a branch-cut square root singularity, going between the branch points at the roots of $\sigma(z)$.

The large $N$ behavior of the eigenvalue spectral density (\ref{eq:52}) can be easily obtained from the discontinuity of $W^+_0(z)$ in the complex plane
\begin{equation}\label{eq:58}
\rho^+_{0}(q)=
\frac{W^+_0(q-i\epsilon)-W^+_0(q+i\epsilon)}{2\pi i}
=\frac{1}{2\pi}|h(q)|\sqrt{-\sigma(q)}
\times 
\textbf{1}_{\sigma(q)<0}\ ,
\end{equation} 
where $\epsilon\rightarrow 0^+$ and $\textbf{1}_{\sigma(q)<0}$ is the indicator function. Writing $\sigma(z)=\prod_{i=1}^p(z-a_i)(z-b_i)$ with $a_i,b_i\in \mathbb{R}$, the model is said to be in a single or multi-cut phase depending on whether $p=1$ or $p\ge 2$. For the single-cut case, given a potential $V(z)$, the functions $h(z)$ and $\sigma(z)$ are easily determined by requiring the asymptotic behavior $W_0^+(z)=1/z+\mathcal{O}(1/z^2)$, which follows from its definition (\ref{eq:52}). 

A central quantity that is closely related to the eigenvalue spectral density is the spectral curve~$y(z)$, defined as
\begin{equation}\label{eq:117}
y(z)^2=
\frac{1}{4}h(z)^2\sigma(z)=
\left( W^+_0(z)-\frac{1}{2}V'(z)\right)^2
\qquad \Longrightarrow \qquad
\rho^+_0(q)=\pm\frac{i}{\pi}y(q \pm i \epsilon)\ .
\end{equation}
The spectral curve is defined on a two-sheeted Riemann surface, corresponding to the two possible signs of the square root. If we denote $\hat{z}$ as the same point as $z$ but in the second sheet, we have~${h(\hat{z})=h(z)}$ and $\sqrt{\sigma(\hat{z})}=-\sqrt{\sigma(z)}$. As we shall briefly recall, the whole perturbative expansion of~$W^+(I)$ can be fixed in terms of the spectral curve $y(z)$. For this reason, a perturbative definition of the matrix model can be given directly in terms of $y(z)$ instead of the potential $V(z)$. 

\paragraph{A universal observable:} Consider now the case in (\ref{eq:51}) with $g=0$ and $I=\lbrace z_1,z_2 \rbrace$. For a matrix model in the single-cut phase $\sigma(z)=(z-a)(z-b)$ one finds the following simple result
\begin{equation}\label{eq:59}
W^+_0(z_1,z_2)=
\frac{1}{2(z_1-z_2)^2}\left[
\frac{ab+z_1z_2-(a+b)(z_1+z_2)/2}{\sqrt{\sigma(z_1)}\sqrt{\sigma(z_2)}}-1
\right]\ .
\end{equation}
We say this is a universal observable given that it does not depend on the details of the spectral curve $y(z)$ or potential $V(z)$, but only on the endpoints $a$ and $b$ of the eigenvalue distribution (and of course implicitly on the type of matrix ensemble). It is easy to show the value of $W_0(z_1,z_2)$ as one coordinate goes across the branch cut is determined from the following relation
\begin{equation}\label{eq:63}
W^+_0(\hat{z}_1,z_2)+W^+_0(z_1,z_2)=
\frac{-1}{(z_1-z_2)^2}
\ .
\end{equation}


\paragraph{General recursion relation:} All the remaining terms in the expansion (\ref{eq:51}) are determined from the following recursion relation
\begin{equation}\label{eq:64}
W^+_g(z,I)=\frac{1}{\sqrt{\sigma(z)}}
\sum_{\lbrace a_i,b_i \rbrace}
{\rm Res}\left[
\frac{\sqrt{\sigma(z')}}{2y(z')}
\frac{F_g(z',I)}{z'-z},z'=\lbrace a_i,b_i \rbrace
\right]\ ,
\end{equation}
where the sum is over the residues at each endpoint $\lbrace a_i,b_i \rbrace$ of the leading eigenvalue distribution~$\rho_0^+(q)$ in (\ref{eq:58}). The function $F_g(z',I)$ is given by
\begin{equation}\label{eq:54}
\begin{aligned}
F_g(z',I)=W^+_{g-1}(z',z',I)&
+\sum_{h,J}'
W^+_h(z',J)W^+_{g-h}(z',I\setminus J)+  \\
& \qquad \qquad
+
\sum_{k=1}^{|I|}
\left[
2W^+_0(z',z_k)+\frac{1}{(z'-z_k)^2}
\right]W^+_g(z',I\setminus z_k)
\ ,
\end{aligned}
\end{equation}
except for the special case
\begin{equation}\label{eq:65}
F_0(z',z_1,z_2)=
2W^+_0(z',z_1)W^+_0(z',z_2)+
\frac{W^+_0(z',z_1)}{(z'-z_2)^2}
+\frac{W^+_0(z',z_2)}{(z'-z_1)^2}\ .
\end{equation}
The sum in the first line (\ref{eq:54}) is over all subsets $J\subseteq I$ and $h=0,\dots,g$ that do not contain a factor of $W^+_0(z)$ or $W_0^+(z_1,z_2)$. Given a spectral curve $y(z)$ it is straightforward to use these relations to compute the perturbative expansion (\ref{eq:51}) to any desired order.

\paragraph{Supergravity matching:} Let us now show the expectation value in \eqref{eq:56} in this Hermitian matrix model matches with the topological expansion of the supergravity partition function (\ref{eq:55}) to all orders. To do this, we need to take the double scaling limit of the matrix model. From the perspective of the loop equations this is quite simple, as it corresponds to taking the limit in which an endpoint of the eigenvalue spectral density goes to infinity. One of the consequences of this is that the spectral density $\rho^+(q)$ is no longer normalizable. Introducing a small parameter $\hbar$ and scaling the spectral density according to $\rho^+(q)\rightarrow \rho^+(q)/\hbar$, the $1/N$ expansion is replaced by an $\hbar$ expansion\footnote{By $\hbar$ here we do not mean the actual Planck constant, but a small parameter that controls the expansion of observables. The reason for calling this $\hbar$ will become clear in the following section.}
\begin{equation}\label{eq:101}
\langle W^+(I) \rangle_c=
\sum_{g=0}^{\infty}
\hbar^{2(g-1)+|I|}W^+_g(I)\ .
\end{equation}
Apart from this difference, the perturbative expansion of a double scaled model works in the same way as for an ordinary model with no double scaling. 
The first trivial observation is that after identifying $\hbar=e^{-S_0}$, the topological expansion of the matrix model is the same as the one appearing in JT supergravity and also its deformations.

As previously mentioned, instead of specifying the potential $V(z)$ in (\ref{eq:56}), a perturbative definition of the matrix model can be given in terms of the spectral curve $y(z)$, or equivalently, the leading eigenvalue spectral density $\rho^+_0(q)$ (\ref{eq:117}). To describe JT supergravity we consider a double scaled model where the parameter $q$ is interpreted as the eigenvalue of the supercharge operator. Therefore we should pick the matrix ensemble such that the disk density of states matches with~\eqref{chargedos}, or equivalently with~\eqref{N1dos} after writing it in terms of~$E=q^2$. Generalizing to the case of supergravity deformed by a gas of defects imposes the following identifications
\begin{eqnarray}\label{eq:62}
\rho_0^+(q)=
\frac{1}{\pi}
\big(
\cosh(2\pi q)+\xi \cosh(2\pi\alpha q)
\big)\ .
\end{eqnarray}
For $\xi=0$ this agrees with the model used in \cite{Stanford:2019vob} to describe JT supergravity without defects. We should think of this spectral density as having a branch-cut along the whole real line, arising from a single-cut model (\ref{eq:58}) with $\sigma(z)=(z-a)(z-b)$ in the limit in which both endpoints go to infinity as $a\to -\infty$ and $b\to + \infty$. This determines the spectral curve, which through the loop equations determines all higher genus corrections from the matrix model side, that we can compare with the supergravity answer.

Starting from (\ref{eq:62}), one can compute the whole perturbative expansion of the matrix model. The leading behavior of two resolvents insertions can be easily obtained by taking the appropriate limit of (\ref{eq:59}) and using (\ref{eq:63})
\begin{equation}\label{eq:60}
W_0^+(z_1,z_2)=
\begin{cases}
\qquad \,\,\, 0 \qquad 
\ , \qquad \lbrace z_1,z_2 \rbrace\,\,{\rm same\,\,sheet}\ , \\
\displaystyle
\,\,\frac{-1}{(z_1-z_2)^2}
\ , \qquad \lbrace z_1,z_2 \rbrace\,\,{\rm different\,\,sheets}\ .
\end{cases}
\end{equation}
All other perturbative contributions are obtained from (\ref{eq:64}), that in this particular case is given by 
\begin{equation}
W^+_g(z,I)=
\sum_{\pm}
{\rm Res}\left[
\frac{F_g(z',I)}{2y(z')(z'-z)},z'=\pm \infty
\right]=
{\rm Res}\left[
\frac{F_g(1/z',I)}{y(1/z')z'(zz'-1)},z'=0
\right]\ .
\end{equation}
Using $W^+(z,I)=1/z+\dots$ in (\ref{eq:54}) and (\ref{eq:65}), one can show $F_g(z,I)=1/z^2+\dots$, which together with $y(\infty)=\infty$ implies the residue vanishes, for arbitrary $g$ and $I$. Altogether, it implies the whole perturbative expansion of the matrix model vanishes, except for the leading contributions to single~(\ref{eq:62}) and double (\ref{eq:60}) trace observables. 

To complete the argument and compare higher genus corrections predicted by supergravity with the ones predicted by the matrix model, we should compute $\langle Z_{{\rm MM}}^+(\beta_1)\dots Z_{{\rm MM}}^+(\beta_n) \rangle$, which involve the Hamiltonian $Q^2$ instead of the supercharge $Q$. This can be extracted from the resolvent $W^+(z)$, using the following identity
\begin{equation}\label{eq:131}
{\rm Tr}\frac{1}{z^2-Q^2}=
\frac{W^+(z)-W^+(-z)}{2z}\ .
\end{equation}
After an appropriate inverse Laplace transform, the left-hand side gives the appropriate factor of~$e^{-\beta Q^2}$ required to construct~(\ref{eq:61}). In this way, one finds
\begin{equation}\label{eq:67}
\begin{aligned}
\langle Z_{{\rm MM}}^+(\beta) \rangle
&\simeq
e^{S_0}\sqrt{\frac{2}{\pi\beta}}
\left[
e^{\pi^2/\beta}+\xi 
e^{\pi^2\alpha^2/\beta}
\right]\ , \\[4pt]
\langle Z_{{\rm MM}}^+(\beta_1)Z_{{\rm MM}}^+(\beta_2) \rangle_c
& \simeq \frac{4}{2\pi}\frac{\sqrt{\beta_1\beta_2}}{\beta_1+\beta_2}\ , \\[4pt]
\langle Z_{{\rm MM}}^+(\beta_1)\dots Z_{{\rm MM}}^+(\beta_n) \rangle_c
& \simeq 0 \ .
\end{aligned}
\end{equation}
Comparing with the supergravity results in~(\ref{eq:55}) we find perfect matching to all orders in perturbation theory, after identifying~$\hbar \to e^{-S_0}$, generalizing the result of~\cite{Stanford:2019vob} to non-zero~$\xi$. It is worth noting that apart from the leading behavior of~$\langle Z_{{\rm MM}}^+(\beta) \rangle$, any Hermitian matrix model with a leading spectral density supported on the whole real line gives the results in~(\ref{eq:67}). The details of~$\rho_0^+(q)$ in~(\ref{eq:62}) are only relevant to the ensure~$\langle Z_{{\rm MM}}^+(\beta) \rangle$ matches with the supergravity result (\ref{eq:55}). The fact that all other non-perturbative corrections vanish in gravity is a non-trivial result required for the match to work.

\subsubsection{Type 0A: Complex Ensemble}

Now we consider the ensemble dual to Type 0A supergravity, which involves adding a topological~$(-1)^\zeta$ term to the sum over spin structures. This theory has a~$(-1)^F$ symmetry and the Hilbert space separates into two sectors, even and odd under this symmetry, with hamiltonian~$H=MM^\dagger$ for arbitrary complex~$M$. When averaging over the supercharge, which in this case involves averaging over the complex matrix $M$, in order to have eigenvalues, we take a potential and observables that depend on the combination~$MM^\dagger$, see~\cite{Morris:1990cq,Dalley:1991qg} for early references. The expectation value we want to compare with gravity is
\begin{equation}\label{eq:77}
\langle Z_{{\rm MM}}^-(\beta_1)\ldots Z_{{\rm MM}}^-(\beta_n) \rangle= 
\frac{2^n}{\mathcal{Z}}
\int dM\,
e^{-N\,{\rm Tr}\,V(MM^\dagger)}
\,{\rm Tr}\,e^{-\beta_1 MM^\dagger}\ldots\,{\rm Tr}\,e^{-\beta_n MM^\dagger} 
\ ,
\end{equation}
where $dM=\prod_{i,j}dM_{ij}dM_{ij}^\ast$. While an arbitrary complex matrix is not be diagonalizable, $MM^\dagger$ is Hermitian and positive definite, meaning it can be diagonalized with eigenvalues we call $E_i\ge 0$. Writing the normalization of the probability distribution $\mathcal{Z}$ in terms of these eigenvalues one finds
\begin{equation}\label{eq:250}
\mathcal{Z}=
\int dM\,e^{-N\,{\rm Tr}\,V(MM^\dagger)}
=\prod_{i=1}^N
\int_{0}^{\infty}dE_i\,\Delta(E_1,\dots,E_N)^2
e^{-NV(E_i)}\ .
\end{equation}
Comparing with the partition function of the Hermitian matrix model (\ref{eq:115}) one finds the same expression, except for the fact that the eigenvalues $E_i$ of $MM^\dagger$ are restricted to the positive real line. Similarly as before, the eigenvalue spectral density and resolvent are respectively defined as
\begin{equation}\label{eq:127}
\rho^-(E)=\frac{1}{N} 
{\rm Tr}\,\delta(E-MM^\dagger)\ ,
\qquad \qquad
W^-(z)={\rm Tr}\frac{1}{z-MM^\dagger}\ .
\end{equation}
While before the spectral density was written naturally in terms of the supercharge operator eigenvalues, now the natural variables are the eigenvalues of the Hamiltonian.

Writing a perturbative expansion for the connected expectation value of $W^-(I)$ equivalent to~(\ref{eq:51}), the corresponding loop equations were derived in~\cite{Stanford:2019vob} starting from an identity analogous to~(\ref{eq:53}). Perhaps unsurprisingly, one finds exactly the same recursion relations~(\ref{eq:64}) and~(\ref{eq:59}), with the only exception being~$W_0^-(z)$, which instead of~(\ref{eq:119}) is given by
\begin{equation}\label{eq:139}
W_0^-(z)=\frac{1}{2}
\Big(
V'(z)-
h(z)
\sqrt{\frac{\sigma(z)}{z}}
\Big)\ ,
\end{equation}
with $h(z)$ and $\sigma(z)$ analytic.\footnote{Compared to \cite{Stanford:2019vob} we are using a slightly different convention for $\sigma(z)$, i.e. $\sigma_{\rm there}(z)=z\sigma_{\rm here}(z)$.} Apart from the extra factor of $1/\sqrt{z}$ in the second term, in this case~$\sigma(z)$ is an odd (instead of even) polynomial with simple real roots. From the discontinuity across the square root branch cut one finds the leading eigenvalue spectral density
\begin{equation}\label{eq:70}
\rho^-_0(E)=\frac{1}{2\pi}|h(E)|
\sqrt{\frac{\sigma(E)}{-E}}\times
\textbf{1}_{\sigma(E)/E<0}\ .
\end{equation}
All in all, the only difference with respect to the Hermitian matrix model is the positivity constraint on the eigenvalues of $MM^\dagger$, which allows for the $1/\sqrt{E}$ behavior of the spectral density. Apart from this, the loop equations are exactly the same after replacing $W^+(z_1,\dots,z_n)\rightarrow W^-(z_1,\dots,z_n)$.

\paragraph{Supergravity matching:} To match with supergravity, we should take into account that now the eigenvalues of $MM^\dagger$ correspond to the eigenvalues of the Hamiltonian in each block, which we conveniently denoted by $E$. We should then match the density of states of the matrix model directly with \eqref{N1dos} (or \eqref{N1dosdef} in the case with deformations). To make the comparison with supergravity, then consider a double scaled model that is perturbatively defined from the following leading spectral density
\begin{equation}\label{eq:72}
\rho^-_0(E)=
\frac{\cosh(2\pi\sqrt{E})+\xi\cosh(2\pi\alpha\sqrt{E})}{\pi \sqrt{2E}}
\ ,
\end{equation}
where for simplicity we consider the case with one defect species. This can be obtained from the general expression in (\ref{eq:70}) by taking $\sigma(E)=(E-b)$ with $b\rightarrow +\infty$ and picking $h(E)$ so that it yields the appropriate $E$ dependence. Using the loop equations and the associated spectral curve
\begin{equation}\label{eq:73}
y(z)=
-\frac{\cos(2\pi\sqrt{-z})+\xi \cos(2\pi \alpha\sqrt{-z})}{ \sqrt{-2z}}\ ,
\end{equation}
one can compute the perturbative expansion of the resolvent $W^-(I)$ to any desired order. From this, a simple Laplace transform
\begin{equation}\label{eq:86}
W^-(z_1,\dots,z_n)=
\frac{(-1)^n}{2^n}
\prod_{i=1}^n
\int_0^{\infty}d\beta_i\,
e^{\beta_iz_i}
\ Z_{{\rm MM}}^-(\beta_1,\dots,\beta_n)\ ,
\end{equation} 
allows us to calculate the expectation value of the matrix operator $Z_{\rm MM}^-(\beta_1,\dots,\beta_n)$ (\ref{eq:61}) in a perturbative expansion 
\begin{equation}
 \langle Z_{{\rm MM}}^-(\beta_1)\ldots Z_{{\rm MM}}^-(\beta_n) \rangle_c=
\sum_{g=0}^{\infty}\hbar^{2(g-1)+n}Z_{{\rm MM},g}^-(\beta_1,\dots,\beta_n)\ ,
\end{equation}
and compare with the supergravity partition function $Z_{g}^-(\beta_1,\dots,\beta_n)$ in (\ref{eq:48}). 

\paragraph{Genus zero:} The leading genus contribution to $Z_{\rm MM}^-(\beta)$ is directly obtained from the eigenvalue spectral density (\ref{eq:72}) 
\begin{equation}\label{eq:162}
Z^-_{{\rm MM},0}(\beta)=2
\int_0^{+\infty}dE\,\rho^-_0(E)e^{-\beta E}=
\sqrt{\frac{2}{\pi \beta}}\left[
e^{\pi^2/\beta}
+\xi e^{\pi^2\alpha^2/\beta}
\right]\ ,
\end{equation}
which agrees (by design) with the supergravity result (\ref{eq:105}). For two insertions consider the universal expression for the resolvent (\ref{eq:59}) with $(a,b)=(0,+\infty)$, which gives
\begin{equation}\label{eq:99}
W_0^-(z_1,z_2)=
\frac{1}{4\sqrt{-z_1}\sqrt{-z_2}(\sqrt{-z_1}+\sqrt{-z_2})^2}
\qquad \Longrightarrow \qquad
Z^-_{{\rm MM},0}(\beta_1,\beta_2)=
\frac{4}{2\pi}\frac{\sqrt{\beta_1\beta_2}}{\beta_1+\beta_2}\ ,
\end{equation}
matching with $Z_0^-(\beta_1,\beta_2)$ in (\ref{eq:71}). The case with $n\ge 3$ is obtained from (\ref{eq:64})
\begin{equation}\label{eq:178}
W^-_0(z,I)=\frac{1}{\sqrt{z}}
{\rm Res}\left[
\frac{\sqrt{z'}}{2y(z')}
\frac{F_0(z',I)}{z'-z},z'=0
\right]
+
\frac{1}{\sqrt{z}}
{\rm Res}\left[
\frac{\sqrt{z'}}{2y(z')}
\frac{F_0(z',I)}{z'-z},z'=+\infty
\right]\ ,
\end{equation}
Setting $I=\lbrace z_1,z_2 \rbrace$ and using (\ref{eq:65}) one can easily compute the residues and show both vanish. Particularly, the first one vanishes due to the $1/\sqrt{-z}$ term in the spectral curve $y(z)$ in (\ref{eq:73}). Using $W_0(z_1,z_2,z_3)=0$ one finds $F_0(z_1,\dots,z_n)$ in (\ref{eq:54}) vanishes for $n\ge 3$, meaning $W_0(z_1,\dots,z_n)=0$. Overall, we conclude $Z^-_{{\rm MM},0}(\beta_1,\dots,\beta_n)=0$, in agreement with the supergravity answer (\ref{eq:71}).

\paragraph{Higher genus:} For $g\ge 1$ one can follow the same procedure and compute $W_g^-(z_1,\dots,z_n)$ using the loop equations to whatever desired order. From the spectral curve $y(z)$ it is useful to define the following coefficients 
\begin{equation}\label{eq:85}
a_k=\left[
\partial_z^k\big(\sqrt{-z}y(z)\big)
\right]_{z=0}\ .
\end{equation}
In appendix \ref{zapp:4} we use the loop equations to derive the following results for $g=1,2,3$\footnote{The genus $g=1,2$ expressions are explicitly derived for arbitrary number of boundaries using induction. Although this can also be done for the $g=3$ result, it becomes increasingly tedious. Instead, we have explicitly checked the~${g=3}$ result in (\ref{eq:87}) for $n=1,2,3,4$.} and an arbitrary spectral curve with $a_0\neq 0$
\begin{equation}\label{eq:87}
\begin{aligned}
Z_{{\rm MM},1}^-(\beta_1,\dots,\beta_n) & =
\frac{1}{2}
\frac{(n-1)!}{4a_0^n}
\prod_{j=1}^n
\sqrt{\frac{\beta_j}{\pi}}\ , \\
Z_{{\rm MM},2}^-(\beta_1,\dots,\beta_n) & =
3
\frac{(n+1)!}{4^{5}a_0^{n+3}}
\left[
(n+2)a_1
+2a_0
\sum_{i=1}^n
\beta_i
\right]
\prod_{j=1}^n
\sqrt{\frac{\beta_j}{\pi}}\ , \\
Z_{{\rm MM},3}^-(\beta_1,\dots,\beta_n) & = 
\frac{1}{5}
\frac{(n+3)!}{4^{9} a_0^{n+6}}
\bigg[
4a_0^2
\bigg(50\sum_{i=1}^n\beta_i^2
+84\sum_{i\neq j}\beta_i\beta_j\bigg)
+168a_0a_1 (n+4)\sum_{i=1}^n\beta_i+
\\
& \hspace{49mm}
+(n+4)\big(
42a_1^2(n+5)-75a_0a_2
\big)
\bigg]
\prod_{j=1}^n\sqrt{\frac{\beta_j}{\pi}} .
\end{aligned}
\end{equation}
Specializing to the spectral curve (\ref{eq:73}) one can easily compute $a_k$ as
\begin{equation}
a_k=(-1)^{k+1}\frac{(2\pi)^{2k}k!}{\sqrt{2}(2k)!}
(1+\xi \alpha^{2k})\ .
\end{equation}
Using this in (\ref{eq:87}) we obtain an explicit match with the supergravity results (\ref{eq:48}) for $g=1,2,3$ and arbitrary $n$.

To prove the equivalence for arbitrary genus $g\in \mathbb{N}$ we use a deformation theorem of \cite{Eynard:2007kz}, which quantifies how the output of the loop equations (\ref{eq:64}) varies as the spectral curve is modified. Following \cite{Maxfield:2020ale}, in Appendix \ref{zapp:4} we use this theorem and show the matching between the supergravity and matrix model expansions to all orders.

\section{Phase transition beyond the topological expansion}
\label{sec:3}

In the previous section we showed how the topological expansion of JT supergravity deformed by a gas of sharp defects matches with the loop equations of appropriately defined matrix models to all orders in perturbation theory. However, the agreement is restricted to $\xi>-1$, as the series expansion (in both cases) diverge for $\xi\rightarrow -1$. The aim of this section is to use the matrix model description of the system to avoid this issue and obtain well-behaved observables when $\xi\le -1$. For reasons we explain below, we do not take $\xi$ arbitrarily negative but restrict ourselves to $\xi\ge-1/\alpha^2$. In the section \ref{sec:4} we explain the problem that arises for $\xi<-1/\alpha^2$ and how it is solved.

Our main tool is the method of orthogonal polynomials, a formalism carefully described in Appendix \ref{zapp:3} for the double scaled models relevant for JT supergravity. Apart from putting together many results scattered in the literature, this Appendix includes some new technical results, like the precise method for computing the matrix model kernel (\ref{eq:204}) and observables (\ref{eq:205}) for double-cut Hermitian matrix models (the string equation had been previously derived in \cite{Crnkovic:1990mr}). Using this approach we show that the divergence of the perturbative expansion at $\xi=-1$ signals a phase transition in the Hermitian (complex) matrix model from a single-cut (hard-edge) to a double-cut (soft-edge) phase (a similar deformation of a toy model in the complex matrix model was previously studied in section~3.C of \cite{Johnson:2019eik}). By properly accounting for the phase transition, we consistently compute observables when $\xi\ge -1/\alpha^2$. Moreover, we show the divergence of the partition function at $\xi=-1$ (see (\ref{eq:48})) is a breakdown of the topological expansion near the phase transition, but the physics of the model is still well-behaved. More precisely, after including non-perturbative contributions (which are invisible in the topological expansion) to the partition function, we show~$Z^\pm(\beta_1,\dots,\beta_n)$ with~${n=1,2}$ has no divergence at~$\xi=-1$.

\subsection{Type 0B JT supergravity}
\label{sec:3.1}

Let us begin by considering Type $0$B JT supergravity with a gas of defects, whose dual description is given by a Hermitian matrix model with leading eigenvalue density in (\ref{eq:62}). Since the loop equations used in the previous section produce a perturbative expansion around a given saddle-point, they are not adequate to understand the behavior of the system near $\xi=-1$ where a phase transition occurs. We shall use the method of orthogonal polynomials instead, which is better suited to capture non-perturbative aspects of the model, such as the change in saddle-point dominance. To avoid derailing the discussion too much, we describe the technical construction of this method in Appendix \ref{zapp:3} and here only describe the main ingredients and tools necessary to understand the output of the formalism after double scaling.

\subsubsection*{Method of orthogonal polynomials}

A single double scaled model is completely specified by a collection of constants $(\mu,t_{2k})$ with $k\in \mathbb{N}$. The parameters $t_{2k}$ and $\mu$ enter in the definition of the potential $V(Q)$ in (\ref{eq:56}) that determines the measure of the matrix model, see (\ref{eq:57}) for the explicit relation. As we shall see, the constant~$\mu$ plays a central role, as its sign indicates the phase of the matrix model. Using the parameters $t_{2k}$, we first need to construct the function $r(x)$ with $x\in \mathbb{R}$, as follows. It is determined by a differential equation called the string equation,  that depends on $t_{2k}$ as \cite{Crnkovic:1990mr}
\begin{equation}\label{eq:206}
    {\rm String\,\,equation:}
    \qquad \quad
    \sum_{k=1}^{\infty}t_{2k}K_{2k}[r(x)]
    +r(x)x=0\ .
\end{equation}
Each parameter $t_{2k}$ controls the contribution to the full model from a single $k$-critical potential, see Appendix \ref{zapp:3} for details. Here $K_{2k}[r(x)]$ is a polynomial of $r(x)$ and its derivatives, constructed from the following recursion relation
\begin{equation}
    K_{2k}[r(x)]=
    \frac{2k}{2k-1}
    \left[
    r(x)\int^x d\bar{x}\,r(\bar{x})\partial_{\bar{x}} K_{2(k-1)}[r(\bar{x})]-
    \frac{\hbar^2}{4}
    \partial_{\bar{x}}^2 K_{2(k-1)}[r({\bar{x}})]
    \right]\ .
\end{equation}
Same as before, the parameter $\hbar$ is identified with $e^{-S_0}$ in the supergravity theory. The leading behavior of $K_{2k}[r(x)]$ is given by $K_{2k}=r(x)^{2k+1}+\mathcal{O}(\hbar^2)$, see  (\ref{eq:202}) for explicit expressions for the first few values of $k$. The string equation (\ref{eq:206}) is solved as a boundary problem for $x\in \mathbb{R}$, see Appendix \ref{zapp:3} for the appropriate boundary conditions as determined by the matrix model.

All observables, such as the correlators of $Z_{\rm MM}^\pm(\beta)$ introduced in the previous section, are ultimately controlled by the function $r(x)$. After determining $r(x)$ for a given set of couplings $t_{2k}$, the next step is to compute the following kernel 
\begin{equation}\label{eq:204}
    \mathcal{K}(q,\bar{q})=
    \sum_{s=\pm1}
    \int_{-\infty}^{\mu}dx
    \Psi_s(x,q)\Psi_s(x,\bar{q})
    =
   \hbar^2\sum_{s=\pm1}
   \frac{\Psi_s(x,q)\overset{\leftrightarrow}{\partial_x}\Psi_s(x,\bar{q})}{q^2-\bar{q}^2}\bigg|_{x=\mu}\ ,
\end{equation}
where $\overset{\leftrightarrow}{\partial_x}=\overset{\rightarrow}{\partial_x}-\overset{\leftarrow}{\partial_x}$. At this point the parameter $\mu$, which has not appeared yet, enters in the range of $x$ integration. The functions $\Psi_s(x,q)$ are obtained from the following eigenvalue problem
\begin{equation}\label{eq:203}
    \mathcal{H}_s\Psi_s(x,q)=q^2\Psi_s(x,q)\ ,
    \qquad \qquad
    \mathcal{H}_s=-(\hbar\partial_x)^2
    +[r(x)^2-s\hbar r'(x)]\ ,
    \qquad 
    s=\pm 1\ ,
\end{equation}
where $s=\pm 1$ labels two independent sectors, related to the orthogonal polynomials of even and odd order. The label $q$ here is obtained from scaling the eigenvalue of the Hermitian matrix and (as explained in the previous section), should be interpreted as the supercharge eigenvalue.\footnote{As explained in Appendix \ref{zapp:3}, the $q_i$ are obtained from the eigenvalues $\lambda_i$ of $Q$ after a proper rescaling in the double scaling limit. In our discussion of the loop equations in the previous section this subtlety was avoided by directly writing $q_i$ as the eigenvalues of $Q$ in (\ref{eq:56}).} Finally, from the kernel $\mathcal{K}(q,\bar{q})$ one can compute the ensemble average of arbitrary insertions of $Z_{\rm MM}^+(\beta)$. We present the general rule in Appendix \ref{zapp:3} for an arbitrary number of insertions, the first two cases are explicitly given by 
\begin{equation}\label{eq:205}
\begin{aligned}
\left\langle 
Z_{\rm MM}^+(\beta) 
\right\rangle & =
2^{\frac{1}{2}}
\int_{-\infty}^{+\infty}
dq\,
\mathcal{K}(q,q)
e^{-\beta q^2}\ , \\
\langle 
Z_{\rm MM}^+(\beta_1)
Z_{\rm MM}^+(\beta_2)
\rangle_c & =2
\int_{-\infty}^{+\infty}
dq\,d\bar{q}
\left[ 
\delta(q-\bar{q})
-
\mathcal{K}(q,\bar{q})
\right]
\mathcal{K}(q,\bar{q})
e^{-\beta_1 q^2}
e^{-\beta_2 \bar{q}^2}\ .
\end{aligned}
\end{equation}
Using the procedure outlined above we can compute these correlators to all orders in $\hbar = e^{-S_0}$. This expansion appears at all levels, staring by the determination of the functions $K_{2k}[r(x)]$. It is a non-trivial consistency check that the topological expansion derived from the string equations is equivalent to the one derived from the topological recursion. The main advantage of this formalism is that it allows the computation of observables beyond their perturbative expansion in $\hbar$ and in particular it will allow us to identify the phase transition.

\subsubsection*{Fixing the double scaled model}

From the loop equations perspective, after picking an ensemble, a matrix model is determined by the spectral curve, the leading order eigenvalue density. In the previous section we determined from gravity the spectral curve associated to Type 0B JT supergravity with a gas of defects. In the approach of orthogonal polynomials, one instead fixes a particular matrix model by picking the couplings $(\mu,t_{2k})$. Since these two description of the matrix model are equivalent, we should determine $(\mu,t_{2k})$ for Type 0B JT supergravity by computing the leading eigenvalue density and compare with the result of the previous section, which we now do. We begin by considering $\xi>-1$. Let us first expand $r(x)$ in $\hbar=e^{-S_0}$ as 
\begin{equation}\label{eq:142}
r(x)=\sum_{m=0}^{\infty}r_m(x)\hbar^m\ .
\end{equation}
Using $K_{2k}[r(x)]=r(x)^{2k+1}+\mathcal{O}(\hbar^2)$ the leading solution of the string equation (\ref{eq:206}) is
\begin{equation}\label{eq:90}
r_0(x)\bigg[
\sum_{k=1}^{\infty}t_{2k}r_0(x)^{2k}+x
\bigg]=0
\qquad \Longrightarrow \qquad
\begin{cases}
\begin{aligned}
\displaystyle
\,\,\sum_{k=1}^{\infty}t_{2k}r_0(x)^{2k}+x&=0
\ , \qquad x\le 0 \ ,\\
\displaystyle
r_0(x)&=0
\ , \qquad x\ge 0\ .
\end{aligned}
\end{cases}
\end{equation}
We have constructed the piecewise solution for $r_0(x)$ so that it is continuous and satisfies the required boundary conditions of the string equation described in Appendix \ref{zapp:3}. Using this leading solution $r_0(x)$, we can construct the operator $\mathcal{H}_s$ in (\ref{eq:203}) and compute its eigenfunctions. To first approximation, we can obtain $\Psi_s(x,q)$ in the WKB approximation, which gives \cite{Johnson:2021owr}\footnote{The two undetermined coefficients in the WKB approximation are fixed by comparing with an exact solution obtained from a toy model which captures the low $q$ behavior of the system (see section 4 of \cite{Johnson:2021owr}). Compared with equation (40) in \cite{Johnson:2021owr} there is an extra factor of $\sqrt{|q|}$ in (\ref{eq:209}). This difference appears because the normalization of the eigenfunctions in the toy model used here is obtained from the completeness relation $\int_{-\infty}^{+\infty}dq\Psi_s^{\rm Toy}(x,q)\Psi_s^{\rm Toy}(\bar{x},q)=\delta(x-\bar{x})$, while \cite{Johnson:2021owr} normalized with respect to the $q^2$ variable.}
\begin{equation}\label{eq:209}
\Psi_s(x,q)=
\sqrt{\frac{|q|}{\pi \hbar}}
\frac{\cos\left[\frac{1}{\hbar}\int_{x_{\rm min}}^{x}d\bar{x}\sqrt{q^2-r_0(\bar{x})^2}-\frac{\pi}{4}(s+1)\right]}{(q^2-r_0(x)^2)^{1/4}}\ ,
\qquad \qquad
x>x_{\rm min}\ .
\end{equation}
This is the WKB solution in the classically allowed region, determined by $x_{\rm min}$ from~${r_0(x_{\rm min})^2=q^2}$. From this expression we can compute the kernel (\ref{eq:204}) in the WKB approximation. For our purposes right now we focus on the spectral density, which from (\ref{eq:205}) it is given by the diagonal components of the kernel (\ref{eq:204}). The leading perturbative contribution can be immediately obtained from~(\ref{eq:209}) as
\begin{equation}\label{eq:92}
\rho_0^+(q)=
\frac{|q|}{\pi}
\int_{-\infty}^{\mu} 
dx
\frac{\Theta(q^2-r_0(x)^2)}{\sqrt{q^2-r_0(x)^2}}\ .
\end{equation}
To solve the integral we must split it around $x=0$ where the solution $r_0(x)$ in (\ref{eq:90}) changes behavior. This distinguishes the cases of $\mu$ positive or negative. We begin by analyzing the case $\mu>0$ since this is relevant for the $\xi>-1$ regime, as will become clear below. In this case the integral gets a contribution from positive $x$ and we find
\begin{equation}\label{eq:93}
\begin{aligned}
\rho_0^+(q) =
\frac{\mu}{\pi}
+
\frac{1}{\pi}
\sum_{k=1}^{\infty}t_{2k}2k
|q|
\int^{|q|}_{0}
\frac{dr_0\,r_0^{2k-1}}{\sqrt{q^2-r_0^2}}=
\frac{\mu}{\pi}
+
\frac{1}{\pi}
\sum_{k=1}^{\infty}t_{2k}
\frac{k!^2}{(2k)!}
(2q)^{2k}\ ,
\end{aligned}
\end{equation}
where we have changed variables to $r_0$, and used (\ref{eq:90}) to compute the Jacobian. From this expression its clear that each coefficient $t_{2k}$ controls the contribution $q^{2k}$ to $\rho_0(q)$, with $\mu$ appearing in the constant term.

Now we can make contact with the previous approach. In this approach the couplings $(\mu,t_{2k})$ are fixed by matching the expansion \eqref{eq:93} to the Taylor expansion in $q$ of the spectral density derived from gravity (\ref{eq:62}). It is a non-trivial fact that this also ensures the matching between supergravity and the loop equations to all orders (away from phase transitions).\footnote{Below we explicitly show the perturbative matching between the supergravity partition function in (\ref{eq:55}) and the matrix model described in this way to leading genus and arbitrary boundaries in the regime $\xi>-1$ (see (\ref{eq:98})). For higher genus contributions, see the arguments given in \cite{Johnson:2021owr}.} The result is that to study deformations of Type 0B JT supergravity the parameters $(\mu,t_{2k})$ of the double scaled model are fixed to
\begin{equation}\label{eq:91}
(\mu,t_{2k})=\left(1+\xi,\frac{\pi^{2k}}{k!^2}(1+ \xi \alpha^{2k})\right)\ .
\end{equation}
For $\xi=0$ this agrees with the undeformed case previously studied in \cite{Johnson:2021owr}. This completely determines the double scaled model. Moreover, we can check $\xi>-1$ implies that $\mu>0$, validating the assumptions that lead to \eqref{eq:93}. 

The implicit constraint equation for $r_0(x)$ in the negative $x$ region (\ref{eq:90}) can be written in terms of modified Bessel functions
\begin{equation}\label{eq:96}
I_0(2\pi r_0)-1
+\xi\big(I_0(2\pi \alpha r_0)-1\big)+x=0\ ,
\qquad x\le 0\ ,
\end{equation}
while for $x>0$ we have $r_0(x)=0$. These results can be generalized to deformations involving several defect species in a trivial way. Finally, from this it is straightforward to check the piecewise function~$r_0(x)$ in~(\ref{eq:90}) is continuous at $x=0$ whenever $\xi\ge-1/\alpha^2$. For that reason, in this section we constraint ourselves to this regime and leave the $\xi<-1/\alpha^2$ case to section \ref{sec:4}. 

\subsubsection{Phase Transition at \texorpdfstring{$\xi=-1$}{xi}}

From (\ref{eq:91}) note that negative $\mu=1+\xi<0$ corresponds precisely to $\xi<-1$, where the topological expansion of the supergravity theory, as well as the matrix model loop equations, break down. However, the formula for the spectral density (\ref{eq:92}) is perfectly well defined when used correctly. For negative~$\mu$ there is no contribution from the $x>0$ region of integration. A similar calculation as in the previous case gives
\begin{equation}\label{eq:94}
\rho_0^+(q)=
\frac{2}{\pi}
\sum_{k=1}^{\infty}
t_{2k}k
|q|
\int^{+\infty}_{q_c}dr_0\,r_0^{2k-1}
\frac{\Theta(q^2-r_0^2)}{\sqrt{q^2-r_0^2}}
\ ,
\qquad {\rm where} \qquad
q_c \equiv r_0(\mu)>0\ .
\end{equation}
The main difference is that there is no constant term and the integral vanishes for $|q|\le q_c$. The threshold value $q_c=r_0(\mu)$ is computed from the implicit constraint (\ref{eq:90}), which in the JT supergravity case (\ref{eq:91}) is given by
\begin{equation}\label{eq:133}
I_0(2\pi q_c)+\xi I_0(2\pi \alpha q_c)=0
\qquad {\rm with} \qquad q_c>0\ ,
\end{equation}
where we used the identification $\mu=1+\xi$. It is straightforward to solve this equation numerically for any desired values of $(\xi,\alpha)$. After solving the series in (\ref{eq:94}), the spectral density can be written in terms of the following integral
\begin{equation}\label{eq:97}
\rho_0^+(q)=
2|q|
\int_{q_c}^{|q|}dr_0
\frac{I_1(2\pi r_0)+\xi \alpha I_1(2\pi \alpha r_0)}
{\sqrt{q^2-r_0^2}}\ ,
\end{equation}
where it is implicitly understood this expression vanishes when $|q|\le q_c$. This density of states characterizes the new phase of gravity that appears at $\xi<-1$ which would be impossible to identify if we restrict ourselves to perturbation theory.   

\begin{figure}
\centering
\includegraphics[scale=0.40]{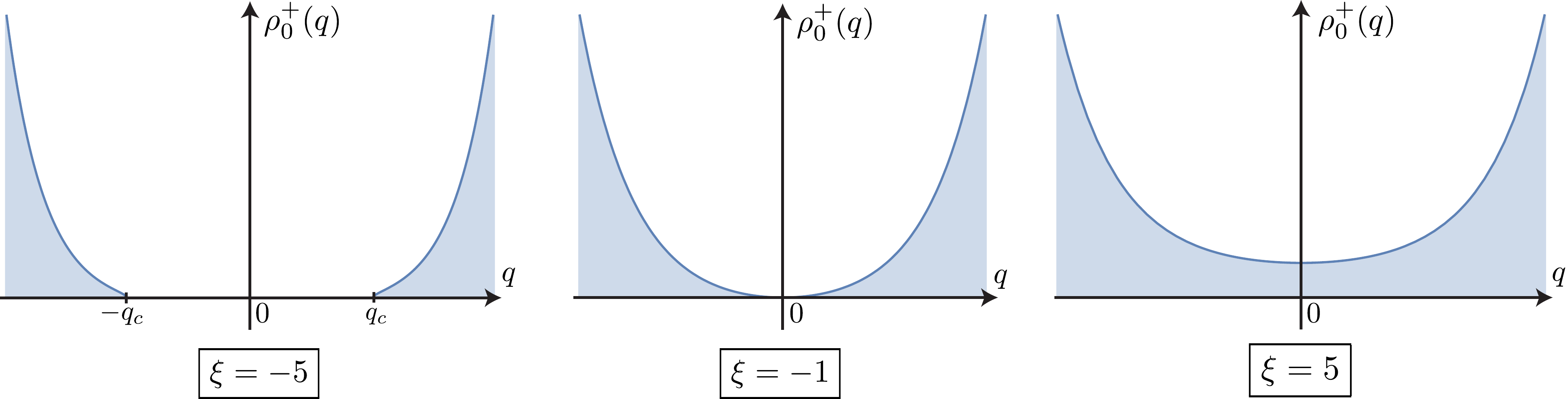}
\caption{Leading genus eigenvalue spectral density $\rho_0^+(q)$ for the Hermitian matrix model relevant for describing deformations of Type 0B JT supergravity with defect angle $\alpha=1/4$ and several values of $\xi$. There is a phase transition occurring at $\xi=-1$, with $\xi<-1$ and $\xi>-1$ in the double and single-cut phase respectively.}
\label{fig:3}
\end{figure}

In figure \ref{fig:3} we plot $\rho_0^+(q)$ for $\alpha=1/4$ and three values of $\xi$ for which $\mu$ is positive, negative and zero. The $\xi=-1$ behavior corresponds to the critical value separating the single from the double-cut phases. This can be compared with the spectral density of the matrix model before double scaling shown in figure \ref{fig:2}. As expected, figure \ref{fig:3} is obtained from figure \ref{fig:2} by zooming in the region $\lambda\sim 0$. For the model with only the $k=1$ contribution, this is in the same universality class as the well-known third order Gross-Witten-Wadia transition \cite{Gross:1980he,Wadia:2012fr} for unitary matrices. For a discussion of this transition for Hermitian matrices, see for example \cite{Bleher:2002th}.

To have a better intuition on the behavior of the model, let us rewrite the integral (\ref{eq:94}) in a way that makes the behavior near $q_c$ manifest
\begin{equation}\label{eq:95}
\rho_0^+(q) =
\sqrt{q^2-q_c^2}
\sum_{k=1}^{\infty}
t_{2k}
\frac{2k}{\pi}
|q|^{2k-1}
\,_2F_1\left[
\frac{1}{2},1-k;\frac{3}{2};
1-\frac{q_c^2}{q^2}
\right]
\ .
\end{equation}
To show this formula, one can change variables to $v=1-(r_0/q)^2$ and realize that the $v$ integral in~\eqref{eq:94} gives an incomplete Beta function, which can be written as a hypergeometric function. 
While before the transition each $k$ critical model contributed with a simple $q^{2k}$ term to the spectral density~(\ref{eq:93}), we now get $\rho^+_0(q)\big|_{k}=|q|P(q)\sqrt{q^2-q_c^2}$, with $P(q)$ an even polynomial of order~${2(k-1)}$. To understand this, note a general two-cut spectral density before double scaling and supported in~${q\in (a_1,b_1)\cup (a_2,b_2)}$ is given by~(\ref{eq:58})
\begin{equation}\label{eq:128}
\rho_0^+(q)=
\frac{1}{2\pi}|h(q)|
\sqrt{-(q-a_1)(q-b_1)(q-a_2)(q-b_2)}\ .
\end{equation}
Taking $a_2=-b_1=q_c$ and $b_2=-a_1=+\infty$ one recovers the same structure given in each term of~(\ref{eq:95}), where $h(q)$ is related to $q P(q)$.

We would like to interpret this phase transition as a large, order one, breaking of supersymmetry. To explain what we mean by this consider first the $\xi>-1$ phase, we expect the lowest energy state to have, on average, a small non-zero energy of order $e^{-S_0}$. Therefore supersymmetry is broken by exponentially small effects. Something similar was observed in $\mathcal{N}=1$ SYK models \cite{Fu:2016vas}. On the other hand, on the $\xi<-1$ phase the ground state has an order one energy given by $q_c^2$. Therefore supersymmetry is broken already to leading order in perturbation theory.

From a gravity perspective, a possible interpretation of the $\xi<-1$ phase is that that the vacuum fermions in supergravity get an effective mass so that they can be integrated out and obtain an effective theory of bosonic dilaton gravity. To check this interpretation we can compute the density of states as a function of energy instead of supercharge eigenvalue. This presents the same behavior as expected from a bosonic theory of dilaton gravity, since it has the correct square root edge and the correct asymptotics at large energies. Of course this correspondence is not valid non-perturbatively, where we can distinguish gravity from supergravity. It would be nice to make further checks of this idea in order to identify the bulk interpretation of the $\xi<-1$ phase better.

Before moving on, let us address some issues that arise for the perturbative expansion of observables of double-cut matrix models, such as the one given in (\ref{eq:128}). As discussed in \cite{Bonnet:2000dz}, a careful analysis shows there are non-analytic terms in $N$ (size of the matrix) that contribute to the expansion of observables. For instance, the leading behavior of two resolvent insertions $W_0^+(z_1,z_2)$ for a symmetric double-cut model supported on $q\in (-b,-a)\cup(a,b)$ is given by (see equation (3.26) in~\cite{Bonnet:2000dz})
\begin{equation}\label{eq:129}
W_0^+(z_1,z_2)=
\frac{1}{2(z_1-z_2)^2}\left[
\frac{(a^2-z_1z_2)(b^2-z_1z_2)}
{\sqrt{\sigma(z_1)}\sqrt{\sigma(z_2)}}
-1
\right]
-
\frac{(-1)^Nab}{2\sqrt{\sigma(z_1)}\sqrt{\sigma(z_2)}}\ ,
\end{equation} 
where $\sigma(z)=(z^2-a^2)(z^2-b^2)$ and the sign of the second term depends on whether~$N$ is even or odd. Does this kind of non-analytic behavior in~$N$ pollutes the double scaled matrix model in the~${\xi<-1}$ regime? The answer is no, as can be seen by noting the second term~(\ref{eq:129}) vanishes for~${(a,b)=(+\infty,q_c)}$, that is the limit we are ultimately interested in 
\begin{equation}\label{eq:130}
W_0^+(z_1,z_2)
\big|_{(a,b)=(+\infty,q_c)}=
\frac{1}{2(z_1-z_2)^2}\left[
\frac{z_1z_2-q_c^2}
{\sqrt{z_1^2-q_c^2}\sqrt{z_2^2-q_c^2}}
-1
\right]
\ .
\end{equation} 
Interestingly, this gives the same result as for a single-cut matrix model (\ref{eq:59}) supported in the complementary region, i.e. $q\in(-q_c,q_c)$. Altogether, this means one should not worry about these kind of non-analytic contributions to the topological expansion of the matrix model in the double-cut phase.

\subsubsection*{Order of the phase transition}

Let us now determine the order of the phase transition from the single to the double-cut phase. From the matrix model perspective, this is obtained by analyzing the free energy defined as $\mathcal{F}=-\ln \mathcal{Z}$ with $\mathcal{Z}$ given in (\ref{eq:115}). However, since we are ultimately interested in gravity we want to determine the order of the transition as characterized by the gravitational free energy, defined as
\begin{equation}\label{eq:135}
F^+(\beta,\xi)=-\frac{1}{\beta}
\ln Z^+(\beta)\ .
\end{equation}
Using the matrix model description, we can expand from both sides of the transition at $\xi=-1$ and determine its order.\footnote{The quantity defined in (\ref{eq:135}) corresponds to the annealed instead of the quenched free energy \cite{Engelhardt:2020qpv}. Both definitions agree for high enough temperatures. Using the methods of \cite{Johnson:2021zuo} it would be interesting to study the behavior of the quenched free energy near the transition and at low temperatures.} Depending on the value of $\xi$, the expectation value of $Z_{\rm MM}^+(\beta)$ to leading order is given by (\ref{eq:162}) or (\ref{eq:136}). To expand these results around the critical value $\xi=-1$ one first needs an expression for $q_c(\xi)$ (\ref{eq:133}). A simple perturbative calculation gives
\begin{equation}
q_c(\xi)=
\frac{1}{\pi}\frac{\sqrt{-1-\xi}}{\sqrt{1-\alpha^2}}
\left[
1+
\frac{3\alpha^2-1}{8(\alpha^2-1)}
(1+\xi)+
\frac{263\alpha^4-250\alpha^2+47}{1152(\alpha^2-1)^2}
(1+\xi)^2
+\mathcal{O}(1+\xi)^3
\right]\ .
\end{equation}
Using this, one can easily expand (\ref{eq:162}) and (\ref{eq:136}) in a perturbative series and obtain
\begin{equation}\label{eq:160}
\sqrt{\frac{\pi \beta}{2}} 
Z_{{\rm MM},0}^+(\beta)\simeq 
\left[
e^{\pi^2/\beta}-e^{\pi^2\alpha^2/\beta}
\right]
+
(1+\xi)
e^{\pi^2\alpha^2/\beta}
+
(1+\xi)^2\times
\begin{cases}
\hspace{9.3mm} 0 \hspace{9.3mm}
\ , \qquad \xi>-1\ , \\
\displaystyle
\frac{\beta}{2\pi^2(1-\alpha^2)} 
\ , \qquad \xi<-1\ ,
\end{cases}
\end{equation}
where we have omitted terms of order $(1+\xi)^3$. Replacing $Z^+(\beta)\rightarrow \langle Z_{{\rm MM},0}(\beta)\rangle$ in the gravitational free energy (\ref{eq:135}) one finds the second derivative at $\xi=-1$ is discontinuous, meaning the transition is of second order as characterized by the gravitational free energy (\ref{eq:135}). 

We would like to stress that this analysis works for $\xi \neq -1$. When $\xi$ is close enough to $-1$, the value at which the phase transition happens, the topological expansion of the matrix model breaks down. See equations \eqref{eq:48} for an example of how higher genus contributions diverge faster as $\xi$ approaches $-1$. Therefore to know what happens at the phase transition, or at least with  exponential accuracy in $S_0$, a non-perturbative analysis is required. This is done in the next section.

\subsubsection*{New perturbative expansion}

Since the leading spectral density (\ref{eq:95}) has a very different structure after the phase transition, the topological expansion of the matrix operator $Z^+_{\rm MM}(\beta_1,\dots,\beta_n)$ changes drastically. In particular, it does not vanish to all orders as is the case when $\xi>-1$, see (\ref{eq:67}). From the approach of the previous section, the reason is that the branch cuts now have finite edges at $q=\pm q_c$ which contribute to the residue involved in the loop equations.

We begin by considering the simplest case. In the $\xi>-1$ phase of JT supergravity, as reviewed in the previous section, the path integral over connected spacetimes with an arbitrary number $n$ of boundaries all vanish for $n>2$, to leading order in the genus expansion. This is due to the presence of fermion zero-modes, as explained in Appendix A of \cite{Stanford:2019vob}.\footnote{Since we can identify $\xi<-1$ as a phase where supersymmetry is broken by large effects, it would be nice to understand whether those fermionic zero-modes get lifted in this phase, giving a gravity understanding of the fact that they are not vanishing for $\xi<-1$.} Instead, in the $\xi<-1$ phase studied in this section these quantities are non-zero. To see this we can use the following useful formula, derived in \cite{Ambjorn:1990ji,Moore:1991ir} and recently applied to the $\xi=0$ matrix model \cite{Johnson:2021owr}
\begin{equation}\label{eq:98}
Z_{{\rm MM},0}^+(\beta_1,\dots,\beta_n)=2^{\frac{n}{2}+1}
\frac{\sqrt{\beta_1\cdots\beta_n}}{2\pi^{n/2}\beta_T}
\left[
\partial_x^{n-2}
e^{-\beta_T r_0(x)^2 }
\right]_{x=\mu}\ ,
\end{equation}
where $\beta_T=\sum_{i=1}^n\beta_i$. See appendix A of \cite{Johnson:2021owr} for an explicit check of the formula for $n=2,3$. While the factor $2^{\frac{n}{2}}$ comes from the definition of $Z_{\rm MM}^+(\beta_1,\dots,\beta_n)$ in (\ref{eq:61}), the additional factor of~$2$ originates from the sum over $s=\pm$ in (\ref{eq:204}). Defining $\partial^{-1}_x=\int_{-\infty}^{\mu}$ and evaluating for the first few values of $n$ one finds
\begin{equation}\label{eq:136}
\begin{aligned}
Z_{{\rm MM},0}^+(\beta) & =
\sqrt{\frac{8\pi}{\beta}}
\int^{+\infty}_{q_c}dr_0
\big[
I_1(2\pi r_0)+\xi \alpha I_1(2\pi\alpha r_0)
\big]
e^{-\beta r_0^2}\ , \\[4pt]
Z_{{\rm MM},0}^+(\beta_1,\beta_2)  & =
\frac{4}{2\pi}\frac{\sqrt{\beta_1\beta_2}}{\beta_1+\beta_2}
e^{-(\beta_1+\beta_2)q_c^2}\ , \\[4pt]
Z_{{\rm MM},0}^+(\beta_1,\beta_2,\beta_3) & =
-\sqrt{\frac{32\beta_1\beta_2\beta_3}{\pi^3}}
\,q_c r_0'(\mu)\,
e^{-q_c^2\sum_{i=1}^3\beta_i}\ , \\[4pt]
Z_{{\rm MM},0}^+(\beta_1,\beta_2,\beta_3,\beta_4) & =
-\sqrt{\frac{64\beta_1\beta_2\beta_3\beta_4}
{\pi^4}}
\left[
r_0'(\mu)^2
+q_cr_0''(\mu)
-2q_c^2r_0'(\mu)^2
\sum_{i=1}^4\beta_i
\right]
e^{-q_c^2\sum_{i=1}^4\beta_i}\ .
\end{aligned}
\end{equation}
The first two cases are already non-zero in the $\xi>-1$ phase, but the results are of course modified. For $n=1$ we have changed variables to $r_0$ and used (\ref{eq:96}) to compute the Jacobian. Equivalently, this same expression can be derived by appropriately integrating the leading spectral density~(\ref{eq:97}). For $n=2$, we find the same universal result obtained also for the single-cut phase (\ref{eq:99}) when~${\xi>-1}$, the only difference being the exponential factor, which arises from the fact there is a threshold value~$q_c$ for the matrix eigenvalues, or equivalently a minimal energy $q_c^2$. 

For $n>2$ it is easy to see from \eqref{eq:98} why the answer vanishes in the $\xi>-1$ phase. The reason is that $\mu$ is positive for this phase and since $r_0(x>0)=0$ the derivatives involved in \eqref{eq:98} all vanish. On the other hand, the results are non-zero when $\xi<-1$ since in this phase the derivatives of $r_0(x)$ at $x=\mu<0$ are non-vanishing. The derivatives of $r_0(x)$ can be easily evaluated and written in terms of $(\xi,\alpha)$ and $q_c$ by differentiating~(\ref{eq:96}). They are given by intricate combinations of modified Bessel functions, for instance
\begin{equation}
r_0'(\mu)=
\frac{-1}{2\pi}
\big[
I_1(2\pi q_c)+\xi \alpha I_1(2\pi \alpha q_c)
\big]^{-1}\ .
\end{equation}
Note that as one approaches the critical value $\xi\rightarrow -1$ this quantity diverges, signaling the transition from the double to the single-cut phase.

So far we focused on contributions at genus zero with multiple boundaries, but in the $\xi>-1$ phase we argued in the previous section all higher genus contributions are also vanishing. This is not the case for the $\xi<-1$ phase as we now show explicitly. For instance, the ${g=1}$ correction to single trace observables is given by
\begin{equation}\label{eq:166}
Z_{{\rm MM},1}^+(\beta)=
\frac{-1}{48a_1^2}
\sqrt{\frac{\beta}{\pi q_c}}
\left[
a_1(28\beta q_c^2-9)
+2a_2 q_c(4\beta q_c^2-3)
\right]
e^{-\beta q_c^2}\ .
\end{equation}
This result can be obtained either from the loop equations applied to the $\xi<-1$ phase, or using the leading $\hbar$ corrections to the string equation. Altogether, the perturbative expansion of the model is very different after the transition to the double-cut phase. It would be interesting to determine whether these results can be recovered from the topological expansion of some kind of supergravity theory.

\subsubsection{Non-perturbative effects}

Our discussion so far has been limited to the perturbative expansion of observables. We now go beyond perturbation theory and compute observables in the matrix model non-perturbatively in~${\hbar=e^{-S_0}}$. This will help clarify the nature of the phase transition, since the perturbative expansions break down at~$\xi=-1$.

\begin{figure}
\centering
\includegraphics[scale=0.57]{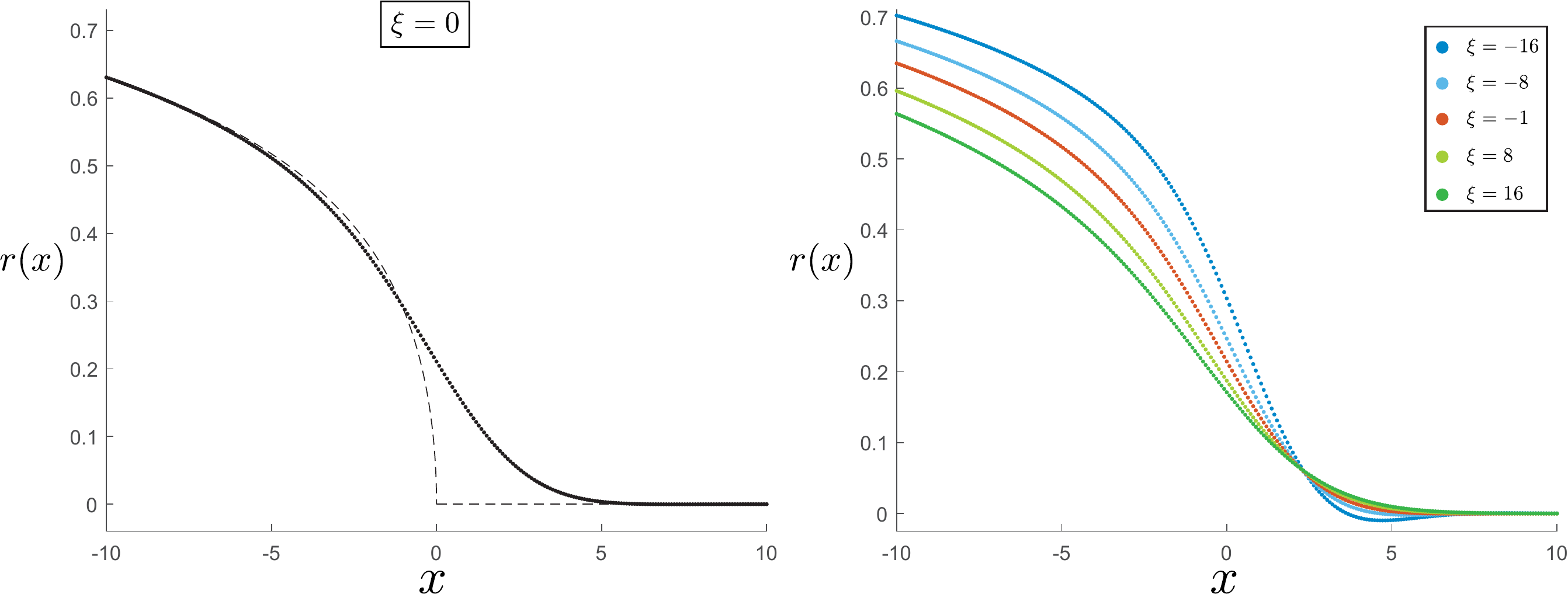}
\caption{Full non-perturbative solution $r(x)$ to the string equation (\ref{eq:143}) with $(k_{\rm max},\hbar,\alpha)=(6,1,1/4)$ and several values of $\xi$. On the left diagram, we plot the $\xi=0$ result previously obtained in \cite{Johnson:2021owr}, with the dashed line corresponding to the perturbative solution $r_0(x)$ in (\ref{eq:90}). In the right diagram, we show the solution at the phase transition $\xi=-1$ (red curve) and several other values of $\xi$.}\label{fig:7}
\end{figure}

The first step is to exactly solve the full string equation (\ref{eq:206}), with $t_{2k}$ given in (\ref{eq:91}), without assuming any perturbative expansion (\ref{eq:142}) for $r(x)$. To do so, we proceed numerically, following~\cite{Johnson:2020exp}. Although we have explored several values of~$\hbar$ and~$\alpha$, all the results shown here take~$\hbar=1$ and~${\alpha=1/4}$, meaning we must constraint~${\xi\ge -1/\alpha^2=-16}$. Since the string equation is formally a differential equation of infinite order we must introduce a truncation~$k_{\rm max}$ in order to make sense of it numerically~\cite{Johnson:2020exp}
\begin{equation}\label{eq:143}
\sum_{k=1}^{k_{\rm max}}
t_{2k}K_{2k}[r(x)]+r(x)x=0\ .
\end{equation}
For the numerical accuracy required here, it is enough to fix $k_{\rm max}=6$. We have explored higher truncations and found no substantial differences in the results. In figure \ref{fig:7} we show several numerical solutions $r(x)$. These are obtained by solving the twelve order differential equation (\ref{eq:143}) with the boundary conditions determined by $r_0(x)$ at $x=\pm \infty$. In the left diagram of figure \ref{fig:7} we plot the full solution together with the perturbative result $r_0(x)$ for the pure JT supergravity case, previously obtained in \cite{Johnson:2021owr}. As shown in that work, all perturbative corrections to $r_0(x)$ vanish in that case, meaning the difference between the dashed and solid lines in figure \ref{fig:7} is entirely generated by non-perturbative effects. Allowing for defects by taking $\xi\neq 0$ deforms the solution $r(x)$ in interesting ways, as shown in the right diagram. Most importantly, note the solution at the phase transition~${\xi=-1}$ (red curve), is perfectly well behaved.

Using this numerical solution we construct the operator $\mathcal{H}_s$ in (\ref{eq:203}) and compute its eigenfunctions. See \cite{Johnson:2020exp,Johnson:2021owr} for details on the numerical methods, particularly regarding the normalization of $\Psi_s(x,q)$. By appropriately integrating in $x$ (\ref{eq:204}) we get the kernel $\mathcal{K}(q,\bar{q})$ which determines all observables. 

\begin{figure}
\centering
\includegraphics[scale=0.55]{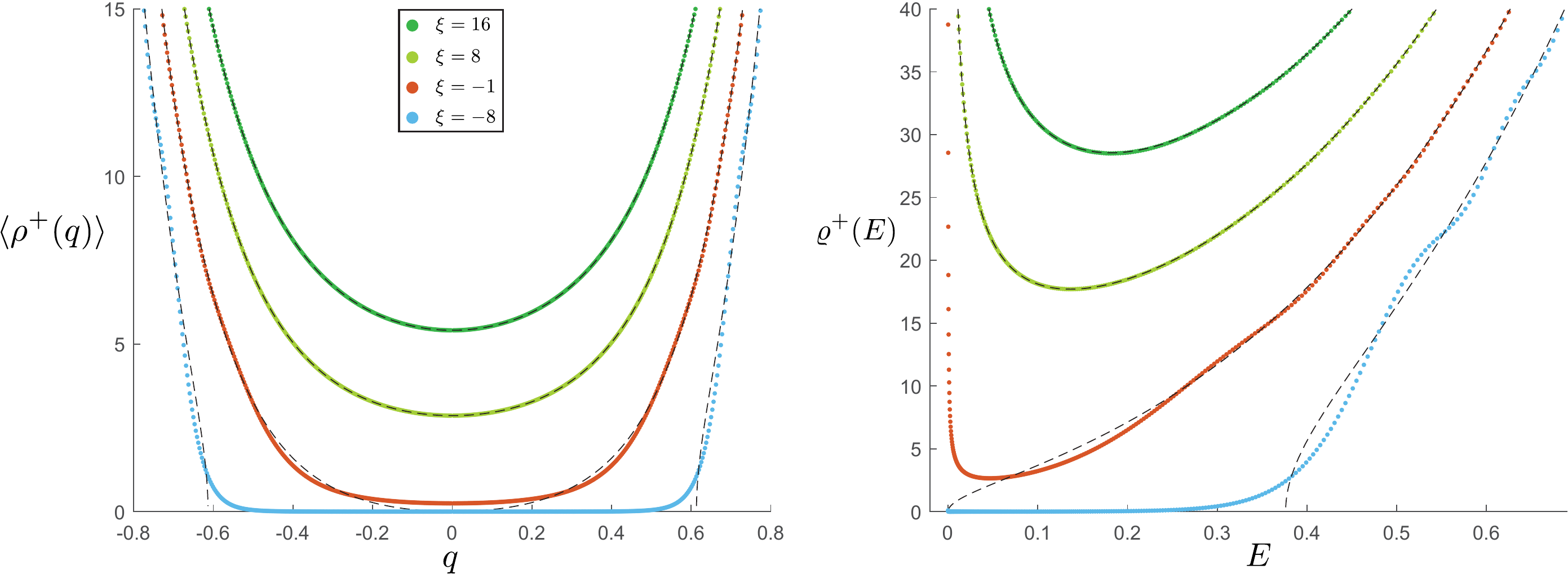}
\caption{On the left diagram we plot the expectation value of the eigenvalue spectral density $\rho^+(q)$ including all perturbative and non-perturbative contributions for several values of $\xi$ and $\alpha=1/4$. On the right diagram, we have the corresponding energy spectral density of Type 0B JT supergravity including sharp defects. Note that in both cases the spectral density is perfectly well defined at the phase transition $\xi=-1$, red curves.}\label{fig:8}
\end{figure}

\paragraph{Spectral density:} Let us start by considering the eigenvalue spectral density $\langle \rho^+(q) \rangle$, obtained from the diagonal components of the kernel. In the left diagram of figure \ref{fig:8} we show the final result for the eigenvalue spectral density for several values of $\xi$. The dashed curves correspond to the leading perturbative answer $\rho^+_0(q)$, given by (\ref{eq:62}) or (\ref{eq:97}) depending on which side of the phase transition we are on. Assuming the matching between matrix model and quantum gravity is valid beyond perturbation theory, we can compute the gravitational energy spectral density $\varrho^+(E)$, defined as
\begin{equation}
\langle Z_{\rm MM}^+(\beta) \rangle =
2^{\frac{1}{2}}\int_{-\infty}^{+\infty}
dq\,
\langle \rho^+(q) \rangle
e^{-\beta q^2}
\qquad \longleftrightarrow \qquad
Z^+(\beta)=\int_0^{+\infty}
dE\,
\varrho^+(E)e^{-\beta E}
\ .
\end{equation}
The final result for $\varrho^+(E)$ is shown in the right diagram of figure \ref{fig:8}, where the dashed line corresponds to the leading perturbative result. 

It is important to point out that neither of the solutions in figure \ref{fig:8} diverge or become ill defined at the phase transition. More precisely, when $\xi=-1$ we obtain a perfectly well defined spectral density for all energies. This is in contrast to perturbative contributions $Z_{{\rm MM},g}^+(\beta)$ and $Z_g^+(\beta)$ which become ill defined as $\xi\rightarrow -1$. This shows the divergence is not real physics but a signature of the breakdown of perturbation theory near the phase transition, ultimately fixed by non-perturbative effects.

\begin{figure}
\centering
\includegraphics[scale=0.54]{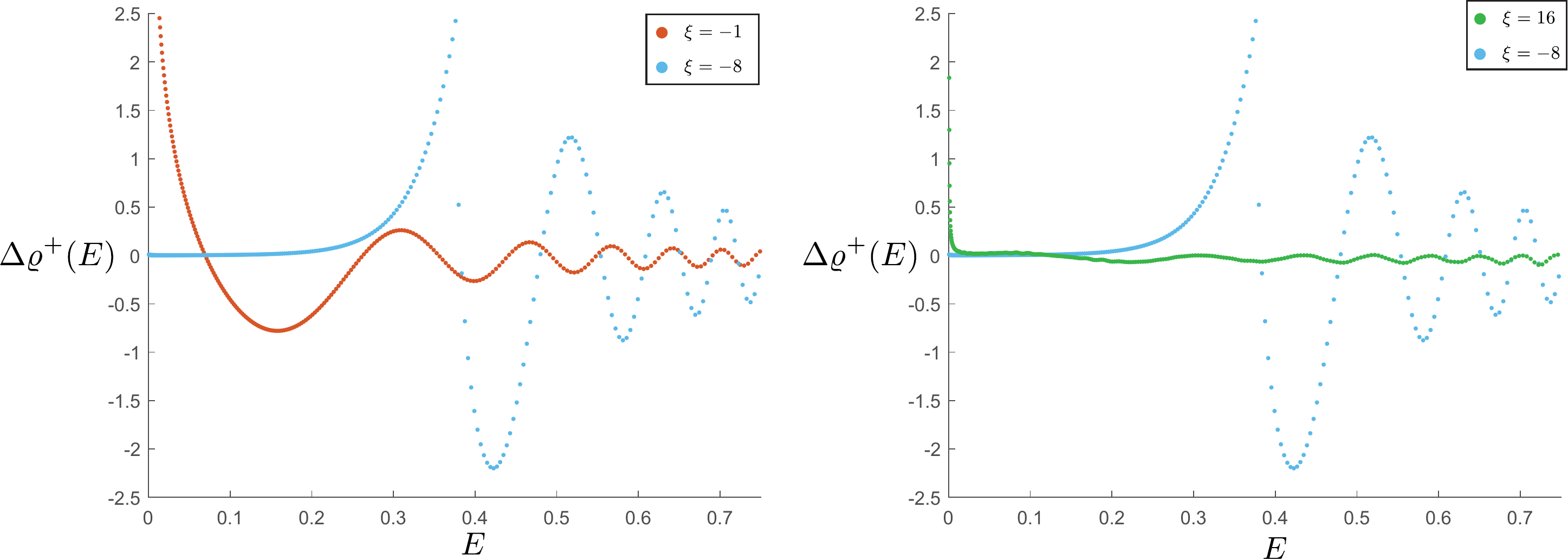}
\caption{In these diagrams we have substracted the perturbative contribution from the full gravitational spectral density $\Delta \varrho^+(E)=\varrho^+(E)-\varrho^+_0(E)/\hbar$ in order to better appreciate how non-perturbative effects change as a function of $\xi$.}
\label{fig:9}
\end{figure}

From figure~\ref{fig:8} we see non-perturbative corrections generate oscillations around the leading answer. Interestingly, the amplitude of the oscillations decrease as $\xi$ grows and increase considerably as one goes beyond the phase transition, from the single to the double-cut (compare $\xi=8,-8$ curves in figure~\ref{fig:8}). To better appreciate this, we can substract the leading behavior from the full answer, as done in figure~\ref{fig:9} for the energy spectral density of the gravity theory~${\Delta \varrho^+(E)=\varrho^+(E)-\varrho_0^+(E)/\hbar}$. In the left diagram of figure~\ref{fig:9} we compare the critical case $\xi=-1$ with the deformed theory in the double-cut phase $\xi=-8$. We observe how the amplitude of the non-perturbative effects is greatly enhanced as we go across the phase transition. The discontinuity and rise of $\Delta \varrho^+(E)$ observed at~${E\sim 0.35}$ when $\xi=-8$ comes from the step function (\ref{eq:94}) in the leading result $\rho^+_0(q)$. In the right diagram of figure~\ref{fig:9} we compare $\Delta \varrho^+(E)$ between two values of $\xi$ in either side of the transition that are further apart from each other. Similarly as non-perturbative effects are enhanced in the double-cut phase as $\xi$ becomes more negative, we observe a suppression as $\xi$ takes larger positive values. 

\paragraph{Spectral form factor:} We now study the expectation value of double trace observables (\ref{eq:205}) which are sensible to the non-diagonal components of the kernel. More precisely, we consider the spectral form factor, defined in terms of the Euclidean partition function as
\begin{equation}\label{eq:207}
S^+(t,\beta)=
\frac{Z^+(\beta+it,\beta-it)+Z^+(\beta+it)Z^+(\beta-it)}{Z^+(2\beta)}
\ .
\end{equation}
To remind the reader, the quantity $Z^+(\beta_1,\beta_2)$ in gravity includes only connected contributions. Therefore to obtain the spectral form fact, corresponding to the full path integral up to a normalization factor, we need to sum the geometries that are both connected (first term) and disconnected (second term). This is an interesting observable to study as it serves as a diagnostic of certain chaotic features of the underlying microscopic description of the supergravity theory \cite{Cotler:2016fpe}. The behavior of~$S^+(t,\beta)$ as a function of $t$ is expected to show an initial dip, followed by a ramp at intermediate times and finally a late time plateau $\lim_{t\rightarrow +\infty}S^+(t,\beta)={\rm const}$. Using (\ref{eq:55}) it is straightforward to write the spectral form factor to all orders in perturbation theory when $\xi>-1$
\begin{equation}\label{eq:208}
\begin{aligned}
S^+(t,\beta)\simeq\frac{2}{Z^+(2\beta)}
\bigg\lbrace
\frac{\sqrt{\beta^2+t^2}}{2\pi \beta} 
+
\frac{e^{2S_0+\frac{2\pi^2\beta}{\beta^2+t^2}}}{\pi \sqrt{\beta^2+t^2}}
\bigg[
1
+&\xi^2 
e^{-\frac{2\pi^2\beta(1-\alpha^2)}{\beta^2+t^2}}+ \\
 & \hspace{12pt} +2\xi 
e^{-\frac{\pi^2\beta(1-\alpha^2)}{\beta^2+t^2}}
\cos\left( 
\frac{\pi^2(1-\alpha^2)t}{\beta^2+t^2}
\right)
\bigg]
\bigg\rbrace\ .
\end{aligned}
\end{equation}
The first term comes from the connected contribution and gives a linear ramp behavior of $S^+(t,\beta)$ for intermediate times. All other terms come from the disconnected piece in (\ref{eq:207}) and generate the initial dip. Crucially, the late time ramp is not captured at all in perturbation theory, meaning it has to be generated by non-perturbative corrections to (\ref{eq:208}). 

\begin{figure}
\centering
\includegraphics[scale=0.51]{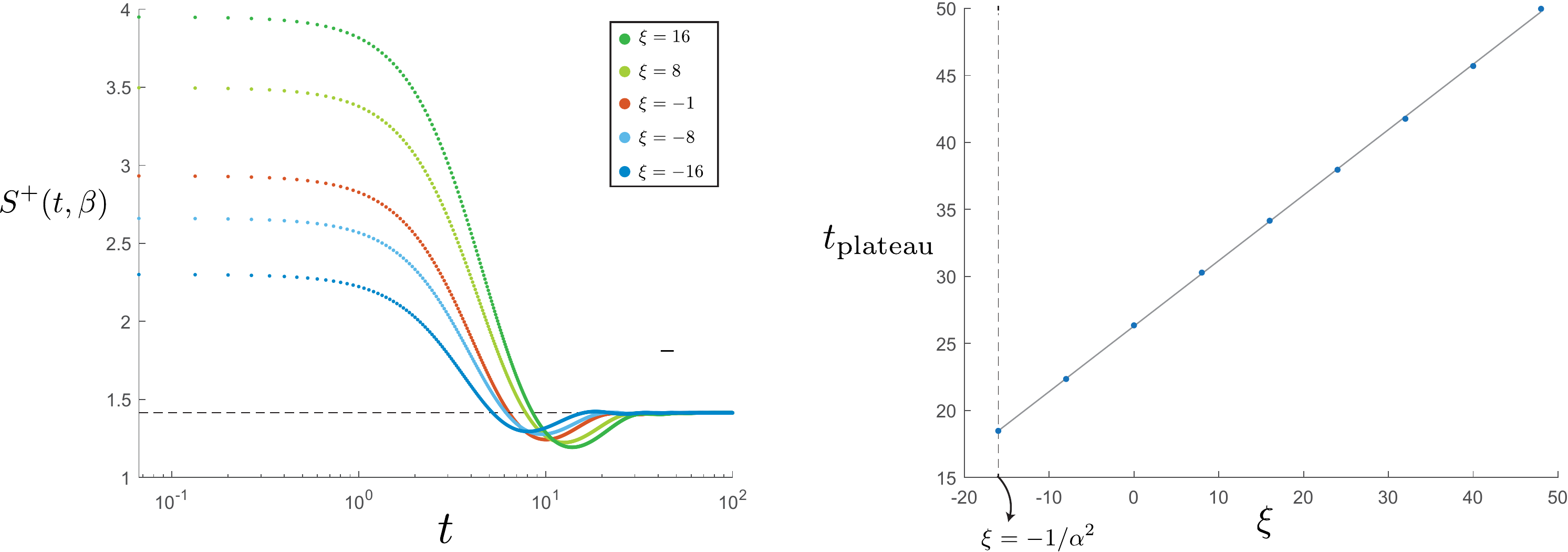}
\caption{On the left diagram we plot the spectral form factor for $\beta=8$ and several values of~${\xi>-1/\alpha^2=16}$, including the red curve with $\xi=-1$. This is obtained by integrating~(\ref{eq:205}) the kernel~${\mathcal{K}(q,\bar{q})}$ numerically computed from the solution to the string equation, shown in figure~\ref{fig:7}. On the right we plot~$t_{\rm plateau}$, the time at which the plateau of height~$\sqrt{2}$ is reached. The gray linear curve provides a good fit of the data.}
\label{fig:17}
\end{figure}

These can be computed using the matrix model description, given in this case by
\begin{equation}
S^+(t,\beta)=
\frac{\langle Z_{\rm MM}^+(\beta+it,\beta-it) \rangle}{\langle Z_{\rm MM}^+(2\beta)\rangle}\ .
\end{equation}
For the undeformed theory this quantity was previously studied in \cite{Johnson:2021owr}. Using the numerical kernel~$\mathcal{K}(q,\bar{q})$ and equation~(\ref{eq:205}) we compute this quantity and plot the final result in the left diagram of figure~\ref{fig:17} for several values of~$\xi\ge -1/\alpha^2$. In every case (including at the phase transition~$\xi=-1$) we obtain the expected $t$ dependence, including the late time plateau that is not captured by~(\ref{eq:208}). Note that in the normalization for the spectral form factor used here~(\ref{eq:207}), the plateau has a height of~$\sqrt{2}$. This comes from from the prefactor in the definition of~$Z_{\rm MM}^+(\beta)$ in~(\ref{eq:61}).

From these results we can extract the time scale $t_{\rm plateau}$ at which $S^+(t,\beta)$ reaches the plateau. As shown in the right diagram of figure \ref{fig:17}, we observe a simple linear dependence in $\xi$. A rough perturbative estimate is given by $t_{\rm plateau}(\xi)\sim Z(2\beta)$, obtained from comparing the connected partition function in (\ref{eq:207}) to the value of the plateau. For $\xi>-1$, before the phase transition, the partition function (\ref{eq:105}) is linear in $\xi$, in agreement with the non-perturbative result in figure \ref{fig:17}. However, this is not the case after the phase transition, where the partition function (\ref{eq:136}) gets a non-linear dependence in $\xi$ through $q_c(\xi)$ obtained from (\ref{eq:133}). 

Altogether, we have shown how the Hermitian matrix model provides a stable non-perturbative definition of deformations of Type 0B JT supergravity by a gas of sharp defects. Most importantly, the divergence of the topological expansion of the Euclidean partition function at the phase transition~${\xi=-1}$ is ultimately fixed by non-perturbative effects.

\subsection{Type 0A JT supergravity}\label{sec:3.2}

We now repeat a similar analysis for the deformations of Type 0A JT supergravity, computing observables beyond $\xi=-1$. A detailed technical description of the method of orthogonal polynomials for complex matrices is given in Appendix \ref{zapp:3}, see also \cite{Dalley:1991qg}. Here we summarize the main ingredients necessary to understand the double scaled model non-perturbatively.

\subsubsection*{Method of orthogonal polynomials}

Similarly as before, a single double scaled model is completely fixed by the couplings~${(\mu,t_k)}$ with~${k\in \mathbb{N}}$ that appear in the potential~$V(MM^\dagger)$ which determines the probability measure~(\ref{eq:77}) (see~(\ref{eq:151}) for the explicit relation). The physics of the model is controlled by a single function~$u(x)$ which is determined from the following string equation~\cite{Dalley:1991qg}
\begin{equation}\label{eq:210}
{\rm String\,\,equation:}
\qquad \qquad
u(x)\mathcal{R}^2-\frac{\hbar^2}{2}\mathcal{R}\mathcal{R}''+\frac{\hbar^2}{4}(\mathcal{R}')^2=0\ ,
\end{equation}
where primes are $x$-derivatives and $\mathcal{R}$ is determined by $t_k$ according to
\begin{equation}
\mathcal{R}[u(x)]=\sum_{k=1}^{\infty}t_kR_{k}[u(x)]\ .
\end{equation}
Here $R_k[u(x)]$ are the Gelfand-Dikii polynomials \cite{Gelfand:1975rn}, functionals of $u(x)$ and its derivatives obtained from the following recursion relation
\begin{equation}
R_{k+1}[u(x)]=
\frac{k+1}{2k+1}
\left[
\int^x d\bar{x}\,u(\bar{x})\partial_x R_k[u(\bar{x})]
+u(x)R_k[u(x)]-\frac{\hbar^2}{2}\partial_x^2R_k[u(x)]
\right]
\ ,
\end{equation}
with $R_0=1$. In this normalization, their $\hbar$ expansion is given by $R_k[u(x)]=u(x)^k+\mathcal{O}(\hbar^2)$, see also~(\ref{eq:152}). After the dust settles, the string equation (\ref{eq:210}) is entirely determined by $t_k$.

Arbitrary observables are then computed form the matrix model kernel $\mathcal{K}(E,\bar{E})$, given by
\begin{equation}\label{eq:211}
\mathcal{K}(E,\bar{E})=
\int_{-\infty}^{\mu}dx
\Psi(x,E)\Psi(x,\bar{E})=
\hbar^2
\frac{\Psi(x,E)\overset{\leftrightarrow}{\partial_x}\Psi(x,\bar{E})}{E-\bar{E}}\bigg|_{x=\mu}\ ,
\end{equation}
where $\overset{\leftrightarrow}{\partial_x}=\overset{\rightarrow}{\partial_x}-\overset{\leftarrow}{\partial_x}$. Similarly as in (\ref{eq:204}), the parameter $\mu$ appears in the $x$ integration variable, with the difference that in this case there is no sum over $s=\pm$. The functions $\Psi(x,E)$ are obtained from the following eigenvalue problem
\begin{equation}\label{eq:212}
\mathcal{H}\Psi(x,E)
=E\Psi(x,E)\ ,
\qquad \qquad
\mathcal{H}= -(\hbar\partial_x)^2+u(x)\ ,
\end{equation}
with $u(x)$ a solution to (\ref{eq:210}). The ensemble average of $Z_{\rm MM}^-(\beta_1,\dots,\beta_n)$ in (\ref{eq:61}) for $n=1,2$ is finally obtained from
\begin{equation}\label{eq:213}
\begin{aligned}
\langle 
Z_{\rm MM}^-(\beta)
\rangle & =2
\int_{0}^{+\infty}
dE\,
\mathcal{K}(E,E)
e^{-\beta E}\ , \\
\langle 
Z_{\rm MM}^-(\beta_1)
Z_{\rm MM}^-(\beta_2)
\rangle_c & =2^2
\int_{0}^{+\infty}
dE\,d\bar{E}
\left[ 
\delta(E-\bar{E})
-
\mathcal{K}(E,\bar{E})
\right]
\mathcal{K}(E,\bar{E})
e^{-\beta_1E}
e^{-\beta_2\bar{E}}\ .
\end{aligned}
\end{equation}
Note that in this case the variable $E$ is positive, as it is obtained by rescaling the eigenvalue of the positive definite matrix $MM^\dagger$.\footnote{As explained in Appendix \ref{zapp:3}, the $E_i$ are obtained from the eigenvalues $\lambda_i$ of $MM^\dagger$ after a proper rescaling in the double scaling limit. In our discussion of the loop equations in the previous section this subtlety was avoided by directly writing $E_i$ as the eigenvalues of $MM^\dagger$ in (\ref{eq:250}).} These expressions are fully non-perturbative. Just like the previous case, the topological expansion comes from doing a perturbative expansion in $\hbar=e^{-S_0}$ when solving the string equation.

\subsubsection*{Fixing the double scaled model}

We can now fix a particular model by picking the values of $(\mu,t_k)$ such that the perturbative expansion of $Z_{\rm MM}^-(\beta_1,\dots,\beta_n)$ reproduces that of the deformation of Type 0A JT supergravity. This can be done by evaluating the leading eigenvalue density for $\xi>-1$. Expanding $u(x)$ in a perturbative expansion $u(x)=\sum_{m=0}^{\infty}u_m(x)\hbar^m$ and inserting in the string equation (\ref{eq:210}), its leading solution is given by
\begin{equation}\label{eq:155}
u_0(x)\left[\,
\sum_{k=1}^{\infty}t_ku_0(x)^k+x
\right]^2=0
\qquad \Longrightarrow \qquad
\begin{cases}
\begin{aligned}
\sum_{k=1}^{\infty}t_ku_0(x)^k+x & = 0\ ,
\qquad x\le 0\ , \\
u_0(x) & = 0 \ ,
\qquad x\ge 0\ .
\end{aligned}
\end{cases}
\end{equation}
This piecewise solution is analogous to (\ref{eq:90}) and ensures continuity as well as the boundary conditions required for the matrix model are satisfied, see Appendix \ref{zapp:3}. From this we can write the differential operator $\mathcal{H}$ in (\ref{eq:212}) and compute its eigenfunctions in the WKB approximation in the same way as done in \cite{Johnson:2021owr}
\begin{equation}
\Psi^{\rm WKB}(x,E)=
\frac{\cos\left[\frac{1}{\hbar}\int_{x_{\rm min}}^{x}d\bar{x}\sqrt{E-u_0(\bar{x})}-\frac{\pi}{4}\right]}{\sqrt{\pi \hbar} (E-u_0(x))^{1/4}}\ ,
\qquad \qquad
x>x_{\rm min}\ ,
\end{equation}
where $x_{\rm min}$ is obtained from $E=u_0(x_{\rm min})$. From this one can write the kernel (\ref{eq:211}), whose diagonal components determine the eigenvalue spectral density. The leading perturbative behavior is given by
\begin{equation}\label{eq:157}
\rho_0^-(E)=
\frac{1}{2\pi}
\int_{-\infty}^{\mu}
dx
\frac{\Theta(E-u_0(x))}{\sqrt{E-u_0(x)}}\ .
\end{equation}
For the case of positive $\mu$, there are two contributions at either side of $x=0$, which can be written as
\begin{equation}\label{eq:163}
\rho_0^-(E)=
\frac{1}{2\pi \sqrt{E}}
\left[
\mu
+
\sum_{k=1}^{\infty}t_k
\frac{(4E)^{k}k!^2}{(2k)!}
\right]\ ,
\end{equation}
where we have changed the integration variable to $u_0$, computed the Jacobian using (\ref{eq:155}) and solved the integral for each separate $k$. Each critical model $t_k$ contributes with $E^{k-1/2}$ to the spectral density. We should compare this general expression to the leading density (\ref{eq:72}) that is required so that the loop equations match with the supergravity topological expansion. This fixes the coefficients to
\begin{equation}\label{eq:156}
(\mu,t_k)=\sqrt{2}
\left(
1+\xi,\frac{\pi^{2k}}{k!^2}
(1+\xi \alpha^{2k})
\right)\ ,
\end{equation}
generalizing the $\xi=0$ case considered in \cite{Johnson:2020heh}. Note that these parameters, which completely determine the double scaled complex matrix model, take essentially the same values as in the Hermitian case (\ref{eq:91}). The implicit constraint for $u_0(x)$ in (\ref{eq:155}) becomes
\begin{equation}\label{eq:158}
I_0(2\pi\sqrt{u_0})-1
+\xi
\big(
I_0(2\pi \alpha \sqrt{u_0})-1
\big)+x/\sqrt{2}=0\ ,
\qquad  x\le 0\ ,
\end{equation}
while $u_0(x)=0$ for $x\ge 0$. Similarly as before, we can use this to check the leading solution $u_0(x)$ in (\ref{eq:155}) is continuous at $x=0$ only when $\xi\ge -1/\alpha^2$. In this section we restrict $\xi$ so that $u_0(x)$ is continuous, leaving the more general case to section \ref{sec:4}.

\subsubsection{Phase transition at \texorpdfstring{$\xi=-1$}{xi}}

\begin{figure}
\centering
\includegraphics[scale=0.38]{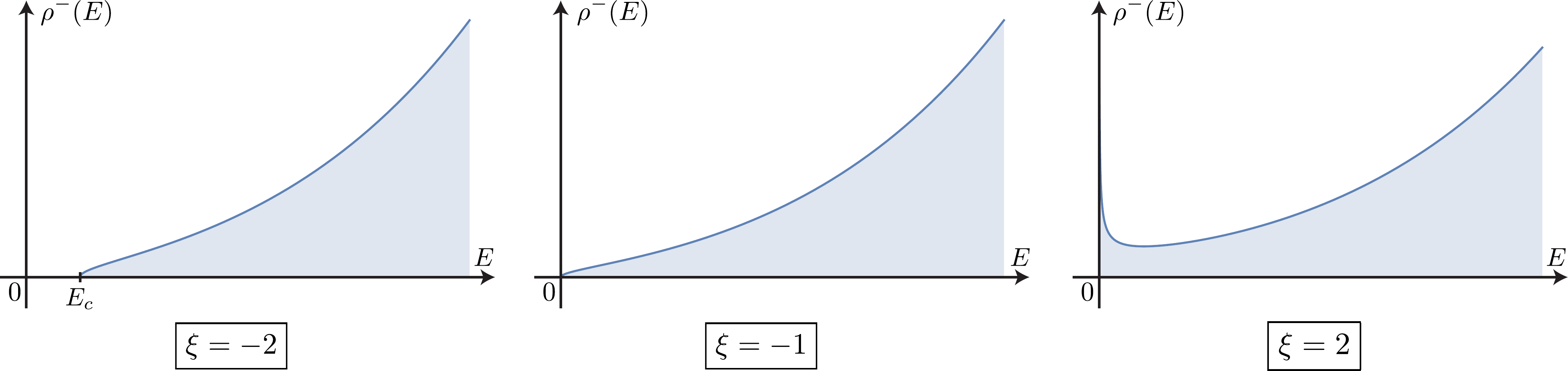}
\caption{Leading eigenvalue spectral density (\ref{eq:157}) for the complex matrix model describing Type~0A JT supergravity with~$\alpha=1/4$. At~$\xi=-1$ there is a phase transition between the hard-edge phase~(${\xi>-1}$) and soft-edge phase~(${\xi<-1}$). The behavior is completely analogous to the one shown in figure~\ref{fig:1} before double scaling, after zooming in to the region $\lambda\sim 0$.}\label{fig:10}
\end{figure}

As $\xi<-1$, the sign of $\mu$ in (\ref{eq:156}) changes and the topological expansion of the supergravity theory breaks down. However, in this formalism the leading spectral density (\ref{eq:157}) is perfectly well defined and can be rearranged into
\begin{equation}\label{eq:175}
\rho_0^-(E)=
\int^{E}_{E_c}
du_0
\frac{I_1(2\pi\sqrt{u_0})+\xi \alpha I_1(2\pi \alpha \sqrt{u_0})}{\sqrt{2u_0(E-u_0)}}
\qquad {\rm where} \qquad
E_c\equiv u_0(\mu)\ .
\end{equation}
The threshold value $E_c$ is obtained from evaluating (\ref{eq:158}) at $x=\mu=\sqrt{2}(1+\xi)$. Note this gives the same condition as for $q_c$ in (\ref{eq:133}) for the Hermitian matrix model, after identifying $E_c\rightarrow q_c^2$. In figure~\ref{fig:10} we plot $\rho_0^-(E)$ for several values of $\xi$. There is a transition at $\xi=-1$ which interpolates between the hard and soft-edge phase (compare with figure \ref{fig:1} for the transition before double scaling). Each critical potential is controlled by $t_k$ and its contribution to the leading spectral density can be better understood by rewriting (\ref{eq:175}) as
\begin{equation}\label{eq:165}
\rho_0^-(E)=
\sqrt{E-E_c}
\sum_{k=1}^{\infty}
\frac{t_k k}{\pi}
 E^{k-1}
\,_2F_1\left[\frac{1}{2},1-k,\frac{3}{2},1-\frac{E_c}{E}\right]\ .
\end{equation}
From this we see each $t_k$ contributes $P(E)\sqrt{E-E_c}$, where $P(E)$ is a polynomial of order $k-1$. This is very different to the behavior of the spectral density before the transition (\ref{eq:163}). Just like in the previous case, this can be interpreted as the new phase breaking supersymmetry by large effects. The new density of states for the $\xi<-1$ phase has the same shape as a bosonic dilaton gravity theory, although this is not necessarily true for higher genus corrections. A discussion on the behavior of the $k=1$ matrix model as $\mu$ goes from positive to negative appeared previously in \cite{Johnson:2019eik}.

\subsubsection*{New perturbative expansion}

Let us analyze how the perturbative expansion of the observable $Z_{\rm MM}^-(\beta_1,\dots,\beta_n)$ in (\ref{eq:61}) changes as one goes across the phase transition at $\xi=-1$. For example, in gravity, or in the $\xi>-1$ phase, all path integrals with $n>2$ boundaries vanish to leading order in the genus expansion, again due to fermion zero-modes. This is not true anymore for the $\xi<-1$ phase where supersymmetry is broken. Higher genus corrections are also different, although this is not as dramatic of a change for Type 0A as for Type 0B, since in the former they do not vanish generically in the $\xi>-1$ phase.

To compute the answer for $n>2$ boundaries at genus zero, we can use the compact formula~(\ref{eq:98}), which was explicitly shown in~\cite{Ambjorn:1990ji} to also hold for complex matrix models
\begin{equation}
Z_{{\rm MM},0}^-(\beta_1,\dots,\beta_n)=
2^{n}
\frac{\sqrt{\beta_1\cdots\beta_n}}{2\pi^{n/2}\beta_T}
\left[
\partial_x^{n-2}
e^{-\beta_T u_0(x) }
\right]_{x=\mu}\ ,
\qquad \qquad
\beta_T\equiv\sum_{i=1}^n\beta_i.
\end{equation}
The factor $2^n$ comes from the definition of the matrix operator in (\ref{eq:61}). To compare with the Hermitian case, it is useful to rewrite this formula in terms of $\bar{u}_0(x)=\sqrt{u_0(x)}$ and $\bar{x}=x/\sqrt{2}$, so that it becomes
\begin{equation}\label{eq:164}
Z_{{\rm MM},0}^-(\beta_1,\dots,\beta_n)=
2^{\frac{n}{2}+1}
\frac{\sqrt{\beta_1\cdots\beta_n}}{2\pi^{n/2}\beta_T}
\left[
\partial_{\bar{x}}^{n-2}
e^{-\beta_T \bar{u}_0(\bar{x})^2 }
\right]_{\bar{x}=\mu/\sqrt{2}}\ .
\end{equation}
From the implicit constraints in (\ref{eq:158}) and (\ref{eq:96}) note the functions $\bar{u}_0(\bar{x})$ and $r_0(x)$ coincide. Moreover, the values of $\mu$ in each case (\ref{eq:156}) and (\ref{eq:91}) are precisely related by a factor of $\sqrt{2}$, so that the Hermitian (\ref{eq:98}) and complex (\ref{eq:164}) observables precisely agree to leading order in genus and arbitrary $n$
\begin{equation}
Z_{{\rm MM},0}^-(\beta_1,\dots,\beta_n)
=Z_{{\rm MM},0}^+(\beta_1,\dots,\beta_n)\ .
\end{equation}
This shows the leading genus matching for $\xi>-1$ also applies after going through the phase transition at $\xi=-1$. From the gravitational perspective this is expected, given that for genus zero the sum over spin structures (which is the feature that differentiates the Type 0A/0B theories) is insensitive to the insertion of $(-1)^\zeta$. In particular, it means the transition for both supergravity theories is of second order, since the calculation around (\ref{eq:135}) only involved the leading behavior of the partition function.

To compute higher genus corrections to observables it is more convenient to use the loop equations of section \ref{sec:3}. Using the new density of states (\ref{eq:165}) in the $\xi<-1$, a careful calculation shows the first genus correction is
\begin{equation}
Z_{{\rm MM},1}^-(\beta)=
\frac{-1}{48a_1^2}
\sqrt{\frac{\beta}{\pi}}
\left[
4a_1\beta+a_2(4\beta E_c-3)
\right]
e^{-\beta E_c}\ ,
\end{equation}
where $a_k=\big[\partial_z^k\big(\sqrt{E_c-z}y(z)\big)\big]_{z=E_c}$ with $y(z)$ the spectral curve associated to (\ref{eq:165}). Comparing with the result for the Hermitian case $Z_{{\rm MM},1}^+(\beta)$ in (\ref{eq:166}) (noting the definition of $a_k$ in each case is different) one finds $Z_{{\rm MM},1}^+(\beta)\neq Z_{{\rm MM},1}^-(\beta)$. This shows the perturbative expansions of each theory only agree for $g=0$, as expected from the gravitational perspective.


\subsubsection{Non-perturbative effects}

\begin{figure}
\centering
\includegraphics[scale=0.57]{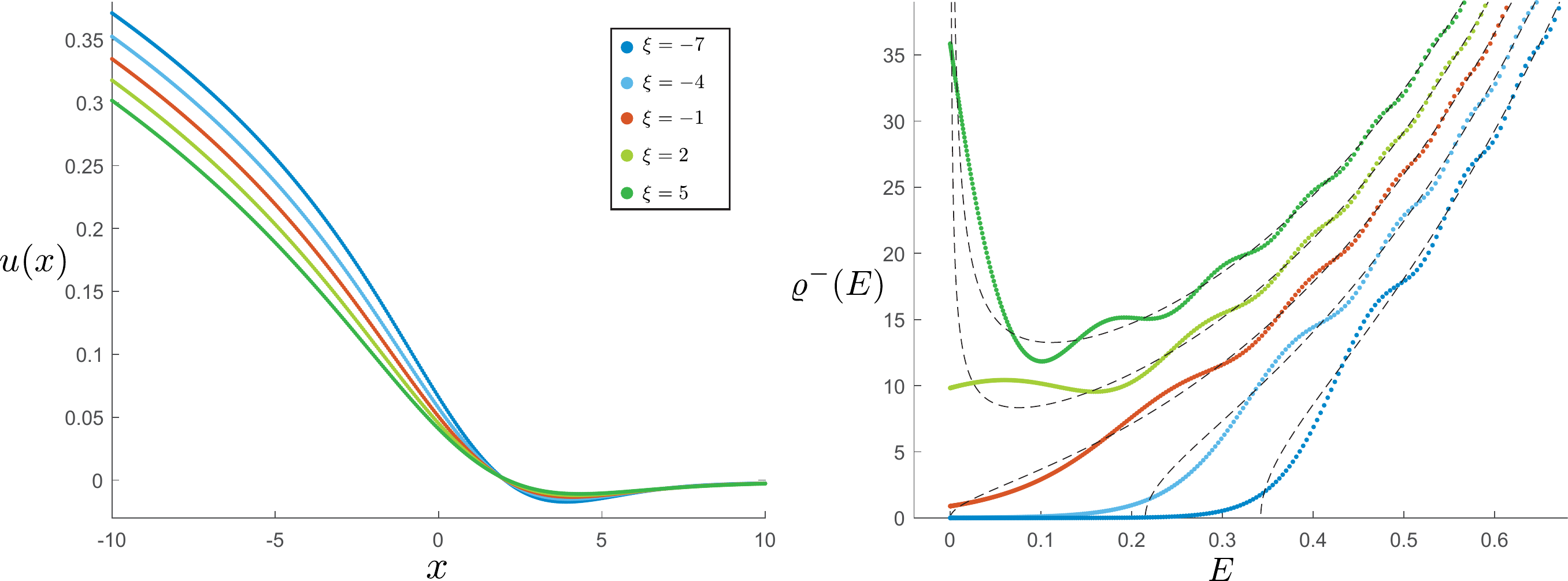}
\caption{On the left diagram we plot the numerical solution $u(x)$ to the string equation of the complex matrix model (\ref{eq:210}) with $(k_{\rm max},\hbar,\alpha)=(6,1,1/4)$ and several values of $\xi$. On the right, we have the corresponding energy spectral density $\varrho^-(E)$ of deformations of Type 0A JT supergravity obtained from the diagonal components of the matrix model kernel $\mathcal{K}(E,\bar{E})$ in (\ref{eq:211}) obtained from the eigenfunctions $\Psi(x,E)$ in~(\ref{eq:212}).}\label{fig:12}
\end{figure}

We now turn our attention to non-perturbative contributions to observables, which become specially important at the phase transition~${\xi=-1}$. To do this, we have to exactly solve the string equation~(\ref{eq:210}) for~$u(x)$ with the coefficients~$t_k$ in~(\ref{eq:156}). Since the string equation is a differential equation of infinite order, we must first introduce a truncation~$k_{\rm max}$ according to~\cite{Johnson:2020exp}
\begin{equation}
\mathcal{R}=\sum_{k=1}^{k_{\rm max}}
t_kR_k[u(x)]\ .
\end{equation}
Same as before we take $(k_{\rm max},\hbar,\alpha)=(6,1,1/4)$, meaning $\xi\ge -1/\alpha^2\ge -16$. The boundary conditions at $x\rightarrow \pm \infty$ are fixed by the leading genus solution $u_0(x)$ in (\ref{eq:155}). In the left diagram of figure \ref{fig:12} we show the numerical solution $u(x)$ to the string equation (\ref{eq:210}) for several values of $\xi$. This includes the critical $\xi=-1$ case (red curve) which shows a smooth and well behaved solution. 

Using these solutions we can construct the operator $\mathcal{H}$ in (\ref{eq:212}) and numerically compute its eigenfunctions $\Psi(x,E)$. See \cite{Johnson:2020exp} for details regarding the numerical method. By appropriately integrating these eigenfunctions, one obtains the matrix model kernel (\ref{eq:211}) that determines all observable in the matrix model (\ref{eq:213}).

\paragraph{Spectral density:} From the first equation in~(\ref{eq:213}), the eigenvalue spectral density is clearly obtained from the diagonal components of the kernel. The energy spectral density of the supergravity theory~$\varrho^-(E)$ (defined from the inverse Laplace transform of~$Z^-(\beta)$) is the same as~$\langle \rho^-(E) \rangle$ (apart from a factor of~$2$ coming from the normalization of~$Z_{\rm MM}^-(\beta)$ in~(\ref{eq:61})). For that reason, in this section we directly work in terms of the supergravity spectral density as computed from the matrix model. 

\begin{figure}
\centering
\includegraphics[scale=0.53]{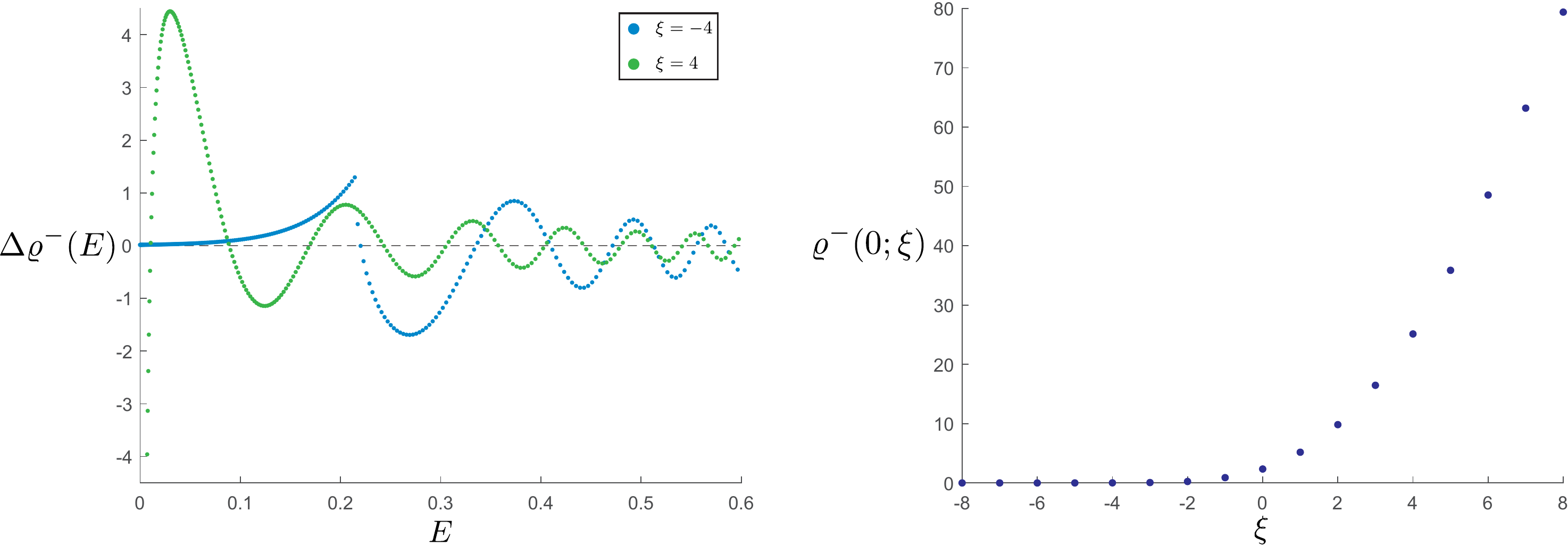}
\caption{The left diagram shows the full gravitational spectral density of figure \ref{fig:12} where we have substracted the perturbative contribution $\Delta \varrho^-(E)=\varrho^-(E)-\varrho^-_0(E)/\hbar$ for $\xi=\pm 4$. On the right we plot the ground state degeneracy $\varrho^-(E=0;\xi)$ as a function of the deformation parameter $\xi$.}\label{fig:13}
\end{figure}

In the right diagram of figure~\ref{fig:12} we plot the final result for~$\varrho^-(E)$ for several values of~$\xi$ at either side of the transition at~$\xi=-1$. The dashed line gives the leading result, with non-perturbative contributions generating fluctuations around it. Similarly as before, the spectral density is completely well defined when we set~$\xi=-1$, showing the divergence in the topological expansion observed in~(\ref{eq:48}) is cured by non-perturbative effects. These are most easily appreciated in the left diagram of figure \ref{fig:13}, were we plot~$\Delta \varrho^-(E)=\varrho^-(E)-\varrho^-_0(E)/\hbar$ for two values of $\xi$ at either side of the transition. Note that the amplitude of the oscillations are of roughly the same order, which is not the case for the Type 0B JT supergravity (compare with figure~\ref{fig:9}).

One of the most interesting aspects of the spectral densities is their low energy behavior, previously studied in \cite{Johnson:2020mwi} for the undeformed JT supergravity. While in the $\xi>-1$ phase the leading vacuum value diverges, non-perturbative effects dramatically change the spectrum so that $\varrho^-(E=0)$ takes a finite value instead. In the right diagram of figure \ref{fig:13} we plot the behavior of~$\varrho^-(0;\xi)$ as a function of $\xi$, which goes to zero as~$\xi<-1$. Although this might be expected from the previous perturbative analysis (see figure \ref{fig:10}), note there are still non-perturbatively small contributions at~${E=0}$ even for $\xi<-1$. This come from the exponential tails of the eigenfunctions $\Psi(x,E)$ that appear in the quantum region, i.e. $x$ values where $E<u(x)$. In the following section, we show how a true non-perturbative energy gap appears when $\xi<-1/\alpha^2$ for deformations of the Type 0A theory.

Before moving on we should mention the behavior of the spectrum when going from positive to negative $\mu$ observed in figures \ref{fig:12} and \ref{fig:13} is analogous to that obtained for the $k=1$ model studied in~\cite{Johnson:2019eik}.

\begin{figure}
    \centering
    \includegraphics[scale=0.55]{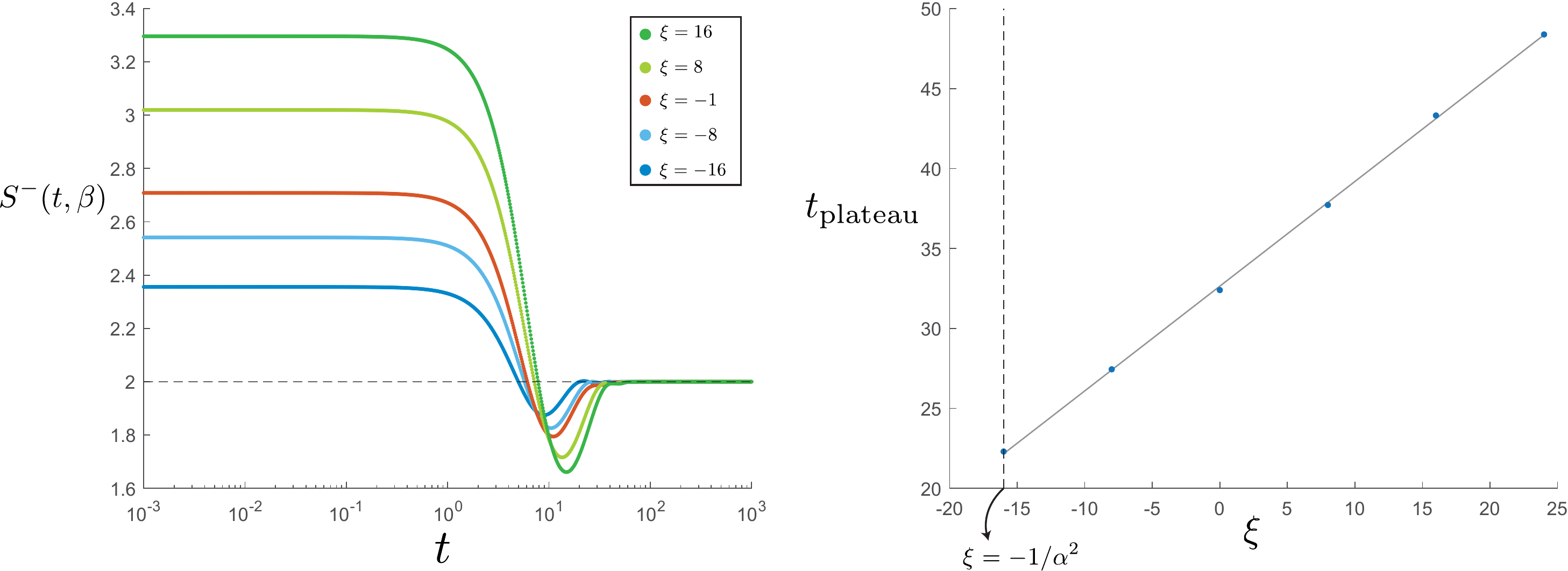}
    \caption{On the left diagram we plot the time dependence of the spectral form factor $(\alpha,\beta,\hbar)=(1/4,10,1)$ and several values of $\xi$. On the right, we have extracted the time $t_{\rm pleateau}$ at which the plateau is reached for several values of $\xi\ge -1/\alpha^2$.}
    \label{fig:18}
\end{figure}

\paragraph{Spectral form factor:} We now turn our attention to the spectral form factor, defined in exactly the same way as before (\ref{eq:207}) after replacing $Z^+(\beta_1,\dots,\beta_n)\rightarrow Z^-(\beta_1,\dots,\beta_n)$. Non-perturbative contributions are crucial in order to capture large $t$ dependence, meaning the spectral form factor must be computed using the matrix model according to
\begin{equation}
S^-(t,\beta)=
\frac{\langle Z_{{\rm MM}}^-(\beta+it,\beta-it) \rangle}{\langle Z_{\rm MM}^-(2\beta) \rangle}\ .
\end{equation}
For the undeformed theory~${\xi=0}$, this quantity was carefully discussed and computed in~\cite{Johnson:2020exp}. Using~(\ref{eq:213}) this becomes certain integrals of the kernel~$\mathcal{K}(E,\bar{E})$ that includes its non-diagonal components. In the left diagram of figure~\ref{fig:18} we show the final result for several values of~$\xi$. As expected, they all exhibit the dip and ramp as well as the late time plateau. In this case, the value of the plateau is~$2$, which comes from the overall factor of~$Z_{\rm MM}^-(\beta)$ in~(\ref{eq:61}). On the right diagram of figure~\ref{fig:18} we plot the time it takes to reach the plateau as a function of~$\xi$. Same as in the Type 0B case, it exhibits a linear behavior.

Altogether, we have shown how the complex random matrix model provides a stable non-perturbative completion of Type 0A JT supergravity deformed by sharp defects. The divergence of the topological expansion at the phase transition $\xi=-1$ is cured by non-perturbative effects.


\section{Another phase transition}
\label{sec:4}

In the previous section we showed how the random matrix description provides a stable non-perturbative completion of the deformations of JT supergravity theories across the transition at~${\xi=-1}$ (blue and green shaded regions of figure~\ref{fig:11}). However, the analysis was restricted to~${\xi\ge -1/\alpha^2}$ given that it is only in this regime that the leading solutions~$r_0(x)$ and~$u_0(x)$ in~(\ref{eq:90}) and~(\ref{eq:155}) are continuous. Building on~\cite{Dalley:1991yi,Dalley:1991xx,Johnson:2020lns}, in this section we show how to appropriately fix the matrix model in order to obtain a stable description of the theories with~${\xi<-1/\alpha^2}$. 


Let us start with a more detailed discussion of the issue, analyzing the string equations of the Hermitian (\ref{eq:206}) and complex (\ref{eq:210}) matrix models. As recently explained in \cite{Johnson:2020lns}, perhaps the first thing one should examine to avoid potential non-perturbative instabilities is whether the leading solutions $u_0(x)$ and $r_0(x)$ are single valued and continuous. Problems can arise at $x=0$, where there is a change of behavior in their piecewise definitions, see (\ref{eq:90}) and (\ref{eq:155}). Given that both functions are equivalent after a simple redefinition (see the discussion around equation (\ref{eq:164})) it is enough to explain the issue for $u_0(x)$, as equivalent statements apply to the Hermitian model.

\begin{figure}
\centering
\includegraphics[scale=0.42]{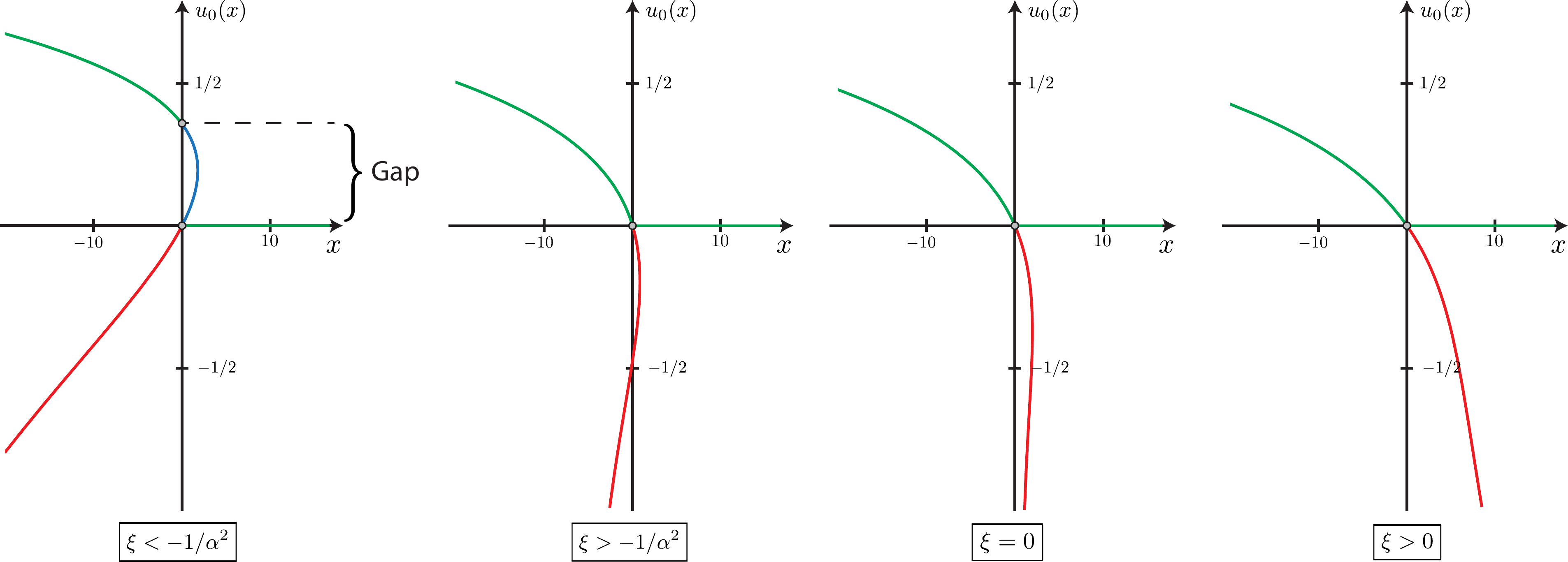}
\caption{Leading genus solution $u_0(x)$ obtained from (\ref{eq:155}) with $t_{k}$ in (\ref{eq:156}) (see also (\ref{eq:158})). For~${\xi<-1/\alpha^2}$ the single valued solution (green sections of the curve) develops a gap and is discontinuous at $x=0$. A completely analogous behavior is obtained for $r_0(x)$ in (\ref{eq:90}) for the Hermitian matrix model.}
\label{fig:4}
\end{figure}

In figure \ref{fig:4} we plot the curve $x(u_0)$ in (\ref{eq:158}) which determines the negative $x$ behavior of $u_0(x)$, as well as $u_0(x)=0$ for $x>0$. The green portions of the curve correspond to the unique single valued solution for $x\in \mathbb{R}$. While for $\xi\ge -1/\alpha^2$ the single valued solution is also continuous, when~${\xi<-1/\alpha^2}$ there is a shift in the value at $x=0$ and $u_0(x)$ becomes discontinuous at the origin (see the leftmost diagram in figure \ref{fig:4}).

Should one worry about the discontinuity of $u_0(x)$? Yes and no, depending on whether one is interested on perturbative or non-perturbative contributions to observables. Since all perturbative effects are controlled by the leading spectral density, we should first analyze its behavior. From the general formula for $\rho_0^-(E)$ in (\ref{eq:157}) we see it is obtained from integrating $u_0(x)$ in $x\in(-\infty,\mu]$. Given that $\mu$ is negative when $\xi<-1/\alpha^2$, the leading spectral density and spectral curve $y(z)$ are not sensitive to the discontinuity at $x=0$. As a result, the perturbative expansion is perfectly well defined and still determined by the spectral curve obtained from (\ref{eq:175}) even when $\xi<-1/\alpha^2$. 

\begin{figure}
\centering
\includegraphics[scale=0.50]{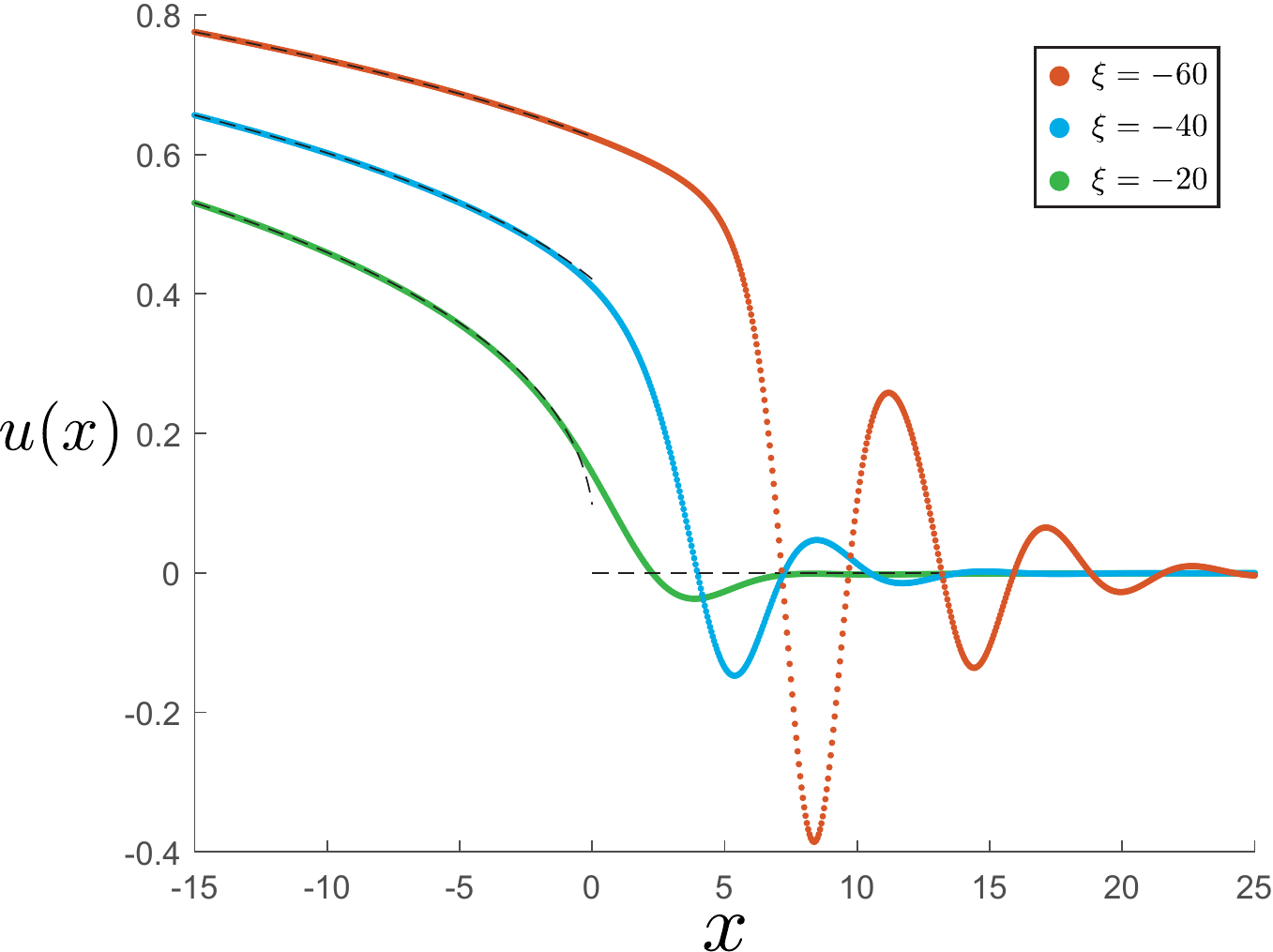}
\caption{Solutions to the complex matrix model string equation (\ref{eq:210}) with $t_k$ in (\ref{eq:156}), ${(k_{\rm max},\hbar,\alpha)=(6,1,1/4)}$ and several values of $\xi<-1/\alpha^2$. The dashed curves give the perturbative result $u_0(x)$ (\ref{eq:155}), which is discontinuous at the origin. As $\xi$ becomes more negative the solutions are harder to obtain and their behavior at the interior becomes more singular. We have checked these results are independent of the size and number of the points of the grid used to solve the differential equation.}\label{fig:14}
\end{figure}

The situation is quite different if one is interested in computing observables including non-perturbative contributions. In that case, one must instead construct the operator~$\mathcal{H}=-(\hbar\partial_x)^2+u(x)$ and compute its spectrum. To first approximation, it might be tempting to replace~$u(x)$ in~$\mathcal{H}$ by~$u_0(x)$ and solve for its eigenfunctions~$\Psi(x,E)$. However, the discontinuity of~$u_0(x)$ at~$x=0$ yields the spectral problem ill defined and it is not possible to proceed. Another approach is to directly solve the full non-perturbative string equation~(\ref{eq:210}) for~$u(x)$ in the regime~${\xi<-1/\alpha^2}$ and then compute the spectrum of~$\mathcal{H}$. Doing so, one finds that as~$\xi$ takes larger negative values the numerical solution~$u(x)$ becomes more unstable and develops large oscillations near~$x=0$, see figure~\ref{fig:14} for several examples. This is because the differential equation is struggling to find a smooth solution with the corresponding boundary conditions at large~$|x|$. All in all, while the discontinuity of~$u_0(x)$ does not affect perturbation theory, it yields a model that seems to be non-perturbatively unstable.

\subsection{Type 0A JT supergravity: Non-perturbative gap}
\label{sec:4.1}

Instead of starting with the Hermitian matrix model corresponding to the Type 0B case, it is convenient to consider the Type 0A theory, as the discussion for the complex matrix model arises more naturally. A similar issue as the one arising here for $\xi<-1/\alpha^2$ was uncovered in~\mcite{Johnson:2020lns} when studying a complex matrix model description of deformations of ordinary JT gravity. Building on~\mcite{Dalley:1991yi,Dalley:1991xx} it was shown there is a natural parameter that arises in complex matrix models which after double scaling results in a slightly different string equation that gives a single-valued and continuous solution~$u_0(x)$. Let us describe how this works.

In the complex matrix model the eigenvalues of $MM^\dagger$ are lower bounded by $\lambda_{\rm min}=0$. The basic idea is to appropriately move away from this lower bound by considering instead
\begin{equation}\label{eq:214}
\lambda_{\rm min}=2E_0\delta^2\ ,
\qquad \qquad E_0\ge 0\ ,
\end{equation}
where $\delta\rightarrow 0$ in the double scaling limit (see Appendix \ref{zapp:3} for the notation being used here). The role of $E_0$ is not to modify the functional form of the potential $V(MM^\dagger)$, but can be instead interpreted as a shift of its argument, so that the matrix partition function in (\ref{eq:250}) becomes
\begin{equation}
\mathcal{Z}=\int dM\,
e^{-N\,{\rm Tr} \, V(MM^\dagger+\lambda_{\rm min})}
=\prod_{i=1}^N
\int_{\lambda_{\rm min}}^{+\infty}
d\lambda_i\, \Delta(\lambda_1,\dots,\lambda_N)^2
e^{-N V(\lambda_i)}\ ,
\end{equation}
where the Vandermonde determinant is insensitive to the constant shift of the eigenvalues.\footnote{As explained in Appendix \ref{zapp:3}, the $E_i$ are obtained from the eigenvalues $\lambda_i$ after a proper rescaling in the double scaling limit. In our discussion of the loop equations in the section \ref{sec:2.3} this subtlety was avoided by directly writing $E_i$ as the eigenvalues of $MM^\dagger$ in (\ref{eq:250}).}


Since our discussion of the method of orthogonal polynomials in Appendix \ref{zapp:3} already included an arbitrary value for $\lambda_{\rm min}$, it is straightforward to work out the effect of $E_0$ in the double scaling limit. The relevant discrete string equations are given in (\ref{eq:145}) with the scaling ansatz for all the parameters indicated in table \ref{table:2} and the potential in (\ref{eq:151}). Using the same approach which lead to the string equation (\ref{eq:210}) but with $E_0\neq 0$ one readily finds
\begin{equation}\label{eq:172}
(u-E_0)
\mathcal{R}^2-\frac{\hbar^2}{2}\mathcal{R}\mathcal{R}''
+\frac{\hbar^2}{4}(\mathcal{R}')^2=0\ ,
\end{equation}
where~$\mathcal{R}=\sum_{k=1}^{\infty}t_kR_k[u(x)]+x$. A matrix model derivation of this equation from a similar setup was first given in~\cite{Dalley:1991xx}. Note the role of~$E_0$ and~$t_k$ in the string equation is very different, given that~$\mathcal{R}$ is independent of~$E_0$.

\begin{figure}
\centering
\includegraphics[scale=0.5]{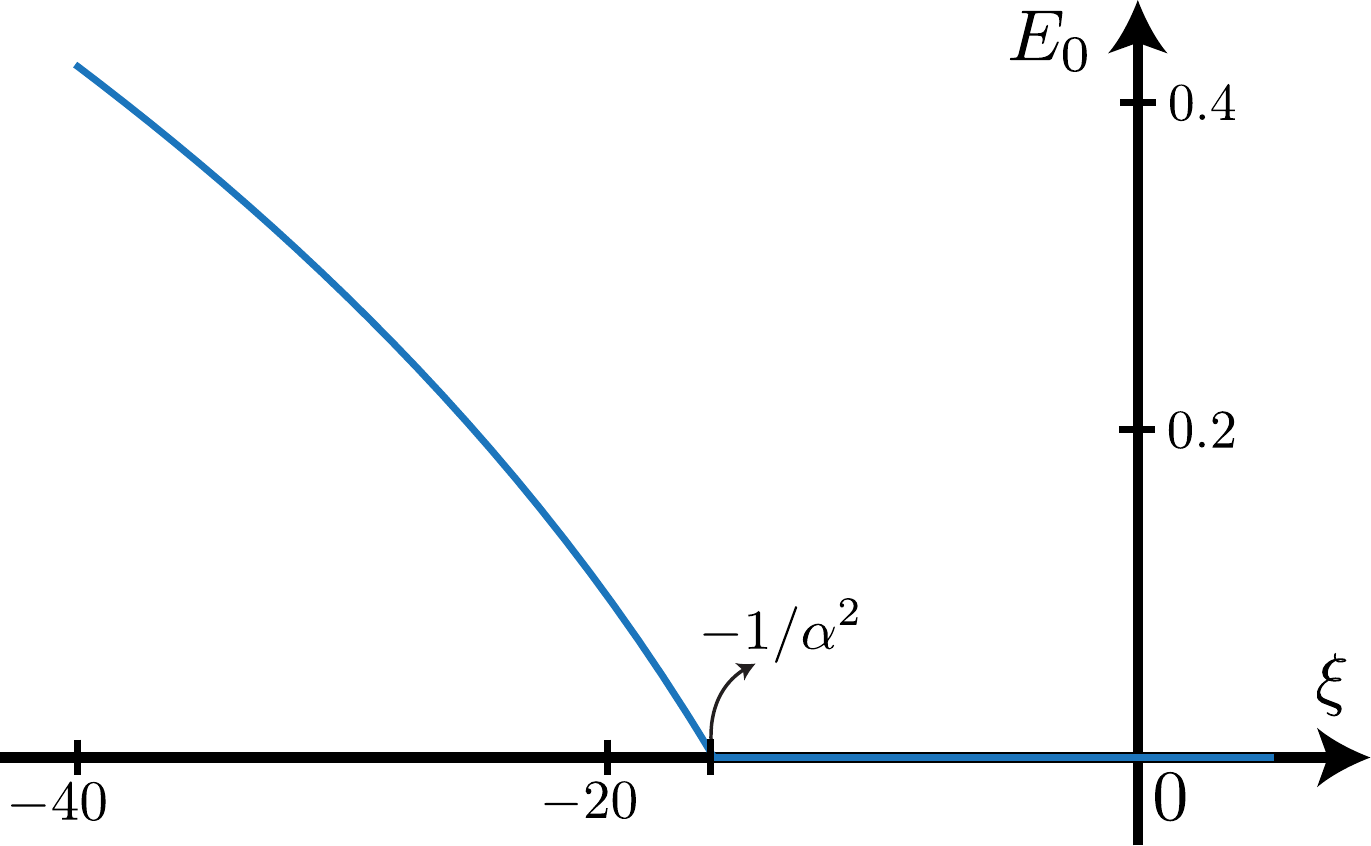}
\caption{Plot of $E_0(\xi)$ obtained from numerically solving (\ref{eq:174}) for $\alpha=1/4$. When $\xi\ge -1/\alpha^2$ one finds $E_0=0$.}\label{fig:15}
\end{figure}

Compared to (\ref{eq:210}), the leading solution $u_0(x)$ now gets modified by $E_0$ in the following way
\begin{equation}\label{eq:173}
\left(
u_0(x)-E_0
\right)
\left[\,
\sum_{k=1}^{\infty}t_ku_0(x)^k+x
\right]^2=0
\qquad \Longrightarrow \qquad
\begin{cases}
\begin{aligned}
\sum_{k=1}^{\infty}t_ku_0(x)^k+x & = 0\ ,
\qquad \,\,\,\, x\le 0\ , \\
u_0(x) & = E_0 \ ,
\qquad x\ge 0\ .
\end{aligned}
\end{cases}
\end{equation}
By appropriately fixing the parameter $E_0$ one can always ensure the solution $u_0(x)$ is continuous at the origin. One should not think of the parameters $E_0$ and $(\mu,t_k)$ as being at the same footing. While fixing $(\mu,t_k)$ is equivalent to choosing a particular double scaled model, the value of $E_0$ is determined by requiring it is consistent, i.e. $u(x)$ is continuous at the origin. In particular, for the matrix model describing deformations of JT supergravity (\ref{eq:158}) $E_0$ is determined from the largest solution to the following equation
\begin{equation}\label{eq:174}
\frac{1}{\sqrt{2}} \sum_{k=1}^{\infty}t_kE_0^k=
\big(I_0(2\pi\sqrt{E_0})-1\big)
+\xi
\big(I_0(2\pi\alpha \sqrt{E_0})-1\big)
=0\ ,
\qquad \qquad
E_0\ge 0\ .
\end{equation}
Perturbatively solving for $E_0(\xi)$ around $\xi=-1/\alpha^2$ one finds
\begin{equation}
E_0(\xi)=-
\frac{4\alpha^2(\xi+1/\alpha^2)}{\pi(1-\alpha^2)}
\left[ 
1+\frac{\alpha^2(4-5\alpha^2)}{9(1-\alpha^2)}
(\xi+1/\alpha^2)+\mathcal{O}(\xi+1/\alpha^2)^3
\right]\Theta(-\xi-1/\alpha^2)\ .
\end{equation}
In figure \ref{fig:15} we plot the full numerical solution $E_0(\xi)$ to (\ref{eq:174}) for $\alpha=1/4$.

\begin{figure}
\centering
\includegraphics[scale=0.60]{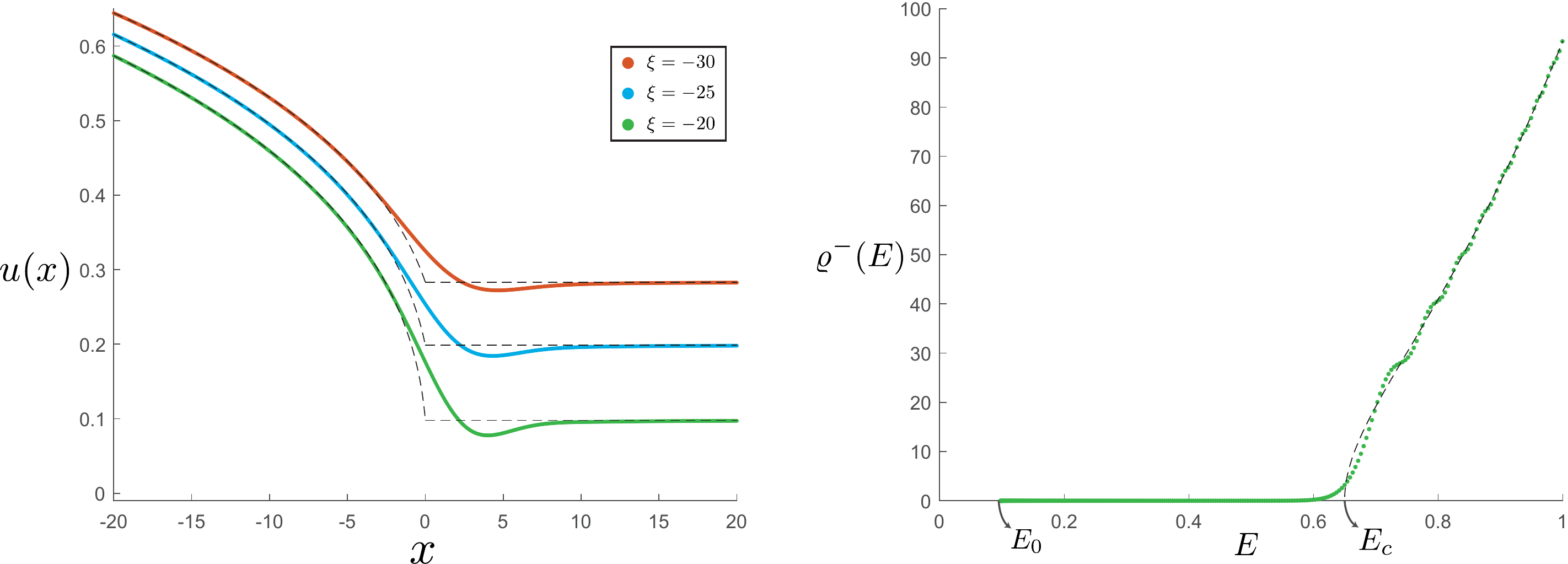}
\caption{In the left diagram we plot the full solution to the string equation (\ref{eq:172}) with ${(k_{\rm max},\hbar,\alpha)=(6,1,1/4)}$ and several values of $\xi<-1/\alpha^2=-16$, which give $E_0\neq 0$. After computing the kernel (\ref{eq:211}) we obtain the Type 0A spectral density for the $\xi=-20$ case. We observe a non-perturbative gap at $E_0\sim 0.1$.}\label{fig:16}
\end{figure}

Taking into account $E_0\neq 0$, we can solve the full string equation (\ref{eq:172}) and obtain the smooth and well behaved solutions shown in the left diagram of figure \ref{fig:16}. Note that the boundary condition for large positive $x$ (determined by $u_0(x)$) is now given by $\lim_{x\rightarrow +\infty}u(x)=E_0$. The dashed curves correspond to the leading solution $u_0(x)$ in (\ref{eq:173}), which is now continuous at the origin. The full solutions are much better behaved than the ones displayed in figure \ref{fig:14}, which were built from a discontinuous $u_0(x)$.

Let us now explain why $E_0$ is identified as a non-perturbative energy gap for the deformations of Type 0A JT supergravity. The eigenvalue spectral density is exactly computed from the diagonal components of the kernel (\ref{eq:211}), where $\Psi(x,E)$ are the eigenfunctions of $\mathcal{H}=-(\hbar\partial_x)^2+u(x)$. The spectrum of the scattering states of the Schrodinger operator $\mathcal{H}$ with potential $u(x)$ determines the matrix model spectrum $\langle \rho^-(E) \rangle$ and therefore the supergravity denity $\varrho^-(E)$. Since the asymptotic behavior for large negative $x$ generates an infinite barrier, the scattering state with minimum energy is determined from $\lim_{x\rightarrow +\infty}u(x)=E_0$. The gravitational spectral density is therefore supported on
\begin{equation}
\varrho^-(E)\neq 0
\qquad \Longleftrightarrow \qquad 
E\in[E_0,+\infty)\ .
\end{equation}
It is important to highlight the difference between~$E_0$ and the threshold value~$E_c$. To start, the implicit constraint which determines each of them in~(\ref{eq:133}) and~(\ref{eq:174}) is not the same. Although~$E_c$ also gives a gap in which the spectral density vanishes, this is only true for the leading contribution~$\varrho_0^-(E)$, obtained from~(\ref{eq:175}). Exponentially small non-perturbative corrections are still present at zero energy, as we can see from the blue curves in the right diagram of figure~\ref{fig:12}. This is a consequence of the asymptotic behavior of the solutions~$u(x)$ in the left diagram of that figure.

To confirm these expectations, we use the numerical solution for $u(x)$ to directly compute the kernel (\ref{eq:211}) and obtain the full spectral density $\varrho^-(E)$ of the supergravity theory. The final result is shown on the right diagram of figure \ref{fig:16}. As expected, the non-perturbative gap of the spectrum at $E_0$ is substantially different from $E_c$, which is only associated to the perturbative density $\varrho_0^-(E)$.

Overall, we have shown how the non-perturbative instability exhibited by the complex matrix model when $\xi<-1/\alpha^2$ is cured by properly accounting for $E_0$. Let us stress $E_0$ is not an external parameter one has to introduce in the system, but one that is entirely fixed by requiring the solution~$u_0(x)$ in~(\ref{eq:173}) is continuous at~$x=0$. Note that in this case the value of~$E_0$ does not affect the perturbative behavior of the model, given that~$E_0$ does not appear in the negative $x$ definition of~$u_0(x)$, which ultimately determines the leading spectral density~$\rho_0^-(E)$ in~(\ref{eq:157}). As we explain now, this is not the case for the Type 0B theories.

\subsection{Type 0B JT supergravity}
\label{sec:4.2}

Let us now consider the Type 0B case, described by the Hermitian matrix model. The same discontinuity at $x=0$ arises for the leading solution $r_0(x)$ in (\ref{eq:90}) when $\xi<-1/\alpha^2$, the diagrams in figure~\ref{fig:4} are essentially the same for $r_0(x)$. We would like to find a parameter analogous to $E_0$ which modifies the string equation (\ref{eq:206}) in such a way that its leading solution is continuous.

Building on the intuition gained from the complex matrix model analysis, one possibility would be to modify the range of the real eigenvalues $\lambda_i\in \mathbb{R}$ of the Hermitian matrix to $|\lambda_i|\ge \lambda_{\rm min}=E_0\delta$. This is similar to (\ref{eq:214}) for the matrix $MM^\dagger$. One can then use the technology described in Appendix~\ref{zapp:3} to try to derive the corresponding string equation. However, one stumbles into a technical problem which does not allow for a derivation of a string equation with $E_0\neq 0$ introduced in this way.\footnote{Let us explain the issue more carefully. The discrete string equation (before double scaling) is obtained from an identity similar to~(\ref{eq:215}) where the integration region is instead~${|\lambda|\ge \lambda_{\rm min}}$. This gives a set of equations analogous to~(\ref{eq:41}), which contain a boundary term involving~${\psi_n(\lambda_{\rm min})}$. The way one can usually get rid of the boundary term is by appropriately combining the two equations in~(\ref{eq:41}) (see~(\ref{eq:145}) for the complex matrix model case). However, since in the Hermitian model the potential is even, the second identity in~(\ref{eq:215}) is always trivially satisfied given that~$S_n$ in~(\ref{eq:36}) vanishes. As a result, one is left with a single equation that contains a boundary term one can not get rid off. We have not been able to sidestep this problem in order to derive a string equation for~$r(x)$ in the double scaling limit when~${|\lambda_i|\ge \lambda_{\rm min}\neq 0}$.} This suggests the parameter~$E_0$ in the Hermitian model has a different origin than that for the complex case.

There is a different approach we can try, which leverages the fact we have a good understanding of the complex matrix model when $E_0\neq 0$. The string equations of the complex and Hermitian models are closely related to the Korteweg-de Vries (KdV) and the modified KdV (mKdV) integrable hierarchies respectively. The Miura transform is a known transformation that relates these hierarchies. Using this, reference \cite{Dalley:1992br} showed there is a mapping between the two matrix models string equations (\ref{eq:206}) and (\ref{eq:210}). Explicitly, given a solution $u(x)$ to the string equation
\begin{equation}\label{eq:220}
u
\mathcal{R}^2-\frac{\hbar^2}{2}\mathcal{R}\mathcal{R}''
+\frac{\hbar^2}{4}(\mathcal{R}')^2=\hbar^2\Gamma^2\ ,
\end{equation}
with $\Gamma\in \mathbb{R}$, one applies the Miura transform $u(x)=r(x)^2-\hbar r'(x)$ and shows $r(x)$ satisfies the following equation
\begin{equation}
    \sum_{k=1}^{\infty}t_{k}
    K_{2k}[r(x)]
    +r(x)x=
    \hbar(\Gamma-1/2)\ ,
\end{equation}
which for $\Gamma=1/2$ is the Hermitian model (\ref{eq:206}) with $t_k\rightarrow t_{2k}$.

Applying this same approach to the string equation (\ref{eq:172}), we should be able to `derive' the string equation for the Hermitian matrix model with $E_0\neq 0$. This calculation was recently performed in Appendix C of \cite{Johnson:2020lns}. By considering the following generalization of the Miura transform~${u(x)=r(x)^2-\hbar r'(x)+E_0}$ where~$u(x)$ a solution to~(\ref{eq:172}) with~$\hbar^2\Gamma^2$ on the right-hand side, one finds~$r(x)$ satisfies the following differential equation\footnote{This is obtained from equation (C.4) of \cite{Johnson:2020lns} after replacing $(v_+(x),\Gamma,t_k)\rightarrow(-r(x),1/2,t_{2k})$.}
\begin{equation}\label{eq:217}
\sum_{k=1}^{\infty}t_{2k}K_{2k}[r(x),E_0]+r(x)x=0\ ,
\end{equation}
with
\begin{equation}\label{eq:216}
    K_{2k}[r(x),E_0]=
    \Big(r(x)+
    \frac{1}{2}\hbar\partial_x
    \Big)
    R_k[r(x)^2-\hbar r'(x)+E_0]\ .
\end{equation}
For $E_0=0$ one can show (\ref{eq:216}) matches with $K_{2k}[r(x)]$ as computed from the general recursion relation (\ref{eq:122}). When $E_0\neq 0$, this provides the generalization of the Hermitian model string equation (\ref{eq:206}) we are after. It would be interesting to find a way of deriving this string equation directly from the Hermitian matrix model description.

We can now analyze the behavior of the leading solution $r_0(x)$ to (\ref{eq:216}), which using ${R_k[u(x)]=u(x)^k+\mathcal{O}(\hbar^2)}$ can be written as
\begin{equation}\label{eq:249}
r_0(x)\bigg[
\sum_{k=1}^{\infty}t_{2k}(r_0(x)^2+E_0)^k+x
\bigg]=0
\quad \Longrightarrow \quad
\begin{cases}
\begin{aligned}
\displaystyle
\,\,\sum_{k=1}^{\infty}t_{2k}(r_0(x)^2+E_0)^k+x&=0
\ , \qquad x\le 0 \\
\displaystyle
r_0(x)&=0
\ , \qquad x\ge 0
\end{aligned}
\end{cases}
\end{equation}
Compared to the leading solution $u_0(x)$ in the complex matrix model (\ref{eq:173}), the parameter $E_0$ now appears in the negative $x$ instead of the positive $x$ region. For the values of $t_{2k}$ we are ultimately interested, given in (\ref{eq:91}), the solution for $x<0$ is determined from the following constraint
\begin{equation}\label{eq:222}
\big(I_0(2\pi\sqrt{r_0(x)^2+E_0})-1\big)
+\xi
\big(I_0(2\pi\alpha\sqrt{r_0(x)^2+E_0})-1\big)
+x=0\ , \qquad x\le 0\ ,
\end{equation}
which reproduces the previous expression (\ref{eq:96}) when $E_0=0$. Same as before, the value of $E_0$ is uniquely fixed by requiring continuity at the origin $r_0(x=0)=0$, which from (\ref{eq:222}) gives exactly the same condition as in the previous case (\ref{eq:174}) (with the numerical solution shown in figure \ref{fig:15}). 

It is quite curious how the parameter $E_0$ in each of the string equations (\ref{eq:172}) and (\ref{eq:217}) corrects the discontinuity of the solution at $x=0$ in a different way. In the complex model the leading solution $u_0(x)$ is only modified in the positive $x$ region (\ref{eq:173}), where one gets $u_0(x)=E_0\ge 0$, which effectively raises the solution for $x>0$ and closes the gap (see leftmost diagram in figure \ref{fig:4}). For the Hermitian model we have the opposite situation. The solution $r_0(x)$ is only modified by $E_0$ in the negative $x$ region (\ref{eq:249}), effectively lowering its value at $x=0^-$ by $E_0$.

\begin{figure}
    \centering
    \includegraphics[scale=0.45]{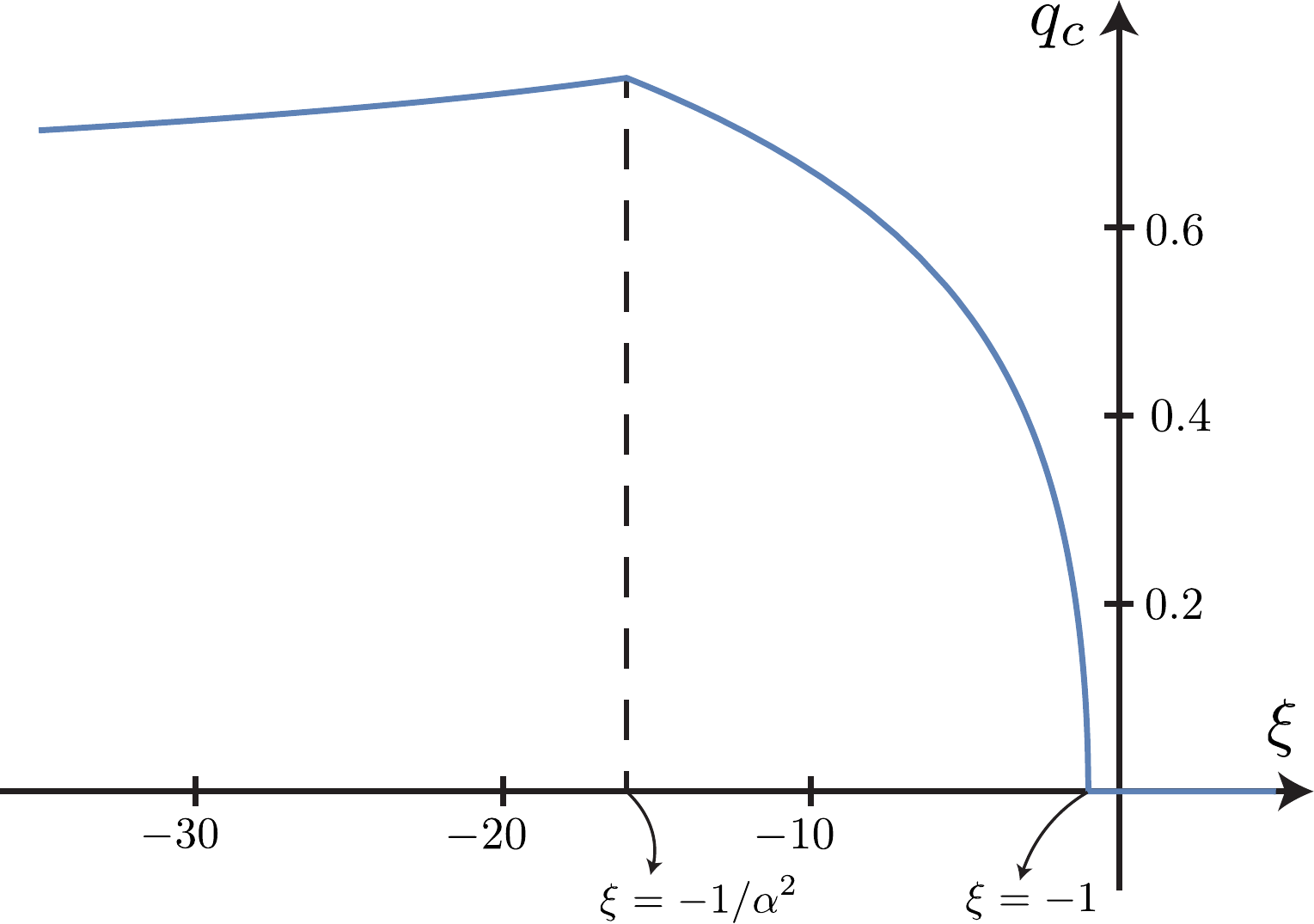}
    \caption{Solution to $q_c(\xi)$ with $\alpha=1/4$ obtained from (\ref{eq:218}), where we observe the transition at $\xi=-1/\alpha^2=-16$.}
    \label{fig:19}
\end{figure}

This distinction turns out being quite important when analyzing the observables of the theory. While for the Type 0A case the perturbative expansion is not affected by $E_0$, this is not true for Type 0B. The leading eigenvalue density of the matrix model (\ref{eq:92}) becomes
\begin{equation}\label{eq:223}
\rho_0^+(q)=
2|q|
\int^{|q|}_{q_c} 
dr_0 r_0
\frac{I_1(2\pi\sqrt{r_0^2+E_0})+\xi \alpha I_1(2\pi\alpha\sqrt{r_0^2+E_0}) }{[(r_0^2+E_0)(q^2-r_0^2)]^{1/2}}\ .
\end{equation}
The threshold parameter $q_c=r_0(\mu)$, which determines the support of this function, is now obtained from the following constraint obtained from (\ref{eq:222})
\begin{equation}\label{eq:218}
I_0(2\pi\sqrt{q_c^2+E_0})
+\xi
I_0(2\pi\alpha\sqrt{q_c^2+E_0})
=0\ , \qquad q_c\ge 0\ ,
\end{equation}
generalizing (\ref{eq:133}) when $E_0\neq 0$. Solving in a perturbative expansion around $\xi=-1/\alpha^2$ one finds
\begin{equation}\label{eq:221}
q_c(\xi)=b_0-
\left[ 
\frac{1}{2\pi} 
\frac{\alpha I_0(2\pi \alpha b_0 )}
{\alpha I_1(2\pi b_0)-I_1(2\pi \alpha b_0 )}
-
\frac{2\alpha^2\Theta(-\xi-1/\alpha^2)}{\pi^2(1-\alpha^2)b_0}
\right]
(\xi+1/\alpha^2)
+\mathcal{O}(\xi+1/\alpha^2)^2\ ,
\end{equation}
where $b_0= q_c(-1/\alpha^2)$. The step function shows the first derivative is discontinuous, as we can confirm from the diagram in figure \ref{fig:19}, obtained from numerically solving (\ref{eq:221}) for $\alpha=1/4$.


\begin{figure}
    \centering
    \includegraphics[scale=0.57]{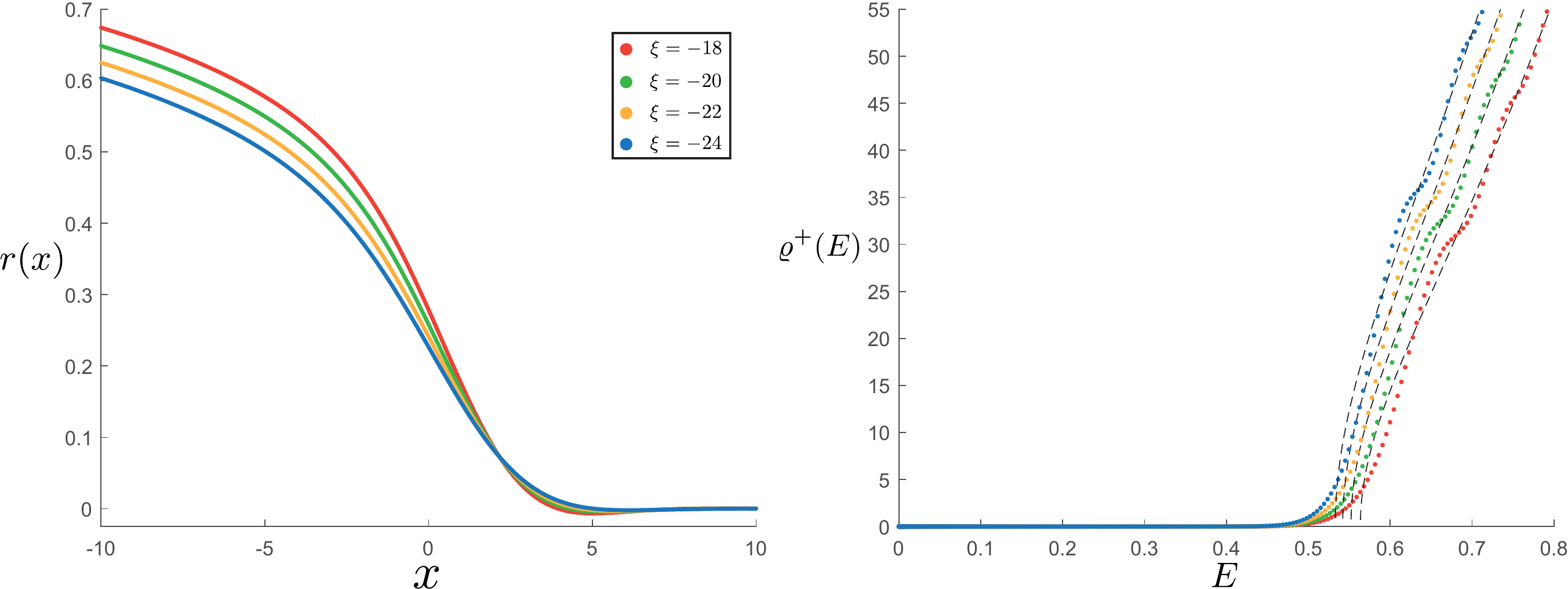}
    \caption{In the left diagram we show the numerical solution to the string equation (\ref{eq:216}) with $(\hbar,k_{\rm max},\alpha)=(1,6,1/4)$ and several values of $\xi<-1/\alpha^2=-16$. From the numerical solution $r(x)$ we compute the full spectral density $\varrho^+(E)$, which exhibits the usual oscillations around the perturbative result given by the dashed line. Compared to figure \ref{fig:16}, in this case there is no non-perturbative gap in $\varrho^+(E)$.}
    \label{fig:20}
\end{figure}

Let us now analyze non-perturbative contributions when $E_0\neq 0$ by directly solving the string equation in (\ref{eq:217}). In the left diagram of figure \ref{fig:20} we give the full solution obtained for several values of $\xi<-1/\alpha^2$. All the solutions are well behaved, meaning they do not contain large oscillations near $x=0$, as we would obtain if $E_0$ was not taken into account (compare with figure \ref{fig:14}) 

From this solution, we can construct the operators~$\mathcal{H}_s$ in~(\ref{eq:203}), compute its eigenfunctions~$\Psi_s(x,q)$ and from the kernel~(\ref{eq:204}) obtain the spectral density. In the right diagram of figure~\ref{fig:20} we show the final result for the energy spectral density of the Type 0B theory~$\varrho^+(E)$ for several values of~$\xi$. The full result has the familiar non-perturbative oscillations around the leading result, obtained from~(\ref{eq:223}) written in terms of $E=q^2$. Compared to the Type 0A case shown in figure~\ref{fig:16}, note that even though~${E_0\neq 0}$, there is no non-perturbative gap. This is expected, given that from~(\ref{eq:249}) the boundary condition for~$r(x)$ is $\lim_{x\rightarrow +\infty}r(x)=0$.

\section{Future directions}
\label{sec:5}

In this paper we computed the Euclidean partition function of certain deformations of $\mathcal{N}=1$ JT supergravity. The associated spectral density, which captures the microsates of the black hole solutions of the theory, exhibit in certain cases negativities which are in conflict with the expected unitarity of the theory. We have shown that whenever a negativity develops, there is a phase transition that removes it, rendering the spectrum positive and well-defined. In order to show this we constructed a dual description between these theories of gravity and random matrix models. The main open question of our work is to give a geometrical understanding of the new phase of gravity that appears. We have already made some comments along these lines in section \ref{sec:3}, so we conclude with other open questions and future directions. 


\subsection*{Beyond sharp defects: String equation from the minimal superstring}
The $(2,p)$ minimal string theory is believed to be dual to a one matrix model (where by minimal string we mean the theory defined by the worldsheet CFT, which is a combination of multicritical matrix models~\cite{Moore:1991ir}). It was an interesting observation in~\cite{Saad:2019lba} (studied further in~\cite{Mertens:2020hbs}) that the large~$p$ limit of the spectral curve of the minimal string/matrix model duality derived from the worldsheet CFT becomes the spectral curve of the JT gravity/matrix model duality. This suggests that the minimal string thought of as worldsheet two dimensional gravity becomes equivalent to JT gravity at large~$p$. 

This was extended in \cite{Turi} to deformations of the minimal string by tachyon operators. In this case the worldsheet theory cannot be solved analytically, but the answer for the spectral curve can be inferred from a bootstrap argument assuming its still dual to a matrix model \cite{Moore:1991ir,Belavin:2008kv}. In the large~$p$ limit, the deformations of the minimal string by a tachyon match precisely to the deformation of JT by a gas of defects. The angle of the defect is related to the scaling dimension of the tachyon while the weight is related to the coupling of the tachyon operator in the action, see \cite{Turi} for more details. A crucial feature of this approach is that it enables the analysis of deformations of JT gravity beyond sharp defects, i.e. one is no longer constrained to $\alpha\in(0,1/2)$ but can have arbitrary defects~${\alpha \in [0,1]}$.

A similar correspondence for the undeformed minimal superstring and JT supergravity was analyzed in \cite{Mertens:2020pfe}. It would be interesting to extend the arguments of \cite{Belavin:2008kv} to find the string equation with deformations, and check whether they match with the deformations of JT supergravity in the appropriate limit. Besides checking this works for sharp defects, it would enable the analysis of deformations of JT supergravity with arbitrary conical defects, just as in \cite{Turi}. It would be interesting as well to compare with the approach of \cite{Kazakov:2021uio}. The comparison in the bosonic case is complicated, but might be more transparent in the supersymmetric case. We leave this to future work.

\subsection*{Adding unorientable surfaces}

In this work we have constrained the topological expansion of $\mathcal{N}=1$ JT supergravity (\ref{eq:3}) such that it only includes the contribution from orientable surfaces. The case in which unorientable surfaces are also allowed was analyzed in \cite{Stanford:2019vob}. In Appendix \ref{zapp:2} we consider the deformations of these theories and show results which are analogous as those obtained for the orientable case. We should mention that since the volumes of the moduli space of unorientable super Riemann surfaces are not as well understood, the discussion in this case is slightly more speculative.


\subsection*{Lessons for non-supersymmetric case}



In the bosonic case, some unphysical negativities can appears when one does not sum over the number of defects. Nevertheles, it was pointed out in \cite{Maxfield:2020ale,Witten:2020wvy} that even after summing over defects some new negativities can appear for certain large deformations. This is the type of negativity we are interested in this paper. It was proposed in \cite{Gao:2021uro} that in the bosonic case, a possible resolution is to declare a jump in the contour of integration of the eigenvalues. In some cases, \cite{Gao:2021uro} found some interesting contours over which the spectral density is positive, although the contours go into the complex plane. The resolution we propose here is different and, if exists, a bosonic version of our proposal would imply a transition to a two-cut phase, although its very hard to test. For the cases we have worked out in supergravity, it would be interesting to see whether it has any implications for the dynamics of end-of-the-world branes following \cite{Gao:2021uro}.

\subsection*{Higher dimensions}
There are several examples in higher dimensional black holes involving a two-dimensional AdS factor near their horizon. Very often, non-perturbative contributions appear as an insertion of defects at the Euclidean horizon of these black holes. Some examples are \cite{Maxfield:2020ale} and \cite{Dabholkar:2014ema}, in three and four dimensions respectively. It would be interesting to find a context in which these negativities are relevant. The main obstacle is that these contributions are usually small. Another issue is that higher dimensional examples involve extended supersymmetry in AdS$_2$. The partition function with defects in examples with extended supersymmetry were studied in \cite{Heydeman:2020hhw, Iliesiu:2021are}, but a matrix model dual (even without deformations) is not currently known.

\section*{Acknowledgements} 

We thank Y. Fan, C. V. Johnson, J. Kruthoff and T. Mertens for valuable discussions and comments on the draft. We are specially grateful to C. V. Johnson for collaboration at the initial stages of this work. GJT would like to thank J. Kruthoff for several discussions on the possibility of similar phase transitions for bosonic models. FR is supported in part by the National Sciences and Engineering Research Council of Canada, and in part by the Simons Foundation. GJT is supported by the Institute for Advanced Study and the National Science Foundation under Grant No. PHY-1911298, and by the Dipal and Rupal Patel funds. 

\appendix
\addtocontents{toc}{\protect\setcounter{tocdepth}{1}}

\section{Weil-Petersson supervolumes}
\label{zapp:1}

The Weil-Petersson supervolumes $V^-_{g,n}(b_1,\dots,b_n)$ that include the difference between even and odd spin structures (relevant for Type 0A JT supergravity) were defined and carefully studied in \cite{Stanford:2019vob}. They can be computed from the following recursion relation 
\begin{equation}\label{eq:34}
\begin{aligned}
bV^-_g(b,B)&=
-\frac{1}{2}\int_0^{\infty}
db'b'
db''b''
D(b'+b'',b)
\bigg[
V^-_{g-1}(b',b'',B)
+
\sum_{h=0}^g
\sum_{B_1\subseteq B}
V^-_{h}(b',B_1)
V^-_{g-h}(b'',B\setminus B_1)
\bigg]+ \\
& \hspace{3mm} -
\sum_{k=1}^{|B|}
\int_0^{\infty}db'b'
\left[
D(b'+b_k,b)
+D(b'-b_k,b)
\right]
V^-_g(b',B\setminus b_k)\ ,
\end{aligned}
\end{equation}
where compared to the notation used in the matin text we are omitting the $n$ index in $V^-_{g,n}(b_1,\dots,b_n)$. We have defined $B=\lbrace b_1,\dots,b_n \rbrace$ as well as
\begin{equation}
D(x,y)=\frac{1}{8\pi}\left[
\frac{1}{\cosh(\frac{x-y}{4})}
-\frac{1}{\cosh(\frac{x+y}{4})}
\right]\ .
\end{equation}
The initial condition for the recursion is $V^-_{1}(b_1)=-1/8$ together with $V^-_0(b_1,\dots,b_n)=0$.

Taking $g=1$ in (\ref{eq:34}) the terms in the first line drop and we are left with the following simple expression
\begin{equation}\label{eq:11}
\begin{aligned}
V^-_1(b,B)=
\sum_{k=1}^{|B|}
\int_0^{\infty}db'b'
\left[
\frac{D(b'+b_k,b)
+D(b'-b_k,b)}{-b}
\right]
V^-_1(b',B\setminus b_k)\ .
\end{aligned}
\end{equation}
From this, we can use induction to show the genus one supervolumes are explicitly given by the first line of~(\ref{eq:10}). For a single boundary $n=1$, the general expression in~(\ref{eq:10}) already matches with~${V_1^-(b_1)=-1/8}$. Assuming the result holds for $n$ boundaries, let us take~${B=\lbrace b_1,\dots,b_n \rbrace}$ in~(\ref{eq:11})
\begin{equation}
V_1^-(b,b_1,\dots,b_n)=
\frac{1}{8}(-1)^n(n-1)!
\sum_{k=1}^{|B|}
\int_0^{\infty}db'b'
\left[
\frac{D(b'+b_k,b)
+D(b'-b_k,b)}{-b}
\right]\ .
\end{equation}
For each value of $k$, the integral can be solved and shown to be equal to minus one, so that the sum gives a factor of $-n$. This proves the explicit formula for the $g=1$ supervolumes given in (\ref{eq:10}). A more involved but conceptually equivalent procedure can be used to prove the $g=2$ and $g=3$ formulas given in (\ref{eq:10}), see \cite{norbury2020enumerative} for details.

\section{Higher genus from loop equations}
\label{zapp:4}

In this appendix we consider the topological expansion of Type 0A JT supergravity with sharp defects, defined by taking the difference between even and odd spin structures. First, let us use the loop equations of the complex matrix model to explicitly derive the $g=1,2,3$ results for the expectation value of~$Z_{\rm MM}^-(\beta_1,\dots,\beta_n)$ in~(\ref{eq:87}). More precisely, we derive the following expressions for the resolvent defined in~(\ref{eq:127})
\begin{equation}\label{eq:88}
\begin{aligned}
W^-_1(z_1,\dots,z_n) & =
\frac{1}{2}
\frac{(-1)^n(n-1)!}{4^{n+1}a_0^n}
\prod_{j=1}^n
\frac{1}{(-z_j)^{3/2}}\ , \\
W^-_2(z_1,\dots,z_n) & =
3
\frac{(-1)^n(n+1)!}{4^{n+5}a_0^{n+3}}
\left[
(n+2)a_1
-3a_0
\sum_{i=1}^n
\frac{1}{z_i}
\right]
\prod_{j=1}^n\frac{1}{(-z_j)^{3/2}}\ , \\
W^-_3(z_1,\dots,z_n) & =
\frac{1}{5}
\frac{(-1)^n(n+3)!}{4^{n+9}a_0^{n+6}}
\bigg[
3a_0^2
\bigg(250\sum_{i=1}^n\frac{1}{z_i^2}+252\sum_{i\neq j}\frac{1}{z_iz_j}\bigg)
-252a_0a_1(n+4)\sum_{i=1}^n\frac{1}{z_i}+ \\
&  \hspace{50mm}
+(n+4)\big(42a_1^2(n+5)-75a_0a_2\big)
\bigg] 
\prod_{j=1}^n\frac{1}{(-z_i)^{3/2}}\ .
\end{aligned}
\end{equation}
Performing the integral transform (\ref{eq:86}) and using the following identities with $q\in\mathbb{N}_0$
\begin{equation}\label{eq:89}
\begin{aligned}
\prod_{j=1}^n
\int_0^{+\infty}d\beta_j\,e^{\beta_j z_j}
\sqrt{\frac{\beta_j}{\pi}}
\bigg(\sum_{i=1}^n \beta_i^q \bigg) & =
\frac{(2q+1)!!}{(-1)^{q}2^{n+q}}
\prod_{j=1}^n\frac{1}{(-z_j)^{3/2}}
\bigg( \sum_{i=1}^n\frac{1}{z_i^q} \bigg)\ , \\
\prod_{j=1}^n
\int_0^{+\infty}d\beta_j\,e^{\beta_j z_j}
\sqrt{\frac{\beta_j}{\pi}}
\bigg(\sum_{i\neq k} \beta_i \beta_k \bigg) & =
\frac{9}{2^{n+2}}
\prod_{j=1}^n\frac{1}{(-z_j)^{3/2}}
\bigg( \sum_{i\neq k }^n\frac{1}{z_iz_k} \bigg)\ ,
\end{aligned}
\end{equation}
one immediately recovers (\ref{eq:87}). 

\paragraph{Genus one:} When applying the loop equations the first step is computing the function $F_g(z',I)$ in (\ref{eq:54}) (with $W^+ \rightarrow W^-$) and then obtaining $W^-_g(z,I)$ from the residue (\ref{eq:64}), in this case at the only branch point of the spectral curve (\ref{eq:73}) at $z'=0$. The function $F_g(z',I)$ in (\ref{eq:54}) for $g=1$ is given by
\begin{equation}\label{eq:74}
F_1(z') = W^-_0(z',z')=\frac{1}{(4z')^2}\ ,
\quad \quad
F_1(z',I) = \sum_{k=1}^{|I|}
\left[2W^-_0(z',z_k)+\frac{1}{(z'-z_k)^2}\right]
W^-_1(z',I\setminus z_k)\ .
\end{equation}
For a single resolvent insertion one computes the residue of $F_1(z')$ in (\ref{eq:54}) and finds
\begin{equation}
W^-_1(z)=
\frac{1}{\sqrt{z}}{\rm Res}\left[
\frac{\sqrt{z'}}{2y(z')(4z')^2(z'-z)}
,z'=0
\right]=
\frac{-1}{32a_0(-z)^{3/2}}\ , 
\end{equation}
in agreement with (\ref{eq:88}) with $n=1$. To prove it for arbitrary $n$ we can use induction. Assuming it holds for $n-1$, take $I=\lbrace z_1,\dots,z_{n-1} \rbrace$ in (\ref{eq:74})
\begin{equation}
F_1(z',I)=
\frac{1}{2}
\frac{(-1)^{n-1}}{4^{n}}
\frac{(n-2)!}{a_0^{n-1}}
\sum_{k=1}^{n-1}
\left\lbrace
\left[
2W^-_0(z',z_k)+\frac{1}{(z'-z_k)^2}
\right]
\frac{1}{(-z')^{3/2}}
\right\rbrace
\prod_{i\neq k}^{n-1}
\frac{1}{(-z_i)^{3/2}}
\ .
\end{equation}
Using the only $z'$ dependence is in the term between curly brackets, one can easily compute the residue (\ref{eq:64}) at the origin $z'=0$ and find
\begin{equation}
\begin{aligned}
W_1^-(z,z_1,\dots,z_{n-1}) & =
\frac{1}{2}
\frac{(-1)^{n-1}}{4^{n}}
\frac{(n-2)!}{a_0^{n-1}}
\sum_{k=1}^{n-1}
\left\lbrace
\frac{-1}{4(-z_k)^{3/2}(-z)^{3/2}a_0}
\right\rbrace
\prod_{i\neq k}^{n-1}
\frac{1}{(-z_i)^{3/2}} \\
&= 
\frac{1}{2}
\frac{(-1)^{n}}{4^{n+1}}
\frac{(n-2)!}{a_0^{n}}
\frac{1}{(-z)^{3/2}}
\prod_{i=1}^{n-1}
\frac{1}{(-z_i)^{3/2}}
\sum_{k=1}^{n-1}
1\ ,
\end{aligned}
\end{equation}
that is precisely the general $n$ expression (\ref{eq:88}).

\paragraph{Genus two:} In this case the function $F_g(z',I)$ (\ref{eq:54}) gets contributions from all three terms. The first two are completely determined by the $g=1$ result in (\ref{eq:88}) and can therefore be written explicitly 
\begin{equation}\label{eq:83}
F_2(z',I)=
\frac{9(-1)^{n+1}n!}{4^{n+4}a_0^{n+1}}
\frac{1}{(-z')^3}
\prod_{i=1}^{n-1}\frac{1}{(-z_i)^{3/2}}
+
\sum_{k=1}^{|I|}
\left[
2W^-_0(z',z_k)+\frac{1}{(z'-z_k)^2}
\right]W^-_2(z',I\setminus z_k)\ ,
\end{equation}
where we have taken $I=\lbrace z_1,\dots,z_{n-1} \rbrace$ and carefully computed the sum over $J\subseteq I$ in (\ref{eq:54}). For~${n=1}$ the second term vanishes and the computation of the residue~(\ref{eq:64}) gives the $n=1$ result in~(\ref{eq:88}). Assuming~(\ref{eq:88}) holds for~${n-1}$ insertions, we can write~(\ref{eq:83}) for~$n$ insertions as
\begin{equation}
\begin{aligned}
F_2(z',z_1,\dots,z_{n-1}) & =3
\frac{(-1)^{n+1}n!}{4^{n+4}a_0^{n+2}}
\prod_{j=1}^{n-1}\frac{1}{(-z_j)^{3/2}}
\Bigg\lbrace
\frac{3a_0}{(-z')^3}
+
\sum_{k=1}^{n-1}
\left[
2W^-_0(z',z_k)+\frac{1}{(z'-z_k)^2}
\right]
 \\
& \qquad \qquad \qquad \quad
\times 
\bigg[
\Big( (n+1)a_1
-3a_0
\sum_{i\neq k}^{n-1}
\frac{1}{z_i}
\Big)
\frac{(-z_k)^{3/2}}{(-z')^{3/2}}
+3a_0
\frac{(-z_k)^{3/2}}{(-z')^{5/2}}
\bigg]
\Bigg\rbrace
\end{aligned}
\end{equation} 
Computing the residue at $z'=0$ according to (\ref{eq:64}) we find
\begin{equation}
\begin{aligned}
W^-_2(z,z_1,\dots,z_{n-1}) & =
3
\frac{(-1)^{n}n!}{4^{n+5}a_0^{n+3}}
\frac{1}{(-z)^{3/2}}
\prod_{j=1}^{n-1}\frac{1}{(-z_j)^{3/2}}
\Bigg\lbrace
6a_1
-6a_0\frac{1}{z}
+
\\ & \qquad \qquad \qquad 
+
(n+4)(n-1)a_1
-3(n+1)a_0\sum_{i=1}^{n-1}\frac{1}{z_i}
-3(n-1)a_0\frac{1}{z} 
\Bigg\rbrace\ ,
\end{aligned}
\end{equation} 
where the terms inside the curly brackets in the first and second lines come from the $g=1$ and $g=2$ contributions in (\ref{eq:83}). This result matches with (\ref{eq:88}) for arbitrary $n$.

\paragraph{Genus three:} Although one can use induction to explicitly derive the $g=3$ result in (\ref{eq:88}), the calculation becomes increasingly tedious. Instead, we have written $F_3(z',I)$ in (\ref{eq:54}) for fixed $I$, computed the residue (\ref{eq:64}) and matched with the general $n$ answer in (\ref{eq:88}) with $n=1,2,3,4$.

\paragraph{Arbitrary genus:} For $g\ge 4$ we follow \cite{Maxfield:2020ale} and use the deformation theorem of \cite{Eynard:2007kz} to prove the equivalence between matrix model and supergravity to all orders in perturbation theory. Since the loop equations of the complex and Hermitian matrix models are the same, they also coincide with the more abstract topological expansion defined in \cite{Eynard:2007kz}.

To do this, it is convenient to consider a slight redefinition of the resolvent matrix model operator~(\ref{eq:127}) by considering the variable~$z(s)=-s^2$ and defining 
\begin{equation}\label{eq:176}
\widehat{W}_{g}^{\rm MM}(s_1,\dots,s_n;\xi)\equiv
\prod_{i=1}^n
(-2s_i)
W^-_g(-s_i^2;\xi)\ ,
\end{equation}
where the $\xi$ dependence is indicated explicitly to keep track of the deformation. We have also added the superscript ${\rm MM}$ to remind ourselves this is computed from the matrix model. The analogous quantity in the supergravity side is obtained by performing a Laplace transform as in (\ref{eq:86}) of the partition function (\ref{eq:39}), so that we get
\begin{equation}\label{eq:182}
\widehat{W}^{\rm SJT}_g(s_1,\dots,s_n;\xi)\equiv 
\prod_{i=1}^n
\int_0^{\infty}d\beta_i
s_i e^{-\beta_i s_i^2}Z^-_{g}(\beta_1,\dots,\beta_n;\xi)\ .
\end{equation}
The right hand side can be written explicitly using the expansion in (\ref{eq:39}) together with the analytic continuation of the supervolumes (\ref{eq:46}). From this we easily get
\begin{equation}\label{eq:181}
\widehat{W}^{\rm SJT}_g(s_1,\dots,s_n;\xi)
=
\sum_{k=0}^{\infty}
\frac{\xi^k}{k!}
\left[
\prod_{i=1}^n
\frac{1}{\sqrt{2}}
\int_0^{\infty}db_i\,b_i\,
e^{-s_ib_i}
V_{g,n+k}^{-}(b_1,\dots,b_n,2\pi i \alpha,\dots,2\pi i \alpha)
\right]
\ ,
\end{equation}
where we have computed the $\beta_i$ integral using the trumpet partition function (\ref{eq:30}).

Our aim is to match (\ref{eq:181}) with (\ref{eq:176}) for arbitrary $g$ and $n$. To do this, we use the deformation theorem of \cite{Eynard:2007kz}, which for our purposes can be stated as follows (see \cite{Maxfield:2020ale} and section 4.3.2 of \cite{Eynard:2015aea}): given an analytic function $f(s)$ and a closed curve $\mathcal{C}$ in the complex plane such that 
\begin{equation}\label{eq:177}
\begin{cases}
\begin{aligned}
\,\,\, &
\frac{\partial}{\partial \xi} y(s;\xi)
=-
\frac{1}{2s}\oint_{\mathcal{C}}
\frac{d\bar{s}}{2\pi i}f(\bar{s})
\widehat{W}_{0}^{\rm MM}(s,\bar{s};\xi)\ , \\
& \frac{\partial}{\partial \xi}
\widehat{W}_{0}^{\rm MM}(s_1,s_2;\xi)=
\oint_\mathcal{C} \frac{d\bar{s}}{2\pi i}f(\bar{s})
\widehat{W}_{0}^{\rm MM}(s_1,s_2,\bar{s};\xi)\ ,
\end{aligned}
\end{cases}
\end{equation}
then one can prove the following identity
\begin{equation}\label{eq:179}
\Longrightarrow \qquad
\frac{\partial^k}{\partial \xi^k}
\widehat{W}_{g}^{\rm MM}(s_1,\dots,s_n;\xi)=
\prod_{j=1}^k
\oint_\mathcal{C} \frac{d\bar{s}_j}{2\pi i}
f(\bar{s}_j)
\widehat{W}_{g}^{\rm MM}(s_1,\dots,s_n,\bar{s}_1,\dots,\bar{s}_k;\xi)\ .
\end{equation}
The second condition in (\ref{eq:177}) is automatically satisfied, given that the $g=0$ two point function is independent of $\xi$ and the three point function vanishes (see (\ref{eq:178})). Same as in~\cite{Maxfield:2020ale}, to ensure the first condition is also satisfied it is enough to take~$\mathcal{C}$ as any closed curve around the origin and~${f(s)=\sin(2\pi \alpha)/\sqrt{2}\pi \alpha}$. Using the two point function in (\ref{eq:99}) the integral picks up the residue at~$s=s_1$ and the right hand side can be shown to agree with $\partial_\xi y(s;\xi)=-\cos(2\pi\alpha s)/\sqrt{2}s $ as obtained from (\ref{eq:73}). 

We can then apply the deformation theorem and use (\ref{eq:179}) to compute the derivatives of (\ref{eq:176}) at $\xi=0$. Moreover, we can then use the Weyl-Petersson supervolumes are related to the undeformed matrix model resolvents as (see equations (5.37) and (5.33) in \cite{Stanford:2019vob})
\begin{equation}
\widehat{W}_{g}^{\rm MM}(s_1,\dots,s_\ell;\xi=0)
=
\prod_{i=1}^\ell
\frac{1}{\sqrt{2}}
\int_0^\infty db_ib_ie^{-s_ib_i}
V^-_{g,\ell}(b_1,\dots,b_\ell)\ .
\end{equation}
Taking $\ell=n+k$ and using this in (\ref{eq:179}) after evaluating at $\xi=0$ one finds
\begin{equation}\label{eq:186}
\left.\frac{\partial^k}{\partial \xi^k}
\widehat{W}_{g}^{\rm MM}(s_1,\dots,s_n;\xi)
\right|_{\xi=0}=
\prod_{i=1}^n
\frac{1}{\sqrt{2}}
\int_0^\infty db_ib_ie^{-s_ib_i}
G(b_1,\dots,b_n,\alpha)
\ ,
\end{equation}
where we have defined
\begin{equation}
G(b_1,\dots,b_n,\alpha)=
\prod_{j=1}^k
\oint_\mathcal{C} \frac{d\bar{s}_j}{2\pi i}
\frac{\sin(2\pi \alpha\bar{s}_j)}{2\pi \alpha}
\int_0^\infty d\bar{b}_j\bar{b}_j
e^{-\bar{s}_j\bar{b}_j}
V^-_{g,n+k}(b_1,\dots,b_n,\bar{b}_1,\dots,\bar{b}_k)\ .
\end{equation}
Since the supervolumes are even polynomials in the geodesic lengths, it is enough to compute the integrals for the monomial $\bar{b}_j^{2m}$, which gives $(2\pi i\alpha)^{2m}$. This shows the function $G(b_1,\dots,b_n,\alpha)$ is precisely the analytic continuation $\bar{b}_j\rightarrow (2\pi i \alpha)$ appearing in (\ref{eq:181}). Altogether, this shows the agreement between the expansions of the complex matrix model and the deformations of Type 0A JT supergravity to all orders in $e^{-S_0}$.

\section{Orthogonal polynomials and double scaling}
\label{zapp:3}

The aim of this Appendix is to introduce a powerful formalism for studying double scaled models, called the method of orthogonal polynomials. Compared to the loop equations, it allows for explicit computation of observables beyond the $1/N$ perturbative expansion, see \mcite{Bessis:1979is,Itzykson:1979fi,Bessis:1980ss} for early references and \cite{Eynard:2015aea} for a review. Apart from putting together many results scattered in the literature, this Appendix includes some new technical results, like the precise method for computing the matrix model kernel and observables for double scaled and double-cut Hermitian matrix models.

\subsection{Finite \texorpdfstring{$N$}{N}}

Consider a set of polynomials $P_n(\lambda)$ labeled $n\in \mathbb{N}_0$ and defined as
\begin{equation}\label{eq:78}
P_n(\lambda)\equiv 
\frac{1}{\mathcal{Z}_n}
\bigg[
\prod_{j=1}^n
\int_{\lambda_{\rm min}}^{+\infty}
d\lambda_j
e^{-NV(\lambda_j)}
\bigg]
\Delta(\lambda_1,\dots,\lambda_n)^2
\prod_{i=1}^n(\lambda-\lambda_i) \ ,
\end{equation}
where $P_0(\lambda)=1$ and $\mathcal{Z}_n$ is defined as the numerator in $P_n(\lambda)$ but without the $\prod_{i=1}^n(\lambda-\lambda_i)$ insertion in the integral. This normalization ensures $P_n(\lambda)$ is a monic polynomial with leading behavior~${P_n(\lambda)=\lambda^n+\mathcal{O}(\lambda^{n-1})}$. In this Appendix, $\lambda$ plays the role of the eigenvalue of $Q$ and $MM^\dagger$ for the Hermitian and complex matrix model (the relation with $q$ and $E$ used in the main text will become clear shortly). This means $\lambda_{\rm min}=-\infty$ and $\lambda_{\rm min}=0$ in each of the models. One can show these polynomials form an orthogonal set (see \cite{Itzykson:1979fi} for details)
\begin{equation}\label{eq:32}
\int_{\lambda_{\rm min}}^{+\infty}
d\lambda\,e^{-NV(\lambda)}
P_n(\lambda)
P_{m}(\lambda)=
h_n\delta_{n,m}\ ,
\qquad \qquad
h_n=\frac{1}{n+1}\frac{\mathcal{Z}_{n+1}}{\mathcal{Z}_n}\ge 0\ ,
\end{equation}
where $h_n$ is the norm of $P_n(\lambda)$. 

In most cases, the explicit formula (\ref{eq:78}) cannot be used to write the polynomials explicitly, since the analytic evaluation of the integrals is too complicated. Instead, it is convenient to study the action of the multiplication by $\lambda$ on $P_n(\lambda)$. Since $\lambda P_n(\lambda)$ is a polynomial of order $n+1$, it can be expanded in terms of lower order polynomials
\begin{equation}\label{eq:103}
\lambda P_n(\lambda)=
P_{n+1}(\lambda)+S_nP_{n}(\lambda)+R_nP_{n-1}(\lambda)+\sum_{i=0}^{n-2}g_i P_i(\lambda)\ ,
\end{equation}
where the coefficient of $P_{n+1}(\lambda)$ is fixed to one since $P_{n+1}(\lambda)$ is monic. Using the orthogonality relation (\ref{eq:32}) one can easily show $g_i=0$. It is convenient to work in terms of an orthonormal set of functions $\psi_n(\lambda)$ with respect to the flat measure in $\lambda$
\begin{equation}\label{eq:43}
\psi_n(\lambda)\equiv 
\frac{1}{\sqrt{h_n}}P_n(\lambda)
e^{-\frac{N}{2} V(\lambda)}
\ ,
\qquad \qquad
\braket{n|m}\equiv
\int_{\lambda_{\rm min}}^{+\infty}d\lambda\,
\psi_n(\lambda)
\psi_m(\lambda)
=\delta_{n,m}\ .
\end{equation}
Using (\ref{eq:103}), the multiplication by $\lambda$ has the following simple action on $\psi_n(\lambda)$\footnote{To write this we have been careful with the normalization, using $h_{n+1}=h_nR_{n+1}$ which follows from computing $\int d\lambda e^{-NV(\lambda)} P_{n+1}(\lambda)\lambda P_n(\lambda)$ using (\ref{eq:32}) for either $P_{n+1}(\lambda)$ or $P_n(\lambda)$.}
\begin{equation}\label{eq:36}
\lambda \psi_n(\lambda)=
\sqrt{R_{n+1}}\psi_{n+1}(\lambda)
+S_n\psi_n(\lambda)
+\sqrt{R_n}\psi_{n-1}(\lambda)\ .
\end{equation}
Given a potential $V(\lambda)$, the coefficients $(R_n,S_n)$ can be computed from the string equations. These are a set of recursion relations derived from the following identity
\begin{equation}\label{eq:215}
\int_{\lambda_{\rm min}}^{+\infty}d\lambda\,
\partial_\lambda
\big(
\psi_n(\lambda)\psi_m(\lambda)
\big)=
-
\psi_n(\lambda_{\rm min})\psi_m(\lambda_{\rm min})\ .
\end{equation}
Using $P_n(\lambda)=\lambda^n+\mathcal{O}(\lambda^{n-1})$ and the orthogonality, the string equations are derived by evaluating the left-hand side for $m=n$ and $m=n-1$
\begin{equation}\label{eq:41}
{\rm String\,\,equations:}
\qquad
\begin{cases}
\begin{aligned}
\braket{n|V'(\lambda)|n-1}&=
\frac{n}{N\sqrt{R_n}}
+
\frac{1}{N}
\psi_{n}(\lambda_{\rm min})
\psi_{n-1}(\lambda_{\rm min})\ , \\
\braket{n|V'(\lambda)|n} & = \frac{1}{N}\psi_n(\lambda_{\rm min})^2\ .
\end{aligned}
\end{cases}
\end{equation}

For any particular potential $V(\lambda)$ one can use (\ref{eq:36}) to write an explicit set of recursion relations for $(R_n,S_n)$ which can then be used to compute all observables in the matrix model. The simplest case is the partition function $\mathcal{Z}_N\equiv \mathcal{Z}$, which can be easily obtained from (\ref{eq:32})
\begin{equation}\label{eq:33}
\mathcal{Z}=N!\prod_{k=0}^{N-1}h_k=
N!\,h_0^N\prod_{k=1}^{N-1}
R_k^{N-k}\ ,
\end{equation}
where in the second equality we have used $h_{n+1}=h_nR_{n+1}$. 

We are interested in extending this to the expectation value of the multi-trace observables $Z_{\rm MM}^{\pm}(\beta_1,\dots,\beta_n)$ in the Hermitian (\ref{eq:56}) and complex (\ref{eq:77}) matrix models. Consider the generating function  of connected correlation functions, defined as
\begin{equation}\label{eq:79}
G(\vec{\alpha})
=\ln\left[
\langle
e^{\sum_{j=1}^p\alpha_j
{\rm Tr}\,F_j(B)}
\rangle
\right]\ ,
\end{equation}
where $\vec{\alpha}=(\alpha_1,\dots,\alpha_p)$ and $F_j(B)$ is an arbitrary function of the matrix $B=Q$ or $B=MM^\dagger$ in each case. Differentiating with respect to $\alpha_j$ and evaluating at $\alpha_j=0$ we find
\begin{equation}\label{eq:45}
\partial^{\,\vec{k}}_{\vec{\alpha}}
\,G(\vec{\alpha})
\big|_{\vec{\alpha}=0}
=
\big\langle 
\prod_{q=1}^p \left({\rm Tr}\,F_q(B)\right)^{k_i} 
\big\rangle_c\ ,
\qquad \qquad
\partial^{\,\vec{k}}_{\vec{\alpha}}\equiv
\prod_{j=1}^p\partial_{\,\alpha_j}^{k_j}\ ,
\end{equation}
where $\vec{k}=(k_1,\dots,k_p)$ and the subscript $c$ indicates it is the connected correlation function. The main complication in computing $G(\vec{\alpha})$ comes from the fact that the~$N$ integrals over $\lambda_i$ in the expectation value are coupled due to the Vandermonde determinant $\det(\lambda_i^{j-1})$ when writing the integrals (\ref{eq:56}) and (\ref{eq:77}) in terms of the appropriate eigenvalues. To simplify this, we can write this determinant as
\begin{equation}
\begin{aligned}
\det\big(\lambda_i^{j-1}\big)^2=
\det\big(P_{j-1}(\lambda_i)\big)^2=
\frac{\mathcal{Z}}{N!}
\det\left(
\psi_{j-1}(\lambda_i)
\right)^2
\prod_{i=1}^N
e^{NV(\lambda_i)}\ .
\end{aligned}
\end{equation} 
In the second equality we have used the determinant is invariant under linear combinations of its columns to rewrite it directly in terms of the polynomials. We then wrote the argument in terms of $\psi_n(\lambda)$ and used (\ref{eq:32}) to write the prefactor in terms of the matrix model partition function $\mathcal{Z}$. Expanding the determinant using its ordinary definition we can write the generating function (\ref{eq:79}) as
\begin{equation}
G(\vec{\alpha})=
\ln\left[
\frac{1}{N!}
\sum_{\sigma,\sigma'\in S_n}
(-1)^{\pi_\sigma+\pi_\sigma'}
\prod_{i=1}^N
\braket{\sigma(i)-1|e^{\sum_{j=1}^p\alpha_j
F_j(\lambda)}|\sigma'(i)-1}
\right]\ ,
\end{equation}
where $\sigma$ and $\sigma'$ are elements of the permutation group $S_N$ with parity $\pi_\sigma$ and $\pi_{\sigma'}$. By relabelling the indices in the sum, the argument of the logarithm is the determinant of an $N$-dimensional matrix $A$ defined as
\begin{equation}\label{eq:109}
G(\vec{\alpha})=
{\rm Tr}\big[\ln(A)\,\big]\ ,
\qquad \qquad
A_{nm}(\vec{\alpha})=
\braket{n|e^{\sum_{j=1}^p\alpha_j
F_j(\lambda)}|m}\ ,
\end{equation}
where $n,m=0,\dots,(N-1)$ and we have used Jacobi's formula to write the determinant as a trace. Differentiating with respect to $\vec{\alpha}$ as in (\ref{eq:45}) we can write expressions for the ensemble average of any multi-trace observable. For instance, arbitrary single and double trace observables are given by
\begin{equation}\label{eq:114}
\begin{aligned}
\left\langle 
{\rm Tr}\,F(B) 
\right\rangle & =\sum_{n=0}^{N-1}
\braket{n|F(\lambda)|n}\ , \\
\langle 
{\rm Tr}\,F_1(B)
{\rm Tr}\,F_2(B)
\rangle_c & =
\sum_{n=0}^{N-1}
\braket{n|F_1(\lambda)F_2(\lambda)|n}
-\sum_{n,m=0}^{N-1}
\braket{n|F_1(\lambda)|m}
\braket{m|F_2(\lambda)|n}\ .
\end{aligned}
\end{equation}
where we used $\partial_\alpha (A^{-1})=-A^{-1}(\partial_\alpha A)A^{-1}$. Using the action of $\lambda$ given in (\ref{eq:36}), these expressions are ultimately written in terms of $(R_n,S_n)$ obtained solving the string equations (\ref{eq:41}). 

It is worth mentioning that all observables in the matrix model are determined by a single object, called the matrix model kernel and defined as
\begin{equation}\label{eq:200}
\mathcal{K}(\lambda,\bar{\lambda})=
\sum_{n=0}^{N-1}
\psi_n(\lambda)\psi_n(\bar{\lambda})\ .
\end{equation}
In terms of $\mathcal{K}(\lambda,\bar{\lambda})$, both expressions in (\ref{eq:114}) are given by
\begin{equation}\label{eq:201}
\begin{aligned}
\left\langle 
{\rm Tr}\,F(B) 
\right\rangle & =
\int_{\lambda_{\rm min}}^{+\infty}
d\lambda\,
\mathcal{K}(\lambda,\lambda)
F(\lambda)\ , \\
\langle 
{\rm Tr}\,F_1(B)
{\rm Tr}\,F_2(B)
\rangle_c & =
\int_{\lambda_{\rm min}}^{+\infty}
d\lambda\,d\bar{\lambda}
\left[ 
\delta(\lambda-\bar{\lambda})
-
\mathcal{K}(\lambda,\bar{\lambda})
\right]
\mathcal{K}(\lambda,\bar{\lambda})
F_1(\lambda)
F_2(\bar{\lambda})\ .
\end{aligned}
\end{equation}

\subsection{Double scaling of Hermitian matrices}

Let us now restrict to a Hermitian matrix model, meaning $\lambda_{\rm min}=-\infty$. In the double scaling limit one takes $N$ large while simultaneously approaching a critical potential
\begin{equation}\label{eq:146}
{\rm Double\,\,scaling\,\,limit:}
\qquad
\begin{cases}
\,\,\,\,N
\,\,\,\,\, \longrightarrow \,\,\,\, \infty \\
\,V(\lambda)\longrightarrow V_{\rm critical}(\lambda)
\end{cases}
\end{equation}
By taking these limits in the right way, one captures universal features associated to the behavior of the system close to criticality (see \cite{Brezin:1989ss,Gross:1989aw,Douglas:1989ve} for some early references). The simplest way to define a critical model is by analyzing the spectral density $\rho^+(\lambda)=\frac{1}{N}{\rm Tr}\,\delta(\lambda-Q)$, whose average in the large~$N$ limit is given by (\ref{eq:58})
\begin{equation}
\rho^+_0(\lambda)=\frac{1}{2\pi}
|h(\lambda)|
\sqrt{-\sigma(\lambda)}
\times
\textbf{1}_{\sigma(\lambda)<0}\ ,
\end{equation}
where $\sigma(\lambda)=\prod_{i=1}^p(\lambda-a_i)(\lambda-b_i)$. For our purposes (see section 6.5 in \cite{Akemann:2011csh}), a critical model is defined as
\begin{equation}\label{eq:138}
V(\lambda)\,\,{\rm critical}
\qquad \Longleftrightarrow \qquad
h(\lambda)=0\ , \qquad \lambda\in \cup_{i=1}^p[a_i,b_i]\ .
\end{equation}
There are two classes of critical points, depending on whether $h(\lambda)$ vanishes at the edge or interior of the cuts $[a_i,b_i]$. Here, we focus on the latter case, whos double scaling was first analyzed in \cite{Crnkovic:1990mr,Douglas:1990xv} (see also \cite{Bleher:2002th,Nappi:1992as,Crnkovic:1992wd,Bleher:2002ys}). 

Given a critical point $\lambda_c$ (that from now on we conveniently fix to the origin $\lambda_c=0$), its detailed behavior is determined by the rate at which the spectral density vanishes. A family of critical models characterized by $\rho_0^+(\lambda)\sim \lambda^{2k}$ is obtained from the following large $N$ spectral densities
\begin{equation}\label{eq:121}
\rho^+_0(\lambda)=
\frac{b_k}{2\pi}
\left(\frac{\lambda}{a}\right)^{2k}
\sqrt{\frac{a^2-\lambda^2}{a^2}}
\times
\textbf{1}_{[-a,a]}\ ,
\qquad {\rm where} \qquad
b_k=\frac{2^{2k+1}(k+1)!(k-1)!}{a(2k-1)!}
\end{equation}
is a normalization constant and $a\in \mathbb{R}_+$ determines the support. The potential required to generate this spectral density can be easily computed by requiring the large $N$ limit of the resolvent $W^+_0(z)$ in (\ref{eq:119}) has the correct behavior $W^+_0(z)=1/z+\mathcal{O}(1/z^2)$, which gives
\begin{equation}\label{eq:111}
V'_{2k}(\lambda)=b_k\sum_{n=0}^k
\binom{1/2}{k-n}
(-1)^{k-n}
\left(\lambda/a\right)^{2n+1}\ .
\end{equation}
This gives a finite $N$ definition of a family of critical models. For fixed $k$ there is an infinite number of critical potentials one could write down which produce the desired behavior $\rho_0^+(\lambda)\sim \lambda^{2k}$. We should think of $V_{2k}(\lambda)$ in (\ref{eq:111}) as a representative of this class. One of the crucial features of the double scaling limit is that the end result is universal, i.e. independent of the particular representative. In the following, we explain in detail how to take the double scaling limit, first analyzing the string equations (\ref{eq:41}) and then the computation of observables such as (\ref{eq:114}).

\subsubsection{String equation}

We now analyze the fate of the string equations (\ref{eq:41}) in the double scaling limit. Noting that the critical potentials (\ref{eq:111}) are even and $\lambda_{\rm min}=-\infty$, one can use (\ref{eq:78}) to show $P_n(-\lambda)=(-1)^nP_n(\lambda)$, which implies the coefficients $S_n$ in (\ref{eq:36}) vanish. This means the second string equation in (\ref{eq:41}) is trivial and we only need to worry about the first one, which can be written as
\begin{equation}\label{eq:50}
\mathcal{S}\equiv
\sqrt{R_n}\braket{n|V'(\lambda)|n-1}-\frac{n}{N}=0\ .
\end{equation}
Using (\ref{eq:36}) one can explicitly compute this for a general even potential $V'(\lambda)=\sum b_{2i}\lambda^{2i-1}$ 
\begin{equation}
\begin{aligned}
\mathcal{S} &=
b_2R_n
+b_4R_n\left[R_{n-1}+R_n+R_{n+1}\right]
+b_6R_n\left[R_{n-1}^2
+R_n^2
+R_{n+1}^2\right.\\[4pt]
&\left.
+2R_{n}\left(R_{n-1}+R_{n+1}\right)
+R_{n-2}R_{n-1}
+R_{n-1}R_{n+1}
+R_{n+1}R_{n+2}
\right]-\frac{n}{N}+\mathcal{O}(b_8R^4)\ .
\end{aligned}
\end{equation}
The number of terms appearing for a particular order $b_{2i}$ is given by $\binom{2i-1}{i-1}$. In practice, the simplest way to compute $\braket{n|\lambda^{2i-1}|n-1}$ is to write a (big enough) finite dimensional matrix representation of $\lambda$, compute the appropriate power and extract the $(n,n-1)$ component.

\begin{figure}
\centering	
\includegraphics[scale=0.38]{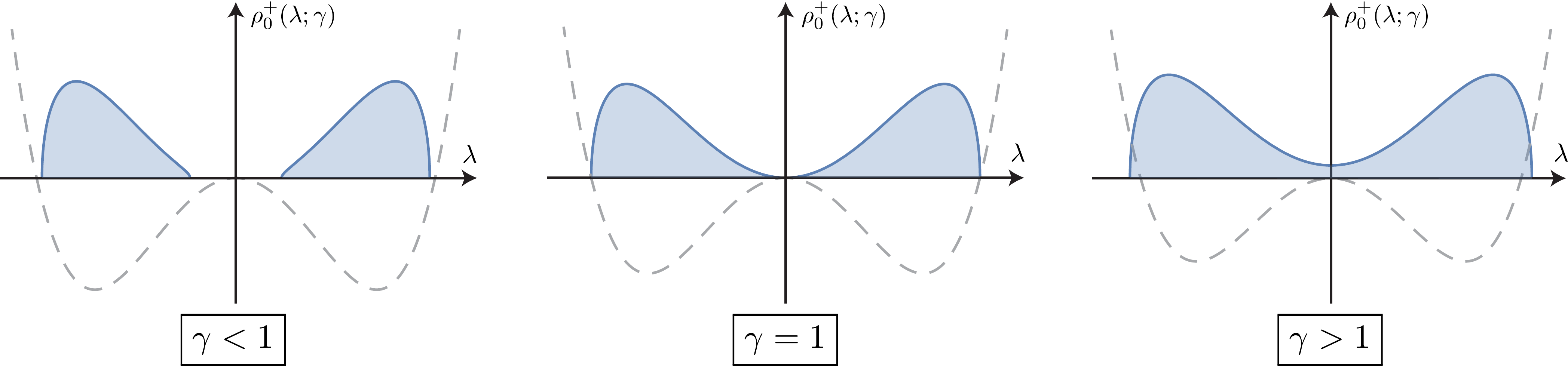}
\caption{Spectral density $\rho^+_0(\lambda;\gamma)$ associated to the critical potential (\ref{eq:111}) with $k=1$ (dashed curve). As $\gamma$ goes to one we approach criticality and there is a phase transition from a double to a single-cut model.}\label{fig:2}
\end{figure}

\paragraph{Single critical model:} Consider the potential $V(\lambda;\gamma)=\frac{1}{\gamma}V_{2k}(\lambda)$, where we have introduced the additional parameter $\gamma$ so that when $\gamma\rightarrow 1$ the system is critical. To get some intuition it is helpful to compute the large $N$ eigenvalue density for general $\gamma$. When $k=1$ a standard calculation yields
\begin{equation}
\rho^+_0(\lambda;\gamma)=
\frac{(2/a)^4}{2\pi \gamma}
\times
\begin{cases}
\displaystyle
\,\,
\left(\lambda^2+(c_0^2-a^2)/2\right)
\sqrt{c_0^2-\lambda^2}\ , \qquad \gamma \ge 1\ ,\\[4pt]
\displaystyle
\quad \,\,
|\lambda|
\sqrt{(c_+^2-\lambda^2)(\lambda^2-c_-^2)}
\,\,\,\,\,\ ,
\qquad \gamma\le 1\ ,
\end{cases}
\end{equation}
where we are omitting the indicator functions that determine the support and have defined
\begin{equation}\label{eq:140}
\left(c_0/a\right)^2=\frac{1+\sqrt{1+3\gamma}}{3}\ ,
\qquad \qquad
\left(c_\pm/a\right)^2=
\frac{1\pm \sqrt{\gamma}}{2}\ .
\end{equation}
Depending on whether $\gamma$ is larger or smaller than one, the system is in a single or double-cut phase respectively (see figure \ref{fig:2}). Precisely at $\gamma=1$ there is a phase transition which signals the criticality of the model. 

To take the double scaling limit we approach criticality and simultaneously take the large $N$ limit in the following way
\begin{equation}\label{eq:113}
\frac{1}{N}=\frac{1}{2}\hbar c_{2k} \delta^{2k+1}\ ,
\qquad \qquad
\gamma=1+c_{2k} \mu  \delta^{2k}\ ,
\end{equation}
where $\delta\rightarrow 0$ and $(\hbar,\mu)$ are the scaling parameters associated to each of these quantities. The coefficient $c_{2k}$ is a normalization constant that is appropriately chosen for notation convenience. Note the sign of $\mu$ is crucial, as it determines whether we approach the critical potential from either the single or double-cut phase. Figuring the right power of $\delta$ in each of the different quantities is a matter of trying different values until one gets a useful ansatz. In the large $N$ limit, $n/N$ becomes a continuum variable $x$, related to $\delta$ in the following way
\begin{equation}\label{eq:82}
\frac{n\gamma}{N}=1+c_{2k} x \delta^{2k}\ .
\end{equation}
It is important to mention that we allow $x$ to take any real value, including infinite. In this way, we can  approach whatever value of $n$ we desire, not necessarily $n\sim N$. The only thing we are missing to compute the double scaling limit of the string equation (\ref{eq:50}) is an ansatz for $R_n$, that we take as
\begin{equation}\label{eq:147}
R_n=1-(-1)^nr(x)\delta+
\sum_{i=2}^{2k+1}\left[
f_i(x)+(-1)^ng_i(x)
\right]\delta^i\ .
\end{equation}
While $r(x)$ controls leading scaling behavior, the functions $\lbrace f_i(x),g_i(x) \rbrace$ determine the subleading contributions with and without the $(-1)^n$ insertion. See \cite{Bleher:2002th,Bleher:2002ys} for a formal justification of this ansatz.

All that is left to do is insert everything into the string equation (\ref{eq:50}) and expand in a power series for $\delta$. The first two orders give
\begin{equation}
\mathcal{S}_k=
\bigg[
\sum_{i=1}^{k+1}
\binom{2i-1}{i-1}b_{2i}
-1
\bigg]
+(-1)^{n+1}r(x)
\bigg[
\sum_{i=1}^{k+1}
\binom{2i-2}{i-1}
b_{2i}
\bigg]
\delta+
\mathcal{O}(\delta^2)\ ,
\end{equation}
where $b_{2i}$ are the coefficients of the critical potential (\ref{eq:111}) defined as $V_{2k}'(\lambda)=\sum_{i=1}^{k}b_{2i}\lambda^{2i-1}$. Both of these terms can be written in terms of a hypergeometric function and vanish when $a$ (defined in~(\ref{eq:111})) is set to $a=2$, which we do from here onwards. Expanding to higher orders in  $\delta$ one obtains more interesting expressions. For instance, for $k=1$ one finds
\begin{equation}
\begin{aligned}
\mathcal{S}_{k=1}=
\left[
4f_2(x)-(x+r(x)^2)
\right]\delta^2
+&\bigg\lbrace
\left[
4f_3(x)-2g_2(x)r(x)
\right]
+\\[4pt]
&\left.+
\frac{(-1)^n}{2}
\left[
4f_2(x)r(x)-
\frac{1}{2}
\hbar^2r''(x)
\right]
\right\rbrace
\delta^3
+\mathcal{O}(\delta^4)\ .
\end{aligned}
\end{equation}
For the functions $\lbrace f_2(x),g_2(x) \rbrace$ we get algebraic equations that can be easily solved. This is not the case for $r(x)$, which instead must satisfy a differential equation, that is nothing more than Painleve~II
\begin{equation}
\lim_{\delta \rightarrow 0}
\frac{1}{\delta^3}
\mathcal{S}_{k=1}=0
\qquad \Longleftrightarrow \qquad
\left[
r(x)^3-\frac{1}{2}\hbar^2r''(x)
\right]
+r(x)x
=0\ .
\end{equation}
In the double scaling limit the string equation for the recursion coefficients $R_n$ becomes an ordinary differential equation for $r(x)$. This also true for higher values of $k$. While the expansion up to order $\delta^{2k}$ gives a set of algebraic equations for $\lbrace f_i(x),g_i(x) \rbrace$, at order $\delta^{2k+1}$ one finds an ordinary differential equation for $r(x)$
\begin{equation}\label{eq:81}
\lim_{\delta \rightarrow 0}
\frac{1}{\delta^{2k+1}}\mathcal{S}_k=0
\qquad \Longleftrightarrow \qquad
K_{2k}+r(x)x=0\ ,
\end{equation}
where $K_{2k}$ is a polynomial in $r(x)$ and its derivatives. For the first few values of $k$ it can be computed explicitly and written as
\begin{align}\label{eq:202}
K_{2}&=r(x)^3-\frac{1}{2}\hbar^2 r''(x)\ , \nonumber \\[4pt]
K_4&=
r(x)^5-\frac{5}{6}\hbar^2r(x)\left(r(x)^2\right)''
+\frac{1}{6}\hbar^4r^{(4)}(x)\ , \nonumber \\[4pt]
K_6&=
r(x)^7-\frac{7}{6} \hbar^2r(x)^2\left(r(x)^3\right)''
-\frac{7}{10}\hbar^4\left[
5r'(x)^2r''(x)+3r(x)r''(x)^2
+4r(x)r'(x)r^{(3)}(x)\right.
\\[4pt]
&\hspace{250pt} \left.
+r(x)^2r^{(4)}(x)
\right]
-\frac{1}{20}\hbar^6r^{(6)}(x)\ ,\nonumber \\[4pt]
K_{2k} & =r(x)^{2k+1}
-\frac{1}{6}(2k+1)\hbar^2 r(x)^{k-1}
\big(r(x)^k\big)''+\dots+
\frac{(-1)^kk!(k-1)!}{2(2k-1)!}\hbar^{2k}r^{(2k)}(x)\ , \nonumber
\end{align}
where we have conveniently fixed $c_{2k}$ in (\ref{eq:113}) to $c_{2k}=(1,3/8,1/8,5/128)$ so that the leading term in $K_{2k}$ is given by $r(x)^{2k+1}$. For arbitrary $k$ one finds the differential polynomials can be computed from the following recursion relation
\begin{equation}\label{eq:122}
K_{2k}=
\frac{2k}{2k-1}
\left[
r(x)\int^x d\bar{x}\,r(\bar{x})K'_{2(k-1)}[r(\bar{x})]-
\frac{\hbar^2}{4}
K_{2(k-1)}''
\right]\ .
\end{equation}
For an explicit derivation of the general $k$ case, see \cite{Bleher:2002th}. This is the Painleve II hierarchy of ordinary differential equations, related to the modified Korteweg–de Vries (mKdV) integrable hierarchy. The boundary conditions for $r(x)$ obtained from the matrix model are given by
\begin{equation}\label{eq:126}
{\rm Boundary\,\,conditions:}
\qquad
\lim_{x\rightarrow -\infty}r(x)=(-x)^{1/2k}
\qquad \qquad
\lim_{x\rightarrow +\infty}r(x)=0\ ,
\end{equation}
which can be obtained from solving (\ref{eq:81}) in the $\hbar\rightarrow 0$ limit.

\paragraph{Multiple critical models:} Given a critical model~$k$, one can perturb it using the other critical models with~$i<k$. To do so, one needs to take a superposition of their respective critical potentials~(\ref{eq:111}) in the following way
\begin{equation}\label{eq:57}
V(\lambda;\gamma)=
\frac{1}{\gamma}
\left[
V_{2k}(\lambda)
+
\sum_{i=1}^{k-1}
\frac{c_{2k}}{c_{2i}}
t_{2i}
\delta^{2(k-i)}
V_{2i}(\lambda)
\right]
\ .
\end{equation}
The perturbation away from the $k$ critical model is controlled by the parameters $\lbrace t_{2i} \rbrace_{i=1}^{k-1}$. We have conveniently chosen extra factors of $\delta$ to get a well defined result in the double scaling limit. To accommodate for $t_{2i}\neq 0$, the ansatz for $\gamma$ and $n$ is slightly modified compared to (\ref{eq:113}) and (\ref{eq:82}), see table \ref{table:1} where we summarize the scaling of all quantities. Taking the double scaling limit of the string equation $\mathcal{S}_k$ (\ref{eq:50}) the same way as before, we find
\begin{equation}\label{eq:118}
\lim_{\delta \rightarrow 0}
\frac{1}{\delta^{2k+1}}\mathcal{S}_k=0
\qquad \Longleftrightarrow \qquad
K_{2k}+
\sum_{i=1}^{k-1}
t_{2i}K_{2i}
+r(x)x=0\ .
\end{equation}
The string equation now gets contributions from the differential polynomials $K_{2i}$ with $i\le k$. Formally taking $k \rightarrow +\infty$ we find
\begin{equation}\label{eq:112}
{\rm String\,\,equation:}
\qquad \qquad
\sum_{k=1}^{\infty}t_{2k}K_{2k}+r(x)x=0\ ,
\end{equation}
where after taking the limit we have relabeled $i\rightarrow k$. This is the most general string equation obtained from the critical potentials with an interior and stable critical point. A particular model is specified by fixing the value of the coefficients $t_{2k}$ that control the inclusion of each critical model. To make contact with JT supergravity these shall be fixed to particular values.

\begin{table}[t]
\setlength{\tabcolsep}{10 pt} 
\centering
\begin{tabular}{ Sc | Sc | Sc  }
\specialrule{.13em}{0em}{0em}
Matrix model  &
Scaling parameter &
Double scaling limit $\delta \rightarrow 0$
\\
\specialrule{.05em}{0em}{0em}
$\displaystyle
N$  &
$\displaystyle
\hbar$ &
$\displaystyle
\frac{1}{N}=\frac{1}{2}\hbar c_{2k}\delta^{2k+1}$ 
\\
\specialrule{.05em}{0em}{0em}
$\displaystyle
\gamma$  &
$\displaystyle
\mu$ &
$\displaystyle
\gamma=1
+\sum_{i=1}^{k-1}\frac{c_{2k}}{c_{2i}} t_{2i}\delta^{2(k-i)}
+c_{2k} \mu \delta^{2k}$ 
\\
\specialrule{.05em}{0em}{0em}
$\displaystyle
n$  &
$\displaystyle
x$ &
$\displaystyle
\frac{n\gamma}{N}=1
+\sum_{i=1}^{k-1}\frac{c_{2k}}{c_{2i}} t_{2i}\delta^{2(k-i)}
+c_{2k} x\delta^{2k}$ 
\\
\specialrule{.05em}{0em}{0em}
$\displaystyle
R_n$  &
$\displaystyle
r(x)$ &
$\displaystyle
R_n=
1-(-1)^nr(x)\delta+
\sum_{i=2}^{2k+1}\left[
f_i(x)+(-1)^ng_i(x)
\right]\delta^i$ 
\\
\specialrule{.05em}{0em}{0em}
$\displaystyle
\lambda$  &
$\displaystyle
q$ &
$\displaystyle
\lambda=q \delta$ 
\\
\specialrule{.05em}{0em}{0em}
$\displaystyle
\psi_{2n}(\lambda)$  &
$\displaystyle
\Psi_+(x,q)$ &
$\displaystyle
\psi_{2n}(\lambda)=(-1)^n \sqrt{\hbar/2}\Psi_+(x,q)
$ 
\\
$\displaystyle
\psi_{2n+1}(\lambda)$  &
$\displaystyle
\Psi_-(x,q)$ &
$\displaystyle
\psi_{2n+1}(\lambda)=(-1)^{n}\sqrt{\hbar/2}\Psi_-(x+\hbar \gamma\delta/2,q)
$ 
\\
\specialrule{.13em}{0em}{0em}
\end{tabular}
\caption{Ansatz for the different parameters in the Hermitian matrix model involving the double scaling of a critical model with potential $V_{2k}(\lambda)$ (\ref{eq:111}) perturbed by other critical models with $i<k$ (see (\ref{eq:57}) for the full potential). The perturbation away from the $k$ model is parametrized by the coefficients $\left\lbrace t_{2i} \right\rbrace_{i=1}^{k-1}$. The double scaling limit $\delta\rightarrow 0$ of the string equation $\mathcal{S}_k$ (\ref{eq:118}) can be worked out using the first four rows of this table. The normalization coefficient $c_{2k}$ is fixed so that when double scaling a single critical model we have $K_{2k}=r(x)^{2k+1}+\mathcal{O}(\hbar^2)$. For the first few values we have $c_{2k}=(1,3/8,1/8,5/128)$.}\label{table:1}
\end{table}

\subsubsection{Computing expectation values}

Let us now explain how to compute the expectation value of observables in the double scaling limit. To start, we `zoom in' the region $\lambda\sim 0$ by rescaling $\lambda= q \delta$. For the $L^2(\mathbb{R})$ functions $\psi_n(\lambda)$ one must distinguish between even and odd $n$ \cite{Crnkovic:1990mr,Bleher:2002th,Crnkovic:1992wd,Bleher:2002ys}, so that one ends up with two independent functions $\Psi_\pm(x,q)$, given in the last row of table \ref{table:1}
\begin{equation}
\psi_n(\lambda)
\qquad 
\xrightarrow[{\rm double\,\,scaling}]{}
\qquad
\Psi_s(x,q)\ ,
\qquad s=\pm\ .
\end{equation}
The countable set $\psi_n(\lambda)$ are replaced by the uncountable $\Psi_s(x,q)$ with $x\in \mathbb{R}$ together with an extra spin degree of freedom. The orthonormality condition (\ref{eq:43}) becomes\footnote{The Dirac delta arises from
\begin{equation}
\delta_{n,m}=\frac{1}{2\pi}
\int_{-\pi}^{\pi}d\phi\,e^{i(n-m)\phi}=
\frac{1}{2\pi}\int_{-\pi}^{\pi}
d\phi
\,e^{i(x-y)\frac{2\phi}{\hbar \delta}}=(\hbar \delta)
\delta(x-y)\ ,
\end{equation}
where in the last step we have changed the integration variable to $2\phi/\hbar\delta$ and used the usual integral representation of the Dirac delta.}
\begin{equation}
\braket{s,x|\bar{x},\bar{s}}=
\int_{-\infty}^{+\infty}
dq\,\Psi_s(x,q)\Psi_{\bar{s}}(\bar{x},q)=
\delta(x-\bar{x})\delta_{s,\bar{s}}\ .
\end{equation}
In the double scaling limit, the action of $\lambda$ in (\ref{eq:36}) is given by
\begin{equation}
q \Psi_s(x,q) =
\big(
{-}s\hbar \partial_x+r(x)
\big)\Psi_{-s}(x,q)\ .
\end{equation}
Multiplying with $q$ once again, we obtain a very useful eigenvalue problem for the functions $\Psi_s(x,q)$
\begin{equation}\label{eq:100}
\mathcal{H}_s\Psi_s(x,q)=q^2\Psi_s(x,q)\ ,
\qquad \qquad
\mathcal{H}_s=-(\hbar\partial_x)^2+
\left[r(x)^2-s\hbar r'(x)\right]\ .
\end{equation}
After solving the string equation (\ref{eq:112}) for $r(x)$ one can construct the differential operator $\mathcal{H}_s$ and obtain $\Psi_s(x,q)$ from its eigenvectors. All observables are then determined from the matrix model kernel (\ref{eq:200}), which in the double scaling limit becomes
\begin{equation}
\mathcal{K}(q,\bar{q})=
\sum_{s=\pm}
\int_{-\infty}^{\mu} 
dx
\Psi_s(x,q)\Psi_s(x,\bar{q})=
\hbar^2\sum_{s=\pm}
\frac{\Psi_s(x,q)\overset{\leftrightarrow}{\partial_x}\Psi_s(x,\bar{q})}{q^2-\bar{q}^2}\bigg|_{x=\mu}
\ ,
\end{equation}
where $\overset{\leftrightarrow}{\partial_x}=\overset{\rightarrow}{\partial_x}-\overset{\leftarrow}{\partial_x}$. In the second equality we have integrated by parts using (\ref{eq:100}) to reduce the integral to a boundary term. This last relation is called the  Christoffel-Darboux formula. In terms of this double scaled kernel, the expressions in (\ref{eq:201}) become
\begin{equation}
\begin{aligned}
\left\langle 
{\rm Tr}\,F(Q) 
\right\rangle & =
\int_{-\infty}^{+\infty}
dq\,
\mathcal{K}(q,q)
F(q\delta)\ , \\
\langle 
{\rm Tr}\,F_1(Q)
{\rm Tr}\,F_2(Q)
\rangle_c & =
\int_{-\infty}^{+\infty}
dq\,d\bar{q}
\left[ 
\delta(q-\bar{q})
-
\mathcal{K}(q,\bar{q})
\right]
\mathcal{K}(q,\bar{q})
F_1(q\delta)
F_2(\bar{q}\delta)\ .
\end{aligned}
\end{equation}
Note there are still factor of $\delta$ on the functions $F_i$ on the right-hand side. This means that the observables that are meaningful in the double scaling limit must be constructed from the rescaled matrix $Q/\delta$. For notation simplicity, we shall keep this extra factor of $\delta$ implicit in the computations in the main text.

\subsection{Double scaling of complex matrices}

Let us now turn our attention to the double scaling of a random matrix model built from complex matrices, first analyzed in~\cite{Morris:1990cq,Dalley:1991qg}. Since the eigenvalues of~$MM^\dagger$ are positive, the minimum eigenvalue is at zero~$\lambda_{\rm min}=0$. Critical models are defined the same way as before~(\ref{eq:138}), with the difference that the expectation value of the eigenvalue density~$\rho_0^-(\lambda)$ (\ref{eq:127})~is given by~(\ref{eq:70}) with~$\sigma(\lambda)$ an odd (instead of even) polynomial with simple roots. A family of critical models characterized by~$\rho_0^-(\lambda)\sim \lambda^{k-1/2}$ can be constructed from
\begin{equation}
\rho_0^-(\lambda)=
\frac{b_k}{2\pi}
\left(
\frac{\lambda}{a}
\right)^k
\sqrt{\frac{a-\lambda}{\lambda}}
\times \textbf{1}_{[0,a]}\ ,
\end{equation}
where $b_k$ defined in (\ref{eq:121}). The potential needed to generate this spectral density can be easily worked out by requiring the resolvent $W_0^-(z)$ in (\ref{eq:139}) behaves like $W_0^-(z)=1/z+\mathcal{O}(1/z^2)$, which gives
\begin{equation}\label{eq:141}
V'_k(\lambda)=b_k\sum_{n=0}^k
\binom{1/2}{k-n}
(-1)^{k-n}\left(\frac{\lambda}{a}\right)^{n}\ .
\end{equation}
The double scaling limit (\ref{eq:146}) is realized by taking $N$ large while appropriately approaching $V_k(\lambda)$.

\subsubsection{String equation}

The string equations (\ref{eq:41}) for this matrix model are more complicated for two reasons: the critical potentials (\ref{eq:141}) are not even (meaning $S_n\neq 0$) and $\lambda_{\rm min}\neq -\infty$, so that we must consider the boundary terms in (\ref{eq:141}). One can get rid of the latter complication by combining the two strings equations and using (\ref{eq:36}). A simple calculation shows the following equations are equivalent
\begin{equation}\label{eq:145}
\begin{aligned}
\mathcal{A} & =
\left(S_n-\lambda_{\rm min}\right)\braket{n|V'(\lambda)|n}
+\sqrt{R_n}\braket{n|V'(\lambda)|n-1}
+\sqrt{R_{n+1}}\braket{n+1|V'(\lambda)|n}
-\frac{2n+1}{N}=0 \ , \\[4pt]
\mathcal{B} & =
R_n
\braket{n|V'(\lambda)|n}
\braket{n-1|V'(\lambda)|n-1}
-
\left[
\sqrt{R_n}\braket{n|V'(\lambda)|n-1}-\frac{n}{N}
\right]^2=0\ ,
\end{aligned}
\end{equation}
which have the advantage of being independent of $\psi_n(\lambda_{\rm min})$.

\begin{figure}
\centering	
\includegraphics[scale=0.38]{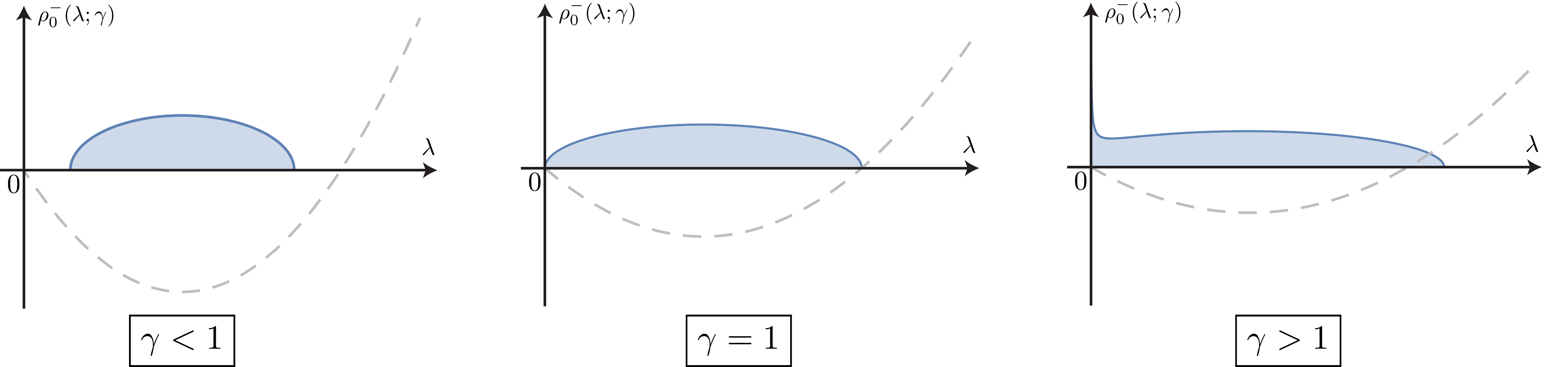}
\caption{Spectral density $\rho^-_0(\lambda;\gamma)$ associated to the critical potential (\ref{eq:141}) with $k=1$ (dashed curve). As $\gamma$ goes to one we approach criticality and there is a phase transition between a hard-edge $\rho_0^-\sim 1/\sqrt{\lambda}$ and soft-edge $\rho_0^-\sim \sqrt{\lambda}$ behavior at $\lambda=0$.}\label{fig:1}
\end{figure}

\paragraph{Single critical model:} Introduce the parameter $\gamma$ and consider the potential ${V(\lambda;\gamma)=\frac{1}{\gamma}V_k(\lambda)}$, so that when $\gamma\rightarrow 1$ the system is critical. It is instructive to compute the large $N$ eigenvalue density for general $\gamma$ and $k=1$, which gives
\begin{equation}
\begin{aligned}
\displaystyle
\rho_0^-(\lambda;\gamma)=
\frac{4(2/a)^2}{2\pi \gamma}
\times
\begin{cases}
\displaystyle
\,\,\,\,
\bigg(\lambda+\frac{c_0^2-a}{2}\bigg)
\sqrt{\frac{c_0^2-\lambda}{\lambda}}\,\,\ ,
\qquad \gamma\ge 1\  , \\[6pt]
\displaystyle
\quad \,
\sqrt{(c_+^2-\lambda)(\lambda-c_-^2)}
\quad \,\,\, \ ,
\qquad \gamma\le 1\ ,
\end{cases}
\end{aligned}
\end{equation}
where $c_0$ and $c_\pm$ are defined in (\ref{eq:140}). Similarly as before, we observe a phase transition at $\gamma=1$ between a hard-edge $\rho_0^-\sim 1/\sqrt{\lambda}$ and soft-edge $\rho_0^-\sim \sqrt{\lambda}$ behavior near $\lambda=0$, see figure \ref{fig:1}.

To take the double scaling we write an ansatz for the parameters $(N,\gamma,n)$ in terms of $(\hbar,\mu,x)$ that is completely analogous to (\ref{eq:113}) and (\ref{eq:82})
\begin{equation}\label{eq:150}
\frac{1}{N}=\sqrt{2}\hbar c_k\delta^{2k+1}\ ,
\qquad \qquad
\gamma=1+c_k\mu \delta^{2k}\ ,
\qquad \qquad
\frac{n\gamma}{N}=1+c_k x \delta^{2k}\ .
\end{equation}
For each $k$ the factor $c_k$ is fixed for normalization convenience of the string equation. For the recursion coefficients $(R_n,S_n)$ we consider the following scaling ansatz \cite{Morris:1990cq,Dalley:1991qg}
\begin{equation}\label{eq:148}
R_n=1-u(x)\delta^2
+\sum_{i=0}^{2k+4}g_i(x)\delta^{2+i}\ ,
\qquad \qquad
S_n=2+u(x)\delta^2
+\sum_{i=0}^{2k+4}g_i(x)\delta^{2+i}\ .
\end{equation}
Comparing with (\ref{eq:147}) note the factors of $(-1)^n$ are not required for the matrix model quantities to have a smooth limit. Replacing into the string equations (\ref{eq:145}) and expanding to linear order in~$\delta$ one finds the parameter $a$ in the critical potentials (\ref{eq:141}) must be fixed to $a=4$.\footnote{By modifying the constant terms in (\ref{eq:148}) it is possible to take other values for $a$.} Expanding~$\mathcal{A}_k$ in (\ref{eq:145}) to higher orders one obtains a set of algebraic equations that can be used to solve for $g_i(x)$
\begin{equation}
\lim_{\delta \rightarrow 0}\frac{1}{\delta^{4k+2}}\mathcal{A}_k=0
\qquad \Longrightarrow \qquad
g_i(x)=\mathcal{G}_i[u(x)]\ ,
\qquad i=0,1,\dots,4k+2\ .
\end{equation}
where $\mathcal{G}_i[u(x)]$ is a functional of $u(x)$ and its derivatives. As an example, for $k=1$ one finds
\begin{equation}
g_0(x)=\frac{1}{2}c_1 x\ ,
\quad
g_1(x)=\frac{1}{8}\sqrt{2}\hbar(c_1+2u'(x))\ ,
\quad
g_2(x)=
\frac{1}{16}\left[
2\hbar^2u''(x)-(c_1x+2u(x))^2
\right]\ .
\end{equation}
With these choices for $g_i(x)$ one expands the $\mathcal{B}_k$ string equation (\ref{eq:145}) in a power series in $\delta$ and finds it is automatically satisfied up to order $\delta^{4k+1}$. Going one order beyond that, one does not find an algebraic equation but an ordinary differential equation entirely given by $u(x)$
\begin{equation}\label{eq:149}
\lim_{\delta \rightarrow 0}\frac{1}{\delta^{4k+2}} \mathcal{B}_k=0
\qquad  \Longrightarrow  \qquad
u(x)\mathcal{R}^2-\frac{\hbar^2}{2}\mathcal{R}\mathcal{R}''
+\frac{\hbar^2}{4}\left(\mathcal{R}'\right)^2=0\ ,
\end{equation}
where primes are derivatives with respect to $x$ and for each critical potential $\mathcal{R}=R_k[u(x)]+x$, with $R_k[u(x)]$ the Gelfand-Dikii polynomials \cite{Gelfand:1975rn}. These are functionals of $u(x)$ and its derivatives determined by the following recursion relation
\begin{equation}
R_{k+1}=
\frac{k+1}{2k+1}
\left[
\int^x d\bar{x}\,u(\bar{x})R_k'[u(\bar{x})]
+u(x)R_k-\frac{\hbar^2}{2}R_k''
\right]\ ,
\end{equation}
with $R_0=1$. Their $\hbar$ expansion has the following general structure
\begin{equation}\label{eq:152}
R_k[u]=u(x)^{k}-
\frac{\hbar^2}{12}k
\left[
\big(u(x)^{k-1}\big)''+(k-1)u(x)^{k-2}u''(x)
\right]
+\dots
+
\frac{2(-1)^{k+1} k!^2}{(2k)!}
(\hbar \partial_x)^{2k-2}
u(x)\ .
\end{equation}
We have explicitly computed and checked (\ref{eq:149}) for $k\le 4$, as first done in \cite{Morris:1990cq,Dalley:1991qg}. Using the method of \cite{Douglas:1989dd}, a derivation for general $k$ is sketched in \cite{Dalley:1991vr} (see also section 5 of \cite{Dalley:1991qg}). The boundary conditions for $u(x)$ are again implied by the large $N$ behavior of the model and given by
\begin{equation}
{\rm Boundary\,\,conditions:}
\qquad
\lim_{x\rightarrow -\infty}u(x)=(-x)^{1/k}
\qquad \qquad
\lim_{x\rightarrow +\infty}u(x)=0\ .
\end{equation}
These are most easily obtained by analyzing the leading $\hbar\rightarrow 0$ behavior of the string equation (\ref{eq:149}).

\begin{table}[t]
\setlength{\tabcolsep}{10 pt} 
\centering
\begin{tabular}{ Sc | Sc | Sc  }
\specialrule{.13em}{0em}{0em}
Matrix model  &
Scaling parameter &
Double scaling limit $\delta \rightarrow 0$
\\
\specialrule{.05em}{0em}{0em}
$\displaystyle
N$  &
$\displaystyle
\hbar$ &
$\displaystyle
\frac{1}{N}=\sqrt{2}\hbar c_k \delta^{2k+1}$ 
\\
\specialrule{.05em}{0em}{0em}
$\displaystyle
\gamma$  &
$\displaystyle
\mu$ &
$\displaystyle
\gamma=1
+\sum_{i=1}^{k-1}\frac{c_{k}}{c_{i}} t_{i}\delta^{2(k-i)}
+c_{k} \mu \delta^{2k}$ 
\\
\specialrule{.05em}{0em}{0em}
$\displaystyle
n$  &
$\displaystyle
x$ &
$\displaystyle
\frac{n\gamma}{N}=1
+\sum_{i=1}^{k-1}\frac{c_{k}}{c_{i}} t_{i}\delta^{2(k-i)}
+c_{k} x\delta^{2k}$ 
\\
\specialrule{.05em}{0em}{0em}
$\displaystyle
R_n$  &
$\displaystyle
r(x)$ &
$\displaystyle
R_n=
1-u(x)\delta^2+
\sum_{i=0}^{2k+4}
g_i(x)\delta^{2+i}
$ 
\\
$\displaystyle
S_n$  &
$\displaystyle
u(x)$ &
$\displaystyle
S_n=
2+u(x)\delta^2+
\sum_{i=0}^{2k+4}
g_i(x)\delta^{2+i}
$ 
\\
\specialrule{.05em}{0em}{0em}
$\displaystyle
\lambda$  &
$\displaystyle
E$ &
$\displaystyle
\lambda=2E \delta^2$ 
\\
\specialrule{.05em}{0em}{0em}
$\displaystyle
\psi_{n}(\lambda)$  &
$\displaystyle
f(x,E)$ &
$\displaystyle
\psi_{n}(\lambda)=
(-1)^n
\big(\hbar/\sqrt{2}\delta\big)^{1/2}
\Psi(x,E)
$ 
\\
\specialrule{.13em}{0em}{0em}
\end{tabular}
\caption{Ansatz for the different parameters in the complex matrix model involving the double scaling of a critical model with potential $V_{k}(\lambda)$ (\ref{eq:141}) perturbed by other critical models with $i<k$ (see (\ref{eq:151}) for the full potential). The perturbation away from the $k$ model is parametrized by the coefficients $\left\lbrace t_{i} \right\rbrace_{i=1}^{k-1}$. The double scaling limit $\delta\rightarrow 0$ of the string equations (\ref{eq:145}) can be worked out using the first four rows of this table. The normalization coefficient $c_{k}$ is fixed so that when double scaling a single critical model the Gelfand-Dikii polynomials are normalized as in (\ref{eq:152}). For the first few values we have $c_{k}=(2,3/2,1,5/8)$.}\label{table:2}
\end{table}

\paragraph{Multiple critical models:} Given a critical model $k$ one can perturb it using the other critical models with $i<k$ by taking the following potential
\begin{equation}\label{eq:151}
V(\lambda;\gamma)=
\frac{1}{\gamma}\left[
V_k(\lambda)+\sum_{i=1}^{k-1}
\frac{c_k}{c_i}
t_i\delta^{2(k-i)}V_i(\lambda)
\right]\ .
\end{equation}
The perturbation away from the $k$ model is controlled by the coefficients $\lbrace t_i \rbrace_{i=1}^{k-1}$, with the extra factors of $\delta$ included to have a sensible limit as $\delta\rightarrow 0$. The ansatz for $n$ and $\gamma$ is slightly modified with respect to (\ref{eq:150}), see table \ref{table:2} where we summarize the scaling of all quantities. Taking the double scaling limit of (\ref{eq:145}) the string equation becomes the same differential equation (\ref{eq:149}) with the difference that $\mathcal{R}$ is now given by $\mathcal{R}=
\sum_{i=1}^{k}t_iR_i[u]+x$. The most general model is obtained by formally taking the limit $k\rightarrow +\infty$, so that we find
\begin{equation}\label{eq:153}
\mathcal{R}=\sum_{k=1}^{\infty}t_k R_k[u]+x\ ,
\end{equation}
where we have relabeled $i\rightarrow k$ after taking the limit. A given model is obtained by taking particular values for the parameters $t_k$.

\subsubsection{Computing expectation values}

To compute the expectation value of observables in the double scaled model we have to `zoom in' to~${\lambda\sim 0}$ and take an appropriate scaling limit of the orthonormal functions~$\psi_n(\lambda)$, see the lower rows of table~\ref{table:2}. A difference with respect to the Hermitian model is that in this case there is no need to distinguish between even and odd~$n$ with two distinct functions~$\Psi_{s=\pm}$ (compare with table~\ref{table:1}) but we instead have
\begin{equation}
\psi_n(\lambda)
\qquad 
\xrightarrow[{\rm double\,\,scaling}]{}
\qquad
\Psi(x,E)\ .
\end{equation}
The precise scaling of $\lambda$ and $\psi_n(\lambda)$ are given in table \ref{table:2}, where the normalization is chosen appropriately so that taking $\delta\rightarrow 0$ of (\ref{eq:43}) and (\ref{eq:36}) we have
\begin{equation}
\braket{x|y} =
\int_0^{+\infty}dE\,\Psi(x,E)\Psi(y,E)=\delta(x-y)\ ,
\end{equation}
as well as
\begin{equation}\label{eq:167}
\mathcal{H}\Psi(x,E)
=E\Psi(x,E)\ ,
\qquad \qquad
\mathcal{H}\equiv -(\hbar\partial_x)^2+u(x)\ .
\end{equation}
The differential operator $\mathcal{H}$ has a completely analogous structure to that of $\mathcal{H}_s$ in (\ref{eq:100}). Solving for $u(x)$ from the string equation (\ref{eq:149}) with (\ref{eq:153}), one can construct $\mathcal{H}$ and obtain $\Psi(x,E)$ from its eigenfunctions. The double scaling of the matrix model kernel (\ref{eq:200}) becomes
\begin{equation}
\mathcal{K}(E,\bar{E})=
\int_{-\infty}^{\mu}dx
\Psi(x,E)\Psi(x,\bar{E})=
\hbar^2
\frac{\Psi(x,E)\overset{\leftrightarrow}{\partial_x}\Psi(x,\bar{E})}{E-\bar{E}}\bigg|_{x=\mu}\ ,
\end{equation}
where $\overset{\leftrightarrow}{\partial_x}=\overset{\rightarrow}{\partial_x}-\overset{\leftarrow}{\partial_x}$. In the second equality we have integrated by parts using (\ref{eq:167}) to reduce the integral to a boundary term, called the  Christoffel-Darboux formula. In terms of the double scaled kernel the expectation value of arbitrary single and double-trace observables (\ref{eq:201}) are given by
\begin{equation}
\begin{aligned}
\langle 
{\rm Tr}\,F(MM^\dagger) 
\rangle & =
\int_{0}^{+\infty}
dE\,
\mathcal{K}(E,E)
F(2E\delta^2)\ , \\
\langle 
{\rm Tr}\,F_1(MM^\dagger)
{\rm Tr}\,F_2(MM^\dagger)
\rangle_c & =
\int_{0}^{+\infty}
dE\,d\bar{E}
\left[ 
\delta(E-\bar{E})
-
\mathcal{K}(E,\bar{E})
\right]
\mathcal{K}(E,\bar{E})
F_1(2E\delta^2)
F_2(2\bar{E}\delta^2)\ .
\end{aligned}
\end{equation}
Similarly as before, observables in the double scaling limit are functions of the rescaled matrix $M\rightarrow \sqrt{2} M\delta$. In the main text we leave this rescaling implicit for notation convenience.

\section{Including unorientable surfaces}
\label{zapp:2}

The inclusion of unorientable surfaces in $\mathcal{N}=1$ JT supergravity was first introduced and methodically studied in \cite{Stanford:2019vob}. As explained in that work, there are eight theories one can define, corresponding to inserting a factor of $e^{-i\pi N'\eta/2}$ in the path integral, where $N'$ is an integer mod 8 and $\eta$ the $\eta$-invariant of Atiyah-Patodi-Singer \cite{atiyah_patodi_singer_1975}. As shown in \cite{Stanford:2019vob}, only the $N'=2,6$ theories have finite and well defined partition functions. From now on, we restrict to these values of $N'$. 

\paragraph{Undeformed theories:} Let us start by briefly reviewing the $\mathcal{N}=1$ JT supergravity theories including unorientable surface but no conical defects. The computation of the Euclidean partition function proceeds similarly as in the orientable case, with the decomposition of surfaces shown in figure~\ref{fig:6}. The fixed genus contributions are analogous to (\ref{eq:35}) and given by
\begin{equation}\label{eq:38}
\begin{aligned}
Z_{{\rm SJT},1/2}^{N'}(\beta)&=\int_0^{\infty}db\,Z_{\rm Trumpet}(\beta,b)
V_{1/2,1}^{N'}(b)\ , \\
Z_{{\rm SJT},g}^{N'}(\beta_1,\dots,\beta_n)&=2^{n-1}
\left[ 
\prod_{i=1}^n
\int_0^{\infty}db_i\,b_i\,
Z_{\rm Trumpet}(\beta_i,b_i)
\right]
V_{g,n}^{N'}(b_1,\dots,b_n)\ ,
\end{aligned}
\end{equation}
where $Z_{\rm Trumpet}(\beta,b)$ is given in (\ref{eq:30}) and $g\in \mathbb{N}/2$. Note the different integration measure for the $g=1/2$ case as well as the additional factor of $2^{n-1}$. Similarly as before, we make the formal definition $V_{0,2}^{N'}(b_1,b_2)=2\delta(b_1-b_2)/b_1$ and for ${(g,n)=(0,1)}$ we have the disk partition function (\ref{eq:102}) instead. For the same reason as for the orientable case, the genus zero volumes with more than two boundaries vanish $V_{g=0,n}^{N'}(b_1,\dots,b_n)=0$, see (\ref{eq:106}). The special $(g,n)=(1/2,1)$ case has been explicitly computed in \cite{Stanford:2019vob} and gives
\begin{equation}\label{eq:230}
V_{1/2,1}^{N'}(b)=\frac{N'-4}{2} \ ,
\qquad \qquad
N'=2,6\ .
\end{equation}
Apart from these results, the remaining supervolumes have not been computed so far. However, building on some results from random matrix models, \cite{Stanford:2019vob} conjectured the vanishing of all the remaining supervolumes 
\begin{equation}
V_{g,n}^{N'}(b_1,\dots,b_n)=0\ ,
\qquad \qquad
N'=2,6\ .
\end{equation}
This provides an extremely simple expressions for the partition function to all orders in perturbation theory. Apart from the $g=1/2$ contribution coming from (\ref{eq:230}), this is analogous to the situation encountered for the Type 0B JT supergravity in (\ref{eq:108}).

As shown in \cite{Stanford:2019vob}, there is a random matrix model which matches the results from this topological expansion to all orders in perturbation theory. The matrix model is most easily described in terms of an arbitrary complex matrix $M$, such that the expectation value of an arbitrary observable $\mathcal{O}$ is defined as
\begin{equation}\label{eq:234}
\langle \mathcal{O} \rangle=
\frac{1}{\mathcal{Z}}
\int dM\,\mathcal{O}\,
\det(MM^\dagger)^\Gamma
e^{-N\,{\rm Tr}\,V(MM^\dagger)}\ ,
\end{equation}
where $\Gamma$ is related to $N'$ according to $\Gamma=(N'-4)/4$. The spectral curve which defines the model is obtained from the leading eigenvalue spectral density, that is fixed to
\begin{equation}\label{eq:232}
\rho_0(E)=
\frac{\cosh(2\pi\sqrt{E})}{2\pi\sqrt{E}}
\ .
\end{equation}
By analyzing the loop equations of this model, the following matching was shown in section 5.3.3 of~\cite{Stanford:2019vob} 
\begin{equation}\label{eq:231}
Z^{N'}_{{\rm SJT}}(\beta_1,\dots,\beta_n)\simeq 
\langle 
Z_{\rm MM}^{N'}(\beta_1)\dots
Z_{\rm MM}^{N'}(\beta_n)
\rangle\ ,
\qquad \qquad 
Z_{\rm MM}^{N'}(\beta)=
2\sqrt{2}\,
{\rm Tr}\,e^{-\beta MM^\dagger}\ ,
\end{equation}
where $\simeq$ means the equality holds to all orders in perturbation theory. Note the different prefactor in the matrix operator $Z_{\rm MM}^{N'}(\beta)$ when compared to (\ref{eq:61}). Non-perturbative aspects of this matrix model were first explored in \cite{Johnson:2020heh} using the method of orthogonal polynomials.

\paragraph{Including conical defects:} Contributions to the path integral coming from conical defects are included the same way as before (\ref{eq:37}), by summing over arbitrary $k$-number of defects. By restricting ourselves to sharp defects $\alpha\in(0,1/2)$, each term $Z_{g,k}^{N'}(\beta_1,\dots,\beta_n;\alpha)$ is written as (\ref{eq:39}). To proceed, we need to assume the analytic continuation (\ref{eq:46}) that allows us to relate the supervolumes with and without defects. Given that the unorientable volumes $V_{g,n}^{N'}(b_1,\dots,b_n)$ are not as well understood, it is not entirely clear whether this analytic continuation still makes sense and yields the desired result. That being said, if one assumes the defects are properly accounted in this way, one gets a result analogous to the Type 0B case with defects (\ref{eq:55})\footnote{Compared to (\ref{eq:39}), these expressions are obtained after rescaling $\xi\rightarrow 2\xi$, same as previously done in (\ref{eq:55}).}
\begin{equation}\label{eq:240}
\begin{aligned}
Z^{N'}(\beta) & \simeq e^{S_0}
\sqrt{\frac{2}{\pi \beta}}
\left[ 
e^{\pi^2/\beta}+\xi e^{\pi^2\alpha^2/\beta} 
\right]+\frac{N'-4}{2\sqrt{2}}\ ,\\
Z^{N'}(\beta_1,\beta_2) & \simeq 
\frac{8}{2\pi}\frac{\sqrt{\beta_1\beta_2}}{\beta_1+\beta_2}\ , \\
Z^{N'}(\beta_1,\dots,\beta_n)& \simeq  0\ .
\end{aligned}
\end{equation}
The leading spectral density of the supergravity theory $\varrho_0(E)$ obtained from the inverse Laplace transform of $Z^{N'}(\beta)$ gives
\begin{equation}\label{eq:233}
\varrho_0(E)=
\frac{\sqrt{2}\cosh(2\pi\sqrt{E})}{\pi\sqrt{E}}
+\xi 
\frac{\sqrt{2}\cosh(2\pi \alpha\sqrt{E})}{\pi\sqrt{E}}\ ,
\end{equation}
which for $\xi=0$ agrees with the undeformed matrix model result (\ref{eq:232}) after taking into account the factor of $2\sqrt{2}$ in (\ref{eq:231}). Same as in the orientable case, the supergravity calculation breaks down when $\xi<-1$ and $\varrho_0(E)$ becomes negative. The constant term in the first line of (\ref{eq:240}) contributes to the spectral density with a subleading (order one) Dirac delta at $E=0$

On the other hand, the random matrix model is well defined for arbitrary values of $\xi$. Using the method of orthogonal polynomials, the leading eigenvalue spectral density $\rho_0(E)$ is given by (\ref{eq:157}), which is explicitly non-negative. To match with the supergravity spectral density in (\ref{eq:233}) one finds the parameters $(\mu,t_k)$ (which entirely define the double scaled model) are given by
\begin{equation}
(\mu,t_k)=\left(1+\xi,\frac{\pi^{2k}}{k!^2}(1+\xi\alpha^{2k})\right)\ .
\end{equation}
This are the same values obtained for the Type 0B theory (\ref{eq:91}) and the Type 0A (\ref{eq:156}) up to an overall factor of~$\sqrt{2}$. The method of orthogonal polynomials applied to (\ref{eq:234}) yields exactly the same results as for the Type 0A case, summarized at the beginning of section \ref{sec:3.2}. The only difference is that the string equation (\ref{eq:210}) gets an additional contribution from $\Gamma$, as given in (\ref{eq:220}). Using the methods described in this work, it is straightforward to compute the observables of this model for arbitrary values of $\xi \in \mathbb{R}$, obtaining a phase diagram analogous to the one sketched in figure \ref{fig:11}.

\pagebreak

\addcontentsline{toc}{section}{References}
\bibliography{sample}
\bibliographystyle{JHEP}

\end{document}